\documentclass[final,3p,times]{elsarticle}

\usepackage{graphicx}
\usepackage{amssymb}

\usepackage[english,francais]{babel}

\usepackage[utf8]{inputenc}
\usepackage[T1]{fontenc}
\usepackage{amsmath}
\usepackage{color}
\usepackage{hyperref}
\usepackage{eqnarray}
\hypersetup{breaklinks=true}
\hypersetup{colorlinks=true}
\hypersetup{urlcolor=blue}
\hypersetup{citecolor=black}
\hypersetup{linkcolor=black}
\usepackage{cleveref}
\newcommand{\bea}{\begin{eqnarray}}
\newcommand{\eea}{\end{eqnarray}}
\newcommand{\be}{\begin{equation}}
\newcommand{\ee}{\end{equation}}
\newcommand{\rr}{\mathbf{r}}
\newcommand{\kk}{\mathbf{k}}

\newcommand{\qq}{\mathbf{q}}

\newcommand{\ii}{\textrm{i}}
\newcommand{\eee}{\textrm{e}}
\newcommand{\dd}{\mathrm{d}}

\DeclareMathOperator\atan{atan}

\DeclareMathOperator\thf{th}

\DeclareMathOperator\sh{sh}
\DeclareMathOperator\re{Re}

\usepackage{float}

\newcommand{\vecl}{\overrightarrow}

\begin{document}

\begin{frontmatter}


% \author{Name\thanksref{label2}}
\selectlanguage{english}
\title{Phase spreading and temporal coherence of a pair-condensed Fermi gas at low temperature}

% If all authors are at the same address, the [label1] can be suppressed

\author{Yvan Castin}
%\ead{yvan.castin@lkb.ens.fr}

\address{Laboratoire Kastler Brossel, ENS-PSL, CNRS, Sorbonne Universit\'e and Coll\`ege de France, Paris, France}
% If your know the dates of reception, and acceptation you can put them now;
%    idem the name of the person presenting your article

%\medskip

%\begin{center}
%{\small Received *****; accepted after revision +++++}
%\end{center}

\begin{abstract}
\selectlanguage{english}
A condensate of pairs in an isolated, homogeneous, unpolarised, finite-size spin 1/2 tridimensional Fermi gas at an extremely low nonzero temperature, undergoes with time a phase change $\hat{\theta}(t)-\hat{\theta}(0)$ with a random component, due at least to the coupling to the thermal phonons of the gas. Thanks to the quantum second Josephson relation connecting $\mathrm{d}\hat{\theta}/\mathrm{d}t$ to the phonon mode occupation numbers, and to linearised kinetic equations giving the evolution of the occupation number fluctuations, we access the behaviour of the phase change variance $\mbox{Var}[\hat{\theta}(t)-\hat{\theta}(0)]$ at times much longer than the phonon collision time. The case where the phonon branch has a convex start is similar to the Bose gas case: the leading collisional processes are the Beliaev-Landau three-phonon processes, and the variance is the sum of a ballistic term $Ct^2$ and of a delayed diffusive term $2D(t-t_0)$, whose analytical expressions are given in the thermodynamic limit. The concave case is much more exotic. It is analysed at time scales much shorter than $T^{-9}$, allowing one to restrict to the 2 phonons $\to$ 2 phonons small-angle Landau-Khalatnikov processes. The total number of phonons is conserved and the phonon mean occupation numbers at equilibrium can exhibit a chemical potential $\mu_\phi < 0$, assumed to be isotropic. The phase change variance is then the sum of a ballistic term $Ct^2$, of a diffusive term $2 D t$, of exotic subsubleading terms $2A(\pi t)^{1/2} + 2B\ln(t^{1/2})$ and of a constant term. The analytic expression of the coefficients $C$, $A$ and $B$ is obtained, as well as the diverging leading behavior of $D$ and of the constant term when $\mu_\phi/k_B T\to 0$. For $\mu_\phi=0$, the variance sub-ballistic part becomes superdiffusive, of the form $d_0 t^{5/3}$, where $d_0$ is known exactly. For a nonzero infinitesimal $\mu_\phi/k_B T$, a law is found, that interpolates between the superdiffusive spreading and the diffusive spreading of the sub-ballistic part. As by-products, new results are obtained on the phonon Landau-Khalatnikov damping rate, in particularly for $\mu_\phi<0$.
\\
\vskip 0.25\baselineskip

\end{abstract} 
\end{frontmatter}

% now the Version fran¡aise abr»g»e, if it exists
\selectlanguage{english}
%\section*{Version fran\c{c}aise abr\'eg\'ee}
% Text of your Version fran¡aise abr»g»e here

\selectlanguage{english}
% main text
%\newpage

\section {Introduction: system and position of the problem} 
\label {sec:intro}

Consider a three-dimensional gas of $N$ spin-1/2 fermions, unpolarized ($N_\uparrow=N_\downarrow$), prepared at instant $ t = 0 $ at thermal equilibrium at temperature $ T $ in a cubic side box $ L $ with periodic boundary conditions. Opposite spin fermions interact in the $ s $ wave with a scattering length $ a $ and a negligible range (contact interaction); fermions of the same spin do not interact. It is known that below a critical temperature $ T_c $, a condensate of $\uparrow\downarrow$ bound pairs of fermions is formed in mode $\phi_0(\rr_1-\rr_2)$ of the pair field; we assume here that $T\ll T_c$. At subsequent instants $ t> 0 $, the gas as a whole is isolated, that is it evolves freely under the effect of interactions without coupling to the environment. On the other hand, the pair condensate is not isolated and interacts with the non condensed modes, populated by thermal excitations, and which form a dephasing bath. We ask what is the corresponding coherence time $ t_c $ of the condensate, that is the width of its coherence function $\langle\hat{a}_0^\dagger(t)\hat{a}_0(0)\rangle$, where $ \hat {a} _0 $ annihilates a condensed pair {and the mean is taken in the system density operator}. As expected, $ t_c $ is the time after which the condensate lost the memory of its initial phase, the variance of the phase shift $\hat{\theta}(t)-\hat{\theta}(0)$ experienced by the condensate being of unit order \cite {CRAS2016}. Here $ \hat {\theta} $ is the condensate phase operator \cite {PRA2013} such that $\hat{a}_0=\eee^{\ii\hat{\theta}}\hat{N}_0^{1/2}$, $ \hat {N} _0 $ being the number of condensed pairs operator.

This problem is fundamental, but by no means academic. It is known how to prepare in the laboratory pair-condensed cold atomic Fermi gases \cite {manipsgen0, manipsgen1, manipsgen2, manipsgen3} in flat-bottom trapping potentials \cite {buxida1, buxida2}. Since the traps used are immaterial, these gases are well isolated and, although interacting, show only small particle losses due to three-body recombinations \cite{PRLdePetrovCSGora}. A Ramsey type interferometric method for measuring the coherence function $\langle\hat{a}_0^\dagger(t)\hat{a}_0(0)\rangle$ \cite {livreModugno} is being implemented on bosonic atom gases \cite {Oberthalerperso}; it can be transposed to the case of fermions by reversible adiabatic bosonization of the atomic Cooper pairs \cite {CRAS2016} (they are transformed into strongly bound dimers by a magnetic field ramp near a Feshbach resonance \cite {Ketterlevortex}). Our question could soon receive an experimental answer. To our knowledge, however, there is no complete theoretical prediction, the study of the spreading of the phase of the condensate of pairs in references \cite {CRAS2016, theseHadrien} being limited to the values of $ k_Fa $ such that the the sound dispersion relation in the superfluid has a convex start, that is $1/k_Fa>-0.14$ from the approximate calculations (Anderson's RPA) of reference \cite {PRA2016} ($k_F=(3\pi^2\rho)^{1/3}$ is the Fermi wave number of the gas of total density $\rho$). This convex case is very similar to that of the weakly interacting bosons, the variance of the phase shift having in general at long times a dominant ballistic term ($ \propto t ^ 2 $) and a retarded diffusive subdominant term ($ \propto t -t_0 $). Phase spreading in the case where the sound dispersion relation has a concave start, which includes the BCS limit of the Fermi gas ($k_Fa\to0^-$) \cite {Strinati}, seemed totally unexplored. The present work fills this gap; it predicts sub-ballistic terms that are more exotic or even grow faster than a simple retarded phase diffusion. The concave case thus hid a new and quite unexpected regime of spreading of the condensate phase. \footnote {This concave case could also in principle be realized with bosonic atoms in the weakly interacting regime, if it was known how to create a short-range interaction potential of width of order the healing length $\xi$ of the condensate \cite {Annalen2017}.} 

{The structure of the paper is simple. We specify in section \ref {sec:regime} the low-temperature regime considered and give a brief reminder of the formalism based on kinetic equations developed over the years by the author and his collaborators, and giving access to the variance of the phase shift $\hat{\theta}(t)-\hat{\theta}(0)$ experienced by the condensate. We treat the case of a convex phonon branch in section \ref {sec:convex}, quite briefly since it is enough to adapt the results of reference \cite {PRAYvanEmiliaAlice} on bosons. The unexplored case of a concave phonon branch is the subject of the long section \ref {sec:concave}: after sorting out the different phononic collision processes at low temperature $ T $ (\S \ref {subsec:edtc}), given the considered very short time scale compared to $ \hbar (mc ^ 2) ^ 8 (k_B T) ^ {- 9} $ ($ m $ is the mass of a fermion and $ c $ the speed of sound at zero temperature), we treat separately the ballistic spreading term of the phase (\S \ref {subsec:sltb}) and the sub-ballistic terms (\S \ref {subsec:edtsb}). The latter have required notable efforts: after writing kinetic equations for four-phonon collisions (\S \ref {subsubsec:lec}), we perform a simple approximation called ``rate approximation"  on their linearized form, of which it neglects non-diagonal terms, and which predicts various exotic contributions to the variance of $\hat{\theta}(t)-\hat{\theta}(0)$, with a non-integer-power or logarithmic law (\S \ref {subsubsec:edvdadt}); we establish what part of truth these approximate predictions contain, returning to the exact linearized kinetic equations (\S \ref {subsubsec:edvslfe}). This study also allowed us to obtain new, much more manageable expressions of the four-phonon Landau-Khalatnikov damping rate, in the form of a double (\ref {eq:036}) then single integral (\ref {eq:086}), instead of a quadruple integral in references \cite {Annalen2017, EPL}. We conclude in section \ref {sec:conclusion}.} 

\section {Regime considered and basic formalism} 
\label {sec:regime}

We are here at a temperature that is extremely low, but not zero. \footnote {{At zero temperature, there is no elementary or collective excitation present in the gas, so no dephasing environment for the condensate since the system is isolated. At a fixed total number $ N $ of particles, our formalism predicts that condensate coherence time is infinite. This is in agreement with a result obtained by Beliaev in the bosonic case \cite {Beliaevcoh}. The fact that the system is not a pure condensate in its ground state, i.e.\ that the condensate is subject to quantum depletion, does not change the case, since the uncondensed fraction does not constitute an excited component from the point of view of the interacting gas (even if it is one in the bosonic case from the point of view of the ideal gas). When $ N $ varies from one realization of the experiment to another, one recovers after average on the realizations the ballistic spreading of the condensate phase predicted for a gas of bosons at zero temperature \cite {Walls,Lewenstein}, see our note \ref {note:flucN}.}} This makes it possible to neglect the gas excitations by breaking pairs, which have a forbidden energy band $ \Delta $ and therefore have an exponentially low density $O(\eee^{-\Delta/k_B T}T^{1/2})$ {for $ k_B T \ll \Delta $}. The pair condensate is then coupled only to a low-density thermal gas of sound excitations (phonons) of energy dispersion relation $\qq\mapsto\epsilon_\qq$, where $ \qq $ is the wave vector, and the equation of evolution of its phase operator \cite {CRAS2016} is reduced (after a temporal smoothing suppressing the fast oscillations with the typical thermal pulsation $\epsilon_{q_{\rm th}}/\hbar\approx k_BT/\hbar$) to 
%alice001 
\be
-\frac{\hbar}{2}\frac{\dd\hat{\theta}}{\dd t} = \mu_0(N) + \sum_{\qq\neq\mathbf{0}} \frac{\dd\epsilon_\qq}{\dd N} \hat{n}_\qq
\label{eq:001}
\ee
{Here} $ \hat {n} _ \qq $ is the number of phonons operator in mode $ \qq $ and $ \mu_0 $ the chemical potential of the gas of $ N $ fermions at zero temperature. {Equation (\ref {eq:001}) has a simple physical interpretation \cite {CRAS2016}: its right-hand side constitutes a microcanonical chemical potential operator at sufficiently low temperature, the second contribution, thermal correction to the first, being the isentropic adiabatic derivative (here at fixed $ \hat {n} _ \qq $) of the phonon energy operator $ \sum_ {\qq \neq \mathbf {0}} \epsilon_ \qq \hat {n} _ \qq $ with respect to the total number of particles $ N $. Equation (\ref {eq:001}) is thus an operatorial quantum generalization of the second Josephson relation relating the time derivative of the order parameter of the gas to its chemical potential. The simplest and most general method for obtaining it uses quantum hydrodynamics, see Annex B of reference \cite {CRAS2016}. We first write the equations of the motion of the phase $ \hat {\phi} (\rr, t) $ and density $ \hat {\rho} (\rr, t) $ field operators of the superfluid gas. In the low temperature limit $ k_B T \ll \Delta, mc ^ 2, \epsilon _ {\rm F} $, where $ \epsilon _ {\rm F} $ is the Fermi energy and $ c> 0 $ is the speed of sound at zero temperature, the spatial fluctuations of density $ \delta \hat {\rho} (\rr, t) $ and phase $ \delta \hat {\phi} (\rr, t) $ are weak, which allows to linearize these equations around the uniform solution $ \hat {\rho} _0 (t) = \hat {N} / L ^ 3 $ and $ \hat {\phi} _0 (t) = \hat {\theta} (t) / 2 $ ($ \hat {N} $ is the total number of particles operator). We then expand $ \delta \hat {\rho} (\rr, t) $ and $ \delta \hat {\phi} (\rr, t) $ on the eigenmodes of the linearized equations, with as coefficients the usual annihilation $ \hat {b} _ \qq $ and creation $ \hat {b} _ \qq ^ \dagger $ operators of phonons of wave vector $ \qq $. After injecting these modal expansions into the expression of $ \dd \hat {\theta} / \dd t $ written to second order in $ \delta \hat {\rho} (\rr, t) $ and $ \mathbf {grad} \, \delta \hat {\phi} (\rr, t) $, we find the diagonal terms $ \propto \hat {n} _ \qq = \hat {b} _ \qq ^ \dagger \hat {b} _ \qq $ of equation (\ref {eq:001}), and non-diagonal terms $ \propto \hat {b} _ \qq \hat {b} _ {- \qq} $ and $ \propto \hat {b}^ \dagger_ \qq \hat {b} ^ \dagger _ {- \qq} $. The non diagonal terms oscillate rapidly on the scale of the collision time between phonons; they can therefore be removed by temporal smoothing, as was done in (\ref {eq:001}), without affecting the spreading of the condensate phase at long times (see Appendix E of reference \cite{justif_lissage})}. 
Finally, the temperature is sufficiently low ($k_BT\ll mc^2, k_B T\ll |\gamma|^{1/2}mc^2$) so that we can ignore the terms other than linear and cubic in the low wavenumber expansion of $ \epsilon_ \qq $: 
%alice002 
\be
\epsilon_\qq\underset{q\to 0}{=}\hbar c q \left[1+\frac{\gamma}{8}\left(\frac{\hbar q}{mc}\right)^2+O(q^4\ln q)\right]
\label{eq:002}
\ee
where the curvature parameter $ \gamma $ is dimensionless. 
{When $ \gamma> 0 $, the phonon branch has a convex start; when $ \gamma <0 $, it has a concave start.} While $ c $ is deduced from the equation of state ($mc^2=\rho\dd\mu_0/\dd\rho$ {as predicted by quantum hydrodynamics}), $ \gamma $ must be measured \cite{He4liq1, He4liq2} or deduced from an exact microscopic theory, which has not yet been done in cold atomic gases {(predictions \cite {PRA2016} of Anderson's RPA are only tentative)}. 

The calculation of the condensate phase spreading is reduced, through equation (\ref {eq:001}), to a pure problem of phonon dynamics. Since the system is in a stationary state, the variance of the phase shift for a time $ t $ is \cite {PRAYvanEmiliaAlice} 
%alice003 
\be
\mbox{Var}[\hat{\theta}(t)-\hat{\theta}(0)]= 2\re\int_0^t\dd\tau\, (t-\tau)C(\tau)
\label{eq:003}
\ee
where $ C (\tau) $ is the correlation function of the rate of change of the phase: 
%alice004 
\be
C(\tau)=\langle\frac{\dd\hat{\theta}}{\dd t}(\tau)\frac{\dd\hat{\theta}}{\dd t}(0)\rangle-\langle\frac{\dd\hat{\theta}}{\dd t}\rangle^2
\label{eq:004}
\ee
the mean $ \langle \ldots \rangle $ being taken in the system density operator. The knowledge of the correlation function $ C (\tau) $ is thus reduced to those $Q_{\qq\qq'}(\tau)=\langle\delta\hat{n}_\qq(\tau)\delta\hat{n}_{\qq'}(0)\rangle$ of the fluctuations $ \delta \hat {n} _ \qq $ of phonon numbers around their mean value $ \bar {n} _ \qq $. These latter are given by kinetic equations describing the unceasing redistribution (of quanta) between the modes of phonons by collision between them, and that one can here linearize to obtain 
%alice005 
\be
\frac{\dd}{\dd t} \delta n_\qq = \sum_{\qq'} M_{\qq\qq'} \delta n_{\qq'}
\label{eq:005}
\ee
{This linearized form inherits the good properties of the complete non-linear kinetic equations: it rests on a solid microscopic ground incorporating the Beliaev-Landau 1 phonon $ \leftrightarrow $ 2 phonons collisions (convex phonon branch) or the Landau-Khalatnikov 2 phonons $ \to $ 2 phonons collisions (concave branch), it describes the exponential-in-time relaxation of the average occupation numbers towards a steady state, thus the thermalization, and obeys strict conservation of the total energy $ E $. In the concave case, it also conserves the total number of phonons $ N_ \phi $. The matrix $ M $ admits an eigenvector of eigenvalue zero for each quantity conserved by the complete kinetic equations, i.e.\ the dimension of its kernel is equal to the number of constants of motion. In the sector of zero angular momentum that matters for phase spreading, the kernel of $ M $ is therefore of dimension one in the convex case and of dimension two in the concave case. These eigenvectors of zero eigenvalue play an important role in the rest of the study: they are associated with undamped linear combinations of $ \delta n_ \qq $ that thereby induce a ballistic spreading of the condensate phase when the corresponding conserved quantities $ E $ (convex case) or $ (E, N_ \phi) $ (concave case) have non-zero variance in the system density operator, i.e.\ vary from one realisation of the experiment to the other.} 

{It remains to solve the linearized kinetic equations (\ref {eq:005}), in their explicit form (\ref {eq:133}) in the convex case and (\ref {eq:034}) in the concave case.} Matrix and vector relations \cite {PRAYvanEmiliaAlice} 
%alice006 
\be
Q(\tau)=\eee^{M\tau} Q(0) \ \ \mbox{and}\ \ C(\tau)=\vec{A}\cdot Q(\tau)\vec{A}
\label{eq:006}
\ee
give the formal solution of the problem; we have introduced here the matrices $M$ and $Q$ of components $M_{\qq\qq'}$ and $Q_{\qq\qq'}$, and the vector $ \vec {A} $ of components 
%alice149 
\be
\label{eq:149}
A_\qq=\frac{2}{\hbar}\frac{\dd\epsilon_\qq}{\dd N}
\ee
{Notice in passing that $ \vec {A} $ lives in the sector of zero angular momentum (its components depend only on the modulus of $ \qq $ and are therefore isotropic), which will subsequently limit the study of $ M $ to this same sector.} 
But the challenge of our work is to obtain explicit results on phase spreading exact to leading order in temperature and in the thermodynamic limit $ N \to + \infty $ for fixed $ \rho $, the temperature being sufficiently low for the interactions between phonons to be well described by quantum hydrodynamics. 

\section {Convex case} 
\label {sec:convex} 

\begin {figure} [t] 
\centerline {\includegraphics [width = 0.4 \textwidth, clip =] {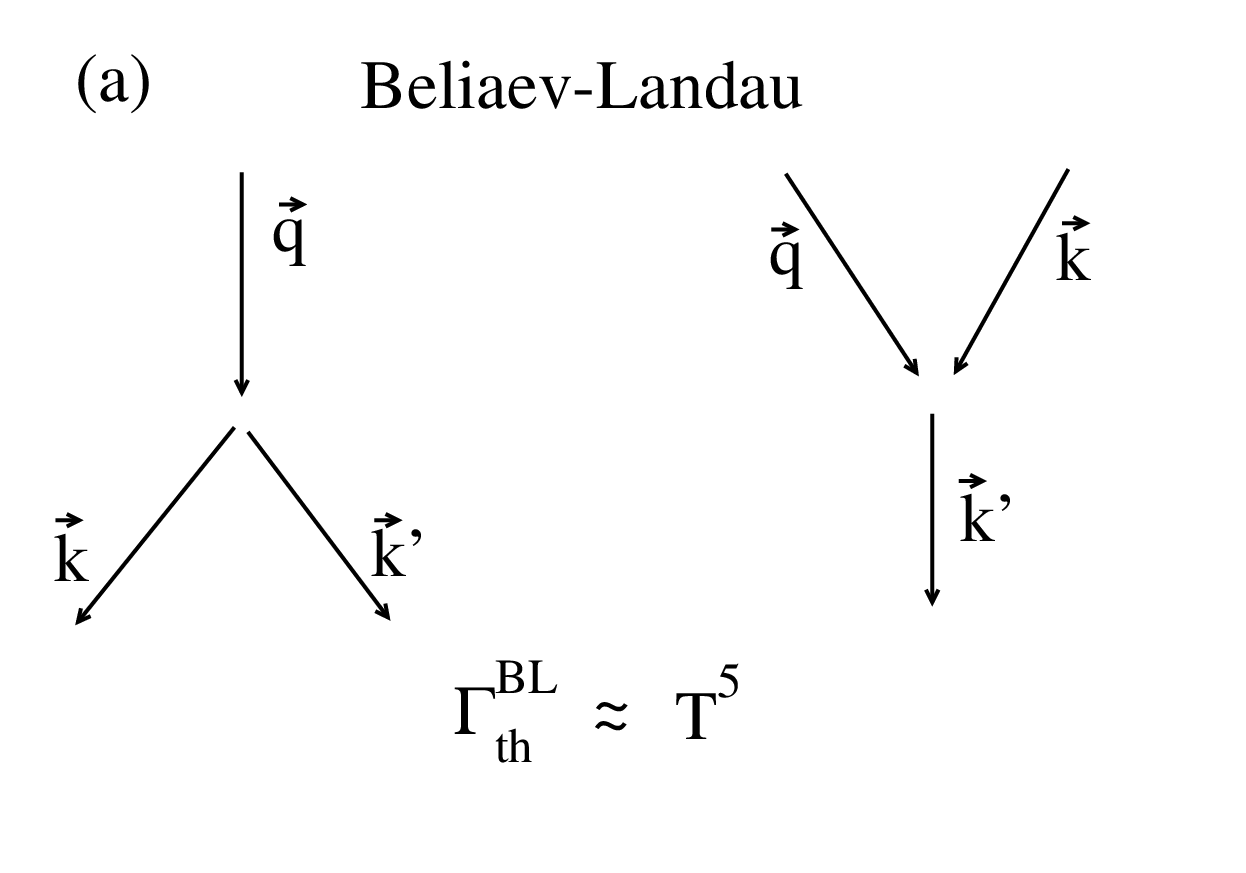} \hspace {1cm} \includegraphics [width = 0.5 \textwidth, clip =] {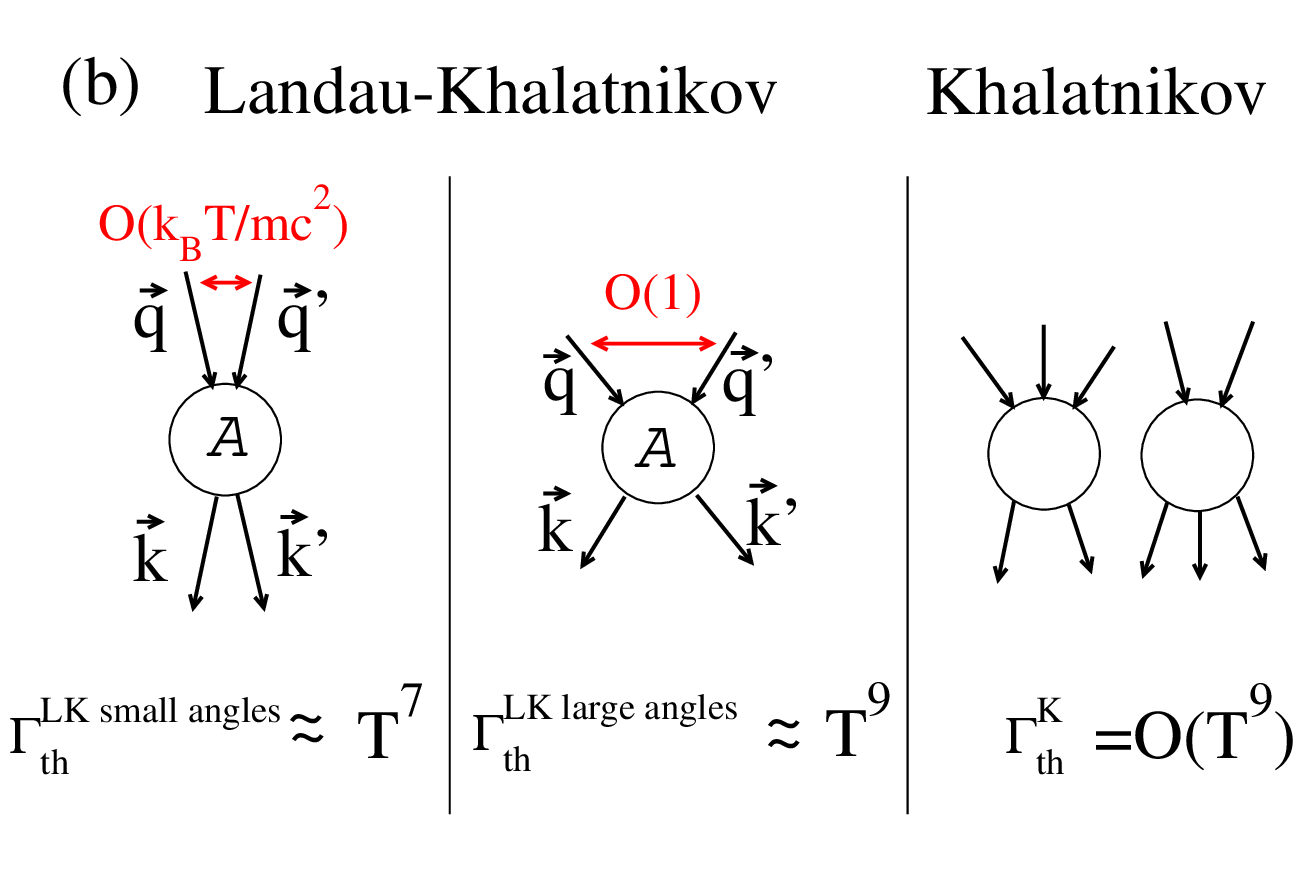} } 
\caption {The dominant interaction processes between low temperature phonons and the scaling laws with temperature of the corresponding decay rate $ \Gamma_ \qq $ of the thermal phonons of wave number $q\approx k_B T/\hbar c$. (a) Case $ \gamma> 0 $ in equation (\ref {eq:002}) (convex starting branch): Beliaev and Landau's three-phonon process ($ T^5 $ law). (b) Case $ \gamma <0 $ (concave starting branch): the low-angle four-phonon (2 phonons $\to$ 2 phonons) processes of Landau-Khalatnikov (angles $O(k_BT/mc^2)$ between the four wave vectors put into play) outweigh ($T^7$ law) those with large angles ($T^9$ law) and those of Khalatnikov with five phonons (2 phonons $\leftrightarrow$ 3 phonons) or those of Beliaev-Landau with three phonons ($ O(T^9)$ law, see note \ref{note:clari}); they are the only ones to be taken into account here. The temperature laws for the typical thermal decay rate are given here in a non-dimensional form, $\Gamma _ {\rm th} \approx T ^ \alpha $. In the convex case, this means $\hbar\Gamma_{\rm th} \approx (mc ^ 2) ^ {- 5/2} \epsilon _ {\rm F}^{-3/2} (k_B T)^5$ where $\epsilon_{\rm F}$ is the Fermi energy of the gas \cite {Annalen2017}; in the concave case, where we necessarily have $ \epsilon _ {\rm F} \approx mc ^ 2 $ \cite {PRA2016}, this means $\hbar\Gamma_{\rm th} \approx (mc ^ 2) ^ {1- \alpha} (k_B T) ^ {\alpha} $.} 
\label {fig:1} 
\end {figure}

When the phonon branch has a convex start ({curvature parameter} $ \gamma> 0 $ {in equation (\ref {eq:002})}) the dominant low-temperature collision processes are those of Beliaev and Landau's three phonons, $\qq\to\kk+\kk'$ and $\qq+\kk\to\kk'$ (see figure \ref {fig:1}a) \cite {BelLan0, BelLan1, BelLan2}. They do not conserve the number of quasi-particles. The phonon gas has therefore a zero chemical potential at equilibrium, 
%alice007 
\be
\bar{n}_\qq = \frac{1}{\eee^{\beta\epsilon_\qq}-1}
\label{eq:007}
\ee
with $ \beta = 1 / k_BT $. This is the same situation as the low-temperature limit of the weakly interacting Bose gas already discussed in reference \cite {PRAYvanEmiliaAlice}, up to global factors in the $ A_ \qq $ (the curvature parameter $\gamma $ depends on $ N $ in the case of fermions, while it is given by $ \gamma = $ 1 in the Bogoliubov spectrum) and on the amplitude of the Beliaev-Landau coupling. A direct transposition of the results of reference \cite {PRAYvanEmiliaAlice} to the case of fermions gives, at the thermodynamic limit and at times long compared to the inverse of the typical collision rate $ \gamma _ {\rm coll} $ between phonons, the usual ballistic-diffusive law \footnote {\label {note:flucN} If $ N $ has normal fluctuations, the coefficient of the ballistic term $ 4t ^ 2 / \hbar ^ 2 $ is worth $\langle[\delta N \partial_N\mu_{\rm mc}(\bar{E},\bar{N})+\delta E\partial_E\mu_{\rm mc}(\bar{E},\bar{N})]^2\rangle$ with $\delta N=N-\bar{N}$ and $\delta E=E-\bar{E}$. In contrast, $ D $ and $ t_0 $ remain unchanged (at the dominant order at the thermodynamic limit).} 
%alice008 
\be
\boxed{
\mbox{Var}[\hat{\theta}(t)-\hat{\theta}(0)] \underset{\gamma_{\rm coll}t\gg 1}{=} \left[\partial_E\mu_{\rm mc}(\bar{E},N)\right]^2 (\mbox{Var}\,E) \frac{4 t^2}{\hbar^2} +2D(t-t_0)+o(1)
}
\label{eq:008}
\ee
with $\mu_{\rm mc}(E,N)$ the microcanonical chemical potential of a gas of $ N $ fermions of total energy $ E $, $ \bar {E} $ being the average of the total energy and $ \mbox {Var} \, E $ the variance supposed to be normal ($\mbox{Var}\,E=O(N)$) of its fluctuations, $ D $ is the diffusion coefficient of the phase and $ t_0 $ the delay time to diffusion. The same transposition leads to the equivalents at low temperature in the canonical ensemble ($\mbox{Var}_{\rm can}E=k_B T^2\partial_T\bar{E}$): 
%alice015 
\bea
\label{eq:015a}
[\partial_E\mu_{\rm mc}(\bar{E},N)]^2\mbox{Var}_{\rm can} E &\underset{T\to 0}{\sim}& \frac{2\pi^2}{15} \frac{u^2}{N} 
\frac{(k_B T)^5}{(\hbar c)^3\rho}\\
\label{eq:015b}
\frac{\hbar N D}{mc^2} &\underset{T\to 0}{\sim}&  c_1 \frac{9\gamma^2(2u-\frac{\partial\ln\gamma}{\partial\ln\rho})^2}{(1+u)^2} \left(\frac{k_B T}{mc^2}\right)^4 \\
\frac{m c^2 t_0}{\hbar} &\underset{T\to 0}{\sim}& \frac{-9 c_2}{16\sqrt{2}c_1} \frac{1}{(1+u)^2} \left(\frac{\hbar\rho^{1/3}}{mc}\right)^3
\left(\frac{k_B T}{mc^2}\right)^{-5}
\label{eq:015c}
\eea
The constants $ c_1 $ and $ c_2 $ are those of reference \cite {PRAYvanEmiliaAlice} ($ c_1 \simeq 0.3036 $, $ c_2 \simeq-0.2033 $), {$ \rho $ is the total density} and $u=\partial\ln c/\partial\ln\rho$ the Gr\"uneisen parameter of the Fermi gas. The result (\ref {eq:015a}) reproduces that of reference \cite {PRAsuperdiff} in the particular case of a weakly interacting Bose gas. The equivalent (\ref {eq:015b}) generalizes the one (specific to the unitary limit $ 1 / a = 0 $) of reference \cite {CRAS2016} and reproduces the one (VIII.59) of reference \cite {theseHadrien} in a slightly different form. Remarkably, the equivalent (\ref {eq:015c}) of the delay time $ t_0 $ coincides with that of the Beliaev-Landau damping time of the thermal phonons $ q = k_B T / \hbar c $ of reference \cite {Annalen2017} up to a numerical factor. The results (\ref {eq:015b}, \ref {eq:015c}) are valid for a large system prepared in any statistical ensemble with normal fluctuations of $ E $ and $ N $ \cite {PRAYvanEmiliaAlice}. {To be complete, we give the expression used at low temperature of the linearized kinetic equations on the isotropic fluctuations $ \delta n_ \qq = \delta n_q $ of the occupation numbers at the thermodynamic limit: 
%alice133 
\begin{multline}
\label{eq:133}
\frac{\dd}{\dd t} \delta n_q = \frac{(1+u)^2\hbar}{4\pi m\rho}\left\{-\delta n_q \left[\int_0^{+\infty}\! \dd k\, k^2 (k+q)^2 (\bar{n}_k^{\ell}-\bar{n}_{k+q}^{\ell}) + \int_0^q \!\dd k\, k^2(q-k)^2 (\bar{n}_k^{\ell}+1/2)\right]\right. \\
\left.
+\int_0^{+\infty}\! \dd k\, k^2(k+q)^2(\bar{n}_{k+q}^\ell-\bar{n}_q^\ell)\delta n_k 
+\int_0^q\!\dd k\, k^2(q-k)^2 (\bar{n}_{q-k}^\ell-\bar{n}_q^\ell)\delta n_k 
+\int_q^{+\infty}\!\dd k\, k^2 (k-q)^2 (1+\bar{n}_{k-q}^\ell+\bar{n}_q^\ell) \delta n_k
\right\}
\end{multline}
where the average thermal values $ \bar {n} _q ^ {\ell} = [\exp (\hbar cq / k_B T) -1] ^ {- 1} $ are those of a linear phonon branch. 
} 

It is useful for the following to recall, in the context of linearized kinetic equations (\ref {eq:005}), where the ballistic term of equation (\ref {eq:008}) comes from \cite { PRAYvanEmiliaAlice}. Due to conservation of the total phonon energy, $\vec{\epsilon}\cdot M=0$ ($ \vec {\epsilon} $ is the vector with coordinates $ \epsilon_ \qq $), i.e.\ $ \vec {\epsilon } $ is a left eigenvector of $ M $ with the eigenvalue zero. According to a result of elementary linear algebra, $ M $ then admits a right eigenvector $ \vec {d} _ \epsilon $ of zero eigenvalue, it is the dual vector of $ \vec {\epsilon} $. Here, it is easy to identify: even if we shift $ \beta $ by $ \delta \beta $ from its physical value $ 1 / k_B T $, the expression (\ref {eq:007}) remains stationary, therefore the corresponding variations $\delta \bar{n}_\qq$ to first order in $ \delta \beta $ must be a stationary solution of the linearized kinetic equations, hence $(d_\epsilon)_\qq\propto-\partial_\beta\bar{n}_\qq=\epsilon_\qq \bar{n}_\qq(1+\bar{n}_\qq)$ up to a normalization factor (we must have $\vec{\epsilon}\cdot\vec{d}_\epsilon=1$) \cite {PRAYvanEmiliaAlice}. In general, the vector $ Q (0) \vec {A} $ admits {a non-zero coefficient $ \vec {\epsilon} \cdot Q (0) \vec {A} $} on $ \vec {d} _ \epsilon $ so that $ C (\tau) $ has a non-zero limit when $ \tau \to + \infty $ and $\mbox{Var}[\hat{\theta}(t)-\hat{\theta}(0)]$ diverges quadratically in time. On the other hand, in the microcanonical ensemble, $ \mbox {Var} \, E = 0 $, all physical fluctuations $ \delta n_ \qq $ are of zero total energy $\sum_\qq \epsilon_\qq \delta n_\qq=0$ and $ \vec {\epsilon} \cdot Q (0) \equiv 0 $: the spreading of the phase is diffusive. 

\section {Concave case} 
\label {sec:concave} 

When the branch of phonons has a concave start ({curvature parameter} $ \gamma <0 $ {in equation (\ref {eq:002})}), the Beliaev-Landau 1 phonon $ \leftrightarrow $ 2 phonons processes no longer conserve energy. The same goes for any process 1 phonon $ \leftrightarrow n $ phonons, $ n> 2 $. The dominant processes are then those of Landau-Khalatnikov with four phonons, 2 phonons $ \to $ 2 phonons, whose effective coupling amplitude $\mathcal{A}(\qq,\qq';\kk,\kk')$ has a known expression in the context of quantum hydrodynamics (a diagram at first order and six diagrams at second-order perturbation theory) \cite {LK1949, Annalen2017}. The corresponding thermal collision rate $\Gamma_{\rm th}^{\rm LK}$, however, is multi-scale in temperature: it varies as $ T ^ 7 $ for collisions at small angles ($ \qq, \qq ', \kk, \kk' $ collinear and in the same direction within angles $O(k_BT/mc^2)$) and as $ T ^ 9 $ for large-angle collisions (see figure \ref {fig:1}b). Here, we study the phase-spreading of the condensate of pairs at the time scale of small angle collisions (\S \ref {subsec:edtc}). While the ballistic term in the variance of $\hat{\theta}(t)-\hat{\theta}(0)$ can be cleverly deduced from the existence of two conserved quantities, the energy $ E_ \phi $ of the phonons and their total number $ N_ \phi $ (\S \ref {subsec:sltb}), the computation of the sub-ballistic contribution requires the explicit writing of the kinetic equations for the four-phonon collisions and the solution of their linearized form around equilibrium (\S \ref {subsec:edtsb}).  

\subsection {Time scale considered} 
\label {subsec:edtc} 

{In this work}, we study the spreading of the condensate phase on a time scale {$ t \approx \hbar (mc ^ 2) ^ 6 (k_B T) ^ {- 7} $} and neglect the collisional processes between phonons appearing at time scales {$ \approx \hbar (mc ^ 2) ^ 8 (k_B T) ^ {- 9} $} or later. In other words, we only take into account small-angle $\qq+\qq'\to\kk+\kk'$ collisions {(see figure \ref {fig:1}b)}. We can not describe either the relaxation of the direction of the wave vectors to an isotropic distribution (the small-angle processes negligibly change the direction \footnote {If each small-angle collision changes the direction of $ \qq $ by a random angle $\delta\theta\approx k_BT/mc^2$, we need $n_{\rm coll}\approx 1/\mbox{Var}\,\delta\theta \approx T^{-2}$ successive collisions (supposedly independent) to thermalize the direction of $ \qq $, which takes the same time (for the same effect) as a single collision at large angles.}) nor the relaxation of the total number of phonons $ N_ \phi $ or more conveniently of the chemical potential $ \mu_ \phi $ phonons to its true thermal equilibrium value (the Landau-Khalatnikov 2 phonons $ \to $ 2 phonons processes conserve $ N_ \phi $, in contrast to the Beliaev-Landau processes and the subdominant 5-phonon processes of figure \ref {fig:1}b). \footnote {\label {note:clari} For 2 phonons $ \leftrightarrow $ 3 phonons processes, of maximum rate at small angles, Khalatnikov finds a decay rate $\Gamma_{\rm th}^{\rm K}\approx T^{11}$. According to him, the total number of phonons $ N_ \phi $ would therefore heat up with a rate $ \propto T ^ {11} $ \cite {JETPKhalatnikov}. However, the Beliaev-Landau processes (which also change $ N_ \phi $) are no longer energetically forbidden at this order, since the Landau-Khalatnikov 2 phonons $ \to $ 2 phonons processes confer to the incoming and outgoing phonons a nonzero energy width $\Gamma_{\rm th}^{\rm LK}\approx T^7$. We find then that $\Gamma_{\rm th}^{\rm BL}\approx T^9$ according to equation (5.6) of reference \cite {phenomamer}. This points to a thermalization rate of $ N_ \phi $ of order $ T ^ 9 $. So we put in figure \ref {fig:1}b a $ O (T ^ 9) $, which suffices here.} At the time scale $ T ^ {- 7} $, there exist therefore apparently stationary distributions of phonons $ \bar {n} _ \qq $ that are anisotropic and of nonzero chemical potential $ \mu_ \phi $, contrary to equation (\ref {eq:007}) \cite {JETPKhalatnikov}. \footnote { One can simply have a chemical potential, or even a temperature, depending on the direction of the wave vector $ \qq $.} For simplicity, and because it is {\sl a priori} naturally the case in a cold atom experiment, we take isotropic mean occupation numbers for phonons, 
%alice009 
\be
\bar{n}_\qq = \frac{1}{\eee^{\beta(\epsilon_\qq-\mu_\phi)}-1}
\label{eq:009}
\ee
but without constraint on $ \mu_ \phi $ (other than $\mu_\phi\leq 0$). It will be convenient in the following to use the notation 
%alice150 
\be
\label{eq:150}
\nu=\beta\mu_\phi
\ee

Knowing the average phonon numbers (\ref {eq:009}) is not enough to determine the spreading of the condensate phase: as we see on equation (\ref {eq:006}), we need to know the covariance matrix $ Q (0) $ of the occupation numbers, so specify the density operator $ \hat {\rho} _ \phi $ of the phonons. For simplicity, we first assume that the unpolarized Fermi gas is prepared in a generalized canonical ensemble, defined by the fixed number $ N $ of fermions, temperature $ T $ and chemical potential $ \mu_ \phi $ of phonons (which sets their average number), hence the Gaussian phonon density operator 
%alice010 
\be
\hat{\rho}_\phi = \frac{1}{\Xi} \eee^{-\beta\sum_{\qq\neq\mathbf{0}}(\epsilon_\qq-\mu_\phi)\hat{n}_\qq} 
\label{eq:010}
\ee
where $ \Xi $ ensures that $ \hat {\rho} _ \phi $ has a unit trace. According to Wick's theorem, the covariance matrix $ Q (0) $ is then diagonal: 
%alice011 
\be
[Q(0)]_{\qq\qq'} \stackrel{\hat{\rho}_\phi\ \mbox{\scriptsize of }(\ref{eq:010})}{=} [Q_{\rm d}(0)]_{\qq\qq'} \equiv \bar{n}_\qq (1+\bar{n}_\qq) \delta_{\qq,\qq'}
\label{eq:011}
\ee
We will see in the following how to treat a more general case. 

\subsection {Get the ballistic term out} 
\label {subsec:sltb} 

Remarkably, we can obtain in the general case the ballistic term of $\mbox{Var}[\hat{\theta}(t)-\hat{\theta}(0)]$ at long times, without explicitly writing the kinetic equations for 2 phonons $ \to $ 2 phonons collisions. It suffices to exploit the existence of two quantities conserved {by temporal evolution} \cite {LeyronasChevy}, here the energy of phonons $E_\phi=\sum_{\qq\neq\mathbf{0}} \epsilon_\qq n_\qq$ and their number {total} $N_\phi=\sum_{\qq\neq\mathbf{0}} n_\qq$. Since the fluctuations $\delta E_\phi=E_\phi-\bar{E}_\phi$ and $\delta N_\phi=N_\phi-\bar{N}_\phi$ are also conserved, the matrix $ M $ of linearized kinetic equations must satisfy the conditions 
%alice016 
\be
\vec{\epsilon}\cdot M=0 \ \ \mbox{and}\ \ \vecl{\mbox{un}}\cdot M=0
\label{eq:016}
\ee
where $ \vec {\epsilon} $ is of coordinates $ \epsilon_ \qq $ as in section \ref {sec:convex} and $ \vecl {\mbox {un}} $ is of coordinates all equal to 1: 
%alice012 
\be
\vec{\epsilon}=(\epsilon_\qq)_{\qq\neq\mathbf{0}} \ \ \mbox{and}\ \ \vecl{\mbox{un}}=(1)_{\qq\neq\mathbf{0}}
\label{eq:012}
\ee
$ M $ thus has two left eigenvectors of zero eigenvalue, and these are the ``only" ones in the isotropic sector of zero angular momentum in which the source term $ Q (0) \vec {A} $ of equation (\ref {eq:006}) lives, since $ M $ exhibits no other conserved quantity: the ``other" isotropic left eigenvectors $ \vec {e} _m $ (linearly independent of $ \vec {\epsilon} $ and $ \vecl {\mbox {un}} $) correspond to eigenvalues {$ \Lambda_m $} of $ M $ with a real part $ <0 $, i.e.\ to exponentially damped kinetic modes {(we will see in \S \ref {subsubsec:edvslfe} that {$ \Lambda_m $} are real)}. We conclude that $ M $ has only two {isotropic} linearly independent right eigenvectors of zero eigenvalue. We find them as in section \ref {sec:convex} by noting that the expression (\ref {eq:009}) of $ \bar {n} _ \qq $ remains a stationary solution of the kinetic equations even if $ \beta $ and $ \nu = \beta \mu_ \phi $ are shifted by $ \delta \beta $ and $ \delta \nu $. The corresponding variations $ \delta \bar {n} _ \qq $ to first order in $ \delta \beta $ and $ \delta \nu $ must constitute a stationary solution of the linearized kinetic equations. As $\partial_\beta\bar{n}_\qq=-\epsilon_\qq\bar{n}_\qq(1+\bar{n}_\qq)$ and $\partial_\nu\bar{n}_\qq=\bar{n}_\qq(1+\bar{n}_\qq)$, we deduce that 
%alice013 
\be
M Q_{\rm d}(0)\vec{\epsilon}=\vec{0} \ \ \mbox{and}\ \ M Q_{\rm d}(0)\vecl{\mbox{un}}=\vec{0}
\label{eq:013}
\ee
In general, $ Q_d (0) \vec {A} $ has non-zero components with coefficients $ \lambda_ \epsilon $ and $ \lambda _ {\rm un} $ on these right eigenvectors of $ M $ of zero eigenvalue, so $\eee^{M\tau}Q_{\rm d}(0)\vec{A}$ and $ C (\tau) $ do not tend to zero at long times. On the other hand, by definition, if we subtract these components, we must obtain a zero limit: 
%alice014 
\be
\eee^{M\tau}[Q_{\rm d}(0)\vec{A}-\lambda_\epsilon Q_{\rm d}(0)\vec{\epsilon}-\lambda_{\rm un} Q_{\rm d}(0) \vecl{\mbox{un}}]\underset{\tau\to+\infty}{\to} 0
\label{eq:014}
\ee
By making the left scalar product of this expression with $ \vec {\epsilon} $ and $ \vecl {\mbox { a}} $, which owing to equation (\ref {eq:016}) gives zero, we end up with the invertible system 
%alice017 
\bea
\label{eq:017a}
\lambda_\epsilon \vec{\epsilon}\cdot Q_{\rm d}(0) \vec{\epsilon} + \lambda_{\rm un} \vec{\epsilon}\cdot Q_{\rm d}(0) \vecl{\mbox{un}} &=& \vec{\epsilon}\cdot Q_{\rm d}(0)\vec{A} \\
\label{eq:017b}
\lambda_\epsilon \vecl{\mbox{un}} \cdot Q_{\rm d}(0) \vec{\epsilon} + \lambda_{\rm un} \vecl{\mbox{un}} \cdot Q_{\rm d}(0) \vecl{\mbox{un}} &=& \vecl{\mbox{un}} \cdot Q_{\rm d}(0)\vec{A}
\eea
whose coefficients, let us note it, are variances or covariances ($\vec{\epsilon}\cdot Q_{\rm d}(0)\vec{\epsilon}=\mbox{Var}\, E_\phi$, etc). We have therefore separated the correlation function of $ \dd \hat {\theta} / \dd t $ into a constant contribution and a contribution of zero limit at long times: 
%alice018 
\be
C(\tau)=C(+\infty)+\Delta C(\tau) \underset{\tau\to+\infty}{\to} C(+\infty)
\label{eq:018}
\ee
which, according to equation (\ref {eq:003}), is equivalent to splitting the variance of the condensate phase shift into a ballistic term and a sub-ballistic term: 
%alice019 
\bea
\label{eq:019a}
\mbox{Var}[\hat{\theta}(t)-\hat{\theta}(0)] &=& C(+\infty) t^2 + \mbox{Var}_{\scriptsize\mbox{s-bal}}[\hat{\theta}(t)-\hat{\theta}(0)] \\
\label{eq:019b}
\mbox{Var}_{\scriptsize\mbox{s-bal}}[\hat{\theta}(t)-\hat{\theta}(0)] &=& 2 \re \int_0^t \dd\tau\, (t-\tau) \Delta C(\tau) \underset{t\to +\infty}{=} o(t^2)
\eea
The expression of the constant $ C (+ \infty) $ so obtained is not very suggestive: 
%alice020 
\be
\label{eq:020}
C(+\infty)=\lambda_\epsilon \vec{A}\cdot Q_{\rm d}(0) \vec{\epsilon} + \lambda_{\rm un} \vec{A}\cdot Q_{\rm d}(0) \vecl{\rm un}
\ee
but a long calculation allows one to put it in a form as illuminating as {in} equation (\ref {eq:008}): 
%alice021 
\be
\label{eq:021}
\boxed{
C(+\infty)=\mbox{Var}\left[(N_\phi-\bar{N}_\phi)\partial_{N_\phi} \frac{2\mu_{\rm mc}^\phi}{\hbar}(\bar{E}_\phi,\bar{N}_\phi)+(E_\phi-\bar{E}_\phi)\partial_{E_\phi} \frac{2\mu_{\rm mc}^\phi}{\hbar}(\bar{E}_\phi,\bar{N}_\phi)\right]
}
\ee
where $\mu_{\rm mc}^\phi(E_\phi,N_\phi)$ is the microcanonical chemical potential of an ideal gas of $ N_ \phi $ phonons of total energy $ E_ \phi $. \footnote {We use the fact that in the limit of a large system, the chemical potential no longer depends on the {statistical} ensemble, provided that the averages of $ E_ \phi $ and $ N_ \phi $ are fixed. Then 
%alice022 
\[
\frac{2}{\hbar}\mu_{\rm mc}^\phi(\bar{E}_\phi(\beta,\mu_\phi),\bar{N}_\phi(\beta,\mu_\phi)) \sim \sum_{\qq\neq\mathbf{0}} A_\qq\bar{n}_\qq(\beta,\mu_\phi)
\]
where the average values {$ \bar {E} _ \phi $, $ \bar {N} _ \phi $ and $ \bar {n} _{\qq} $} are taken in the {canonical} density operator (\ref {eq:010}). It remains to take the derivative of this relation with respect to $ \beta $ and $ \mu_ \phi $ to obtain $\partial_{N_\phi} \mu_{\rm mc}^\phi$ and $\partial_{E_\phi} \mu_{\rm mc}^\phi$. We find in particular that $\lambda_\epsilon=\partial_{E_\phi} \frac{2\mu_{\rm mc}^\phi}{\hbar}$ and $\lambda_{\rm un}=\partial_{N_\phi} \frac{2\mu_{\rm mc}^\phi}{\hbar}$.} The expression of the contribution of zero limit can be made nicely symmetrical thanks to equation (\ref {eq:016}): 
%alice023 
\be
\label{eq:023}
\Delta C(\tau)=\Delta\vec{A}\cdot \eee^{M\tau} Q_{\rm d}(0) \Delta\vec{A} \underset{\tau\to+\infty}{\to} 0
\ee
where 
%alice024 
\be
\label{eq:024}
\Delta\vec{A} = \vec{A}-\lambda_\epsilon \vec{\epsilon} -\lambda_{\rm un} \vecl{\rm un}
\ee
and also presents a simple physical interpretation: it is the correlation function of $ \dd \hat {\theta} / \dd t $ when the phonon gas is prepared in the ``average" microcanonical ensemble $(\bar{E}_\phi,\bar{N}_\phi)$. The vector $Q_{\rm d}(0)\Delta\vec{A}$ actually lives in the subspace of zero energy fluctuations and zero number of phonons fluctuations, see equations (\ref {eq:017a}, \ref {eq:017b}). The contribution $ \Delta C (\tau) $ and the sub-ballistic term (\ref {eq:019b}) are therefore, at the thermodynamic limit that we consider, independent of the (supposedly normal) fluctuations of $ N_ \phi $ and $ E_ \phi $ thus of the statistical ensemble in which the phonon gas is prepared, at fixed $ \bar {N} _ \phi $ and $ \bar {E} _ \phi $, contrarily to $ C (+ \infty) $ and to the ballistic term in (\ref {eq:019a}). This fact had already been established for a weakly-interacting Bose gas in reference \cite {PRAYvanEmiliaAlice}. The expression (\ref {eq:021}) of $ C (+ \infty) $, {though obtained for the Gaussian density operator (\ref {eq:010}),} actually applies to a gas with a fixed number $ N $ of fermions but prepared at the thermodynamic limit in any statistical ensemble (any statistical mixture of fixed $N_\uparrow=N_\downarrow=N/2$ microcanonical ensembles), provided that the fluctuations of the energy and the number of phonons remain normal (of variance $ O (N) $), that the momentum distribution of the phonons remains isotropic and that the temperature is sufficiently low. As in the reasoning of reference \cite {PRAsuperdiff}, this follows directly from the hypothesis of ergodization of the phonon gas by phonon-phonon collisions. \footnote {The right-hand side of equation (\ref {eq:001}) is, as we have seen, an approximation at low density of excitations of a chemical potential operator of the Fermi gas, the sum in the right-hand side being the adiabatic derivative (with fixed occupation number operators $ \hat {n} _ {\qq} $) {of the Hamiltonian of phononic excitations $ \hat {H} _ \phi $} \cite {CRAS2016}. Under the ergodic hypothesis in the presence of Landau-Khalatnikov 2 phonons $ \to $ 2 phonons collisions, the temporal mean of equation (\ref {eq:001}) at long times $\gamma_{\rm coll} t\to +\infty$ no longer depends on each occupation number individually, but on the constants of motion $\hat{H}_\phi=\sum_{\qq\neq\mathbf{0}} \epsilon_\qq \hat{n}_\qq$ and $\hat{N}_\phi=\sum_{\qq\neq\mathbf{0}} \hat{n}_\qq$ {through a microcanonical mean (eigenstate thermalisation hypothesis of the many-body eigenstates \cite {ETH1, ETH2, ETH3})} so that 
%alice025 
\[
\hat{\theta}(t)-\hat{\theta}(0) \underset{t\to+\infty}{\sim} -\left[\mu_0(N) +\mu_{\rm mc}^\phi (\hat{H}_\phi,\hat{N}_\phi)\right]\frac{2 t}{\hbar}
\]
It remains to linearize $\mu_{\rm mc}^\phi(\hat{H}_\phi,\hat{N}_\phi)$ in the fluctuations around the mean $ (\bar {E} _ \phi, \bar {N} _ \phi) $ and take the variance to find the result (\ref {eq:021}) given equation (\ref {eq:019a}). This approach would easily account for fluctuations in the total number of fermions (as long as $ N_ \uparrow = N_ \downarrow $).} 

\subsection {Study of the sub-ballistic term} 
\label {subsec:edtsb} 

We focus now on the sub-ballistic term (\ref {eq:019b}), more difficult and richer. Once the linearized kinetic equations are written for 2 phonons $ \to $ 2 phonons collisions at small angles (\S \ref {subsubsec:lec}), we derive the spreading of the condensate phase, first heuristically by replacing them with a diagonal form (rate approximation, \S\ref{subsubsec:edvdadt}) then quantitatively by returning to their exact form (\S\ref{subsubsec:edvslfe}).

\subsubsection{The kinetic equations} 
\label{subsubsec:lec}

We must {first} write the kinetic equations giving the {mean} evolution of the phonon occupation numbers $ n_ \qq $ under the effect of Landau-Khalatnikov 2 phonons $ \to $ 2 phonons collision processes. The corresponding effective interaction Hamiltonian between phonons, deduced from quantum hydrodynamics \cite {LK1949}, is written 
%alice026 
\be
\label{eq:026}
\hat{H}_{\rm eff}^{\rm LK} = \frac{1}{4L^3} \sum_{\qq_1,\qq_2,\qq_3,\qq_4} \mathcal{A}(\qq_1,\qq_2;\qq_3,\qq_4) \hat{b}^\dagger_{\qq_3}\hat{b}^\dagger_{\qq_4} \hat{b}_{\qq_1}\hat{b}_{\qq_2} \delta_{\qq_1+\qq_2,\qq_3+\qq_4}
\ee
where {$ \hat {b} _ \qq $ is the annihilation operator of a phonon in the mode $ \qq $ and} the real coupling amplitude $\mathcal {A}$ is known explicitly \cite {Annalen2017}. Since $ \mathcal {A} $ is invariant by exchanging its first two arguments and exchanging its last two arguments (phonons are bosons), we have pulled out the symmetry factor $1/4$. It is moreover invariant by exchange of the first two arguments with the last two, which ensures the hermiticity of $ \hat {H} _ {\rm eff} ^ {\rm LK} $ and the microreversibility of the collision processes. Let's consider the collision process $ \qq + \qq '\to \kk + \kk' $ of figure \ref {fig:1}b and the inverse process $ \kk + \kk '\to \qq + \qq' $, then let's calculate the incoming and outgoing fluxes in the $ \qq $ phonon mode by applying the Fermi golden rule to the Hamiltonian $ \hat {H} _ {\rm eff} ^ {\rm LK} $: 
%alice027 
\begin{multline}
\label{eq:027}
\frac{\dd}{\dd t} n_\qq = -\frac{1}{2} \sum_{\kk,\kk',\qq'} \frac{2\pi}{\hbar} \frac{[\mathcal{A}(\qq,\qq';\kk,\kk')]^2}{L^6} \delta_{\kk+\kk',\qq+\qq'}
\delta(\epsilon_\kk+\epsilon_{\kk'}-\epsilon_\qq-\epsilon_{\qq'}) \\
\times [n_\qq n_{\qq'} (1+n_\kk)(1+n_{\kk'}) -n_\kk n_{\kk'}(1+n_\qq)(1+n_{\qq'})]
\end{multline}
We recognize the energy-conserving Dirac $ \delta $ function, the Kronecker $ \delta $ of momentum conservation and the $ 1 + n $ bosonic amplification factors accompanying the addition of a phonon in an already populated mode. The sum over the wave vectors (other than $ \qq $) involved in the collision is assigned a factor of $1/2$ to avoid double counting of the final (or initial) state $ (\kk, \kk ') \equiv (\kk', \kk) $. Finally, the squared amplitude $ \mathcal {A}^2 $ is factorizable by microreversibility. After passage to thermodynamic limit \footnote {{In a typical cold atom experiment, the Landau-Khalatnikov damping rate of the phonons in the concave case is much lower than that of Beliaev-Landau in the convex case \cite { EPL}. Since the collisional width $ \hbar \Gamma_q $ of the phonons is greatly reduced, the discrete nature of their energy levels in the box potential, therefore the finite size effects, are more important. A condition of reaching the thermodynamic limit for Landau-Khalatnikov processes is given in note 5 of reference \cite {EPL}. The minimum box sizes required, of the order of 100 $ \mu $m, remain however feasible in the laboratory \cite {buxida1, buxida2}.}} and linearization of equation (\ref {eq:027}) around the isotropic stationary solution (\ref {eq:009}), restricting to isotropic fluctuations $ \delta n_q $ (in the sector of zero angular momentum) as {allowed by the structure} of (\ref {eq:006}) {and the rotational invariance of $ M $, and as prompted by the remark after (\ref {eq:149})}, we get 
\footnote {\label{note:ast} The expression is greatly simplified by a trick already used by Landau and Khalatnikov: $ (i) $ we eliminate the factors $ 1 + \bar {n} $ thanks to the relation $ 1 + \bar {n} = \eee ^ {\beta (\epsilon- \mu_ \phi)} \bar {n} $, $ (ii) $ we factorize as much as possible also using energy conservation, $ (iii) $ we make exponentials disappear by recognizing $ \eee ^ {\beta (\epsilon- \mu_ \phi)} - 1 = 1 / \bar {n} $ or using the relation inverse of that of $ (i) $.} 
%alice028 
\begin{multline}
\label{eq:028}
\frac{\dd}{\dd t} \delta n_q = -\frac{\pi}{\hbar} \int \frac{\dd^3k \dd^3q'}{(2\pi)^6} [\mathcal{A}(\qq,\qq';\kk,\kk')]^2
\delta(\epsilon_\kk+\epsilon_{\kk'}-\epsilon_\qq-\epsilon_{\qq'}) \\
\times\left[
\frac{\delta n_q}{\bar{n}_q} \bar{n}_k \bar{n}_{k'} (1+\bar{n}_{q'}) + \frac{\delta n_{q'}}{\bar{n}_{q'}} \bar{n}_{k} \bar{n}_{k'} (1+\bar{n}_{q})
-\frac{\delta n_{k}}{\bar{n}_{k}} \bar{n}_{q} \bar{n}_{q'} (1+\bar{n}_{k'}) -\frac{\delta n_{k'}}{\bar{n}_{k'}} \bar{n}_{q} \bar{n}_{q'} (1+\bar{n}_{k})\right]
\end{multline}
once the wave vector $ \kk '$ was eliminated in terms of others, 
%alice029 
\be
\label{eq:029}
\kk'=\qq+\qq'-\kk
\ee
It remains in the form (\ref {eq:028}) a difficult angular integral on the directions $ \hat {\kk} $ and $ \hat {\qq} '$ of $ \kk $ and $ \qq' $. In the limit $ \varepsilon = k_B T / mc ^ 2 \to 0 $ for $ \nu = \mu_ \phi / k_B T $ fixed, it is fortunately dominated by small angles, $ \theta_ \kk = \widehat { (\qq, \kk)} $ and $ \theta _ {\qq '} = \widehat {(\qq, \qq')} $ of order $ \varepsilon $. To zeroth order in $ \varepsilon $, equation (\ref {eq:029}) projected on $ \hat {\qq} $ is reduced to 
%alice030 
\be
\label{eq:030}
k'=q+q'-k
\ee
which is a sufficient approximation {for $ k '$, independent of angles, in {the means} $ \bar {n} _ {k '} $ {and the fluctuations} $ \delta n_ {k'} $ of occupation numbers (but not in the Dirac $ \delta $ function of energy conservation) and which restricts the integration on the wave number $ k $ to the interval $ [0, q + q '] $. It remains to calculate the angular mean of $ \mathcal {A} ^ 2 \delta $ to leading order in $ \varepsilon $. The procedure is detailed in reference \cite {Annalen2017}: we introduce the reduced wave numbers 
%alice031 
\be
\label{eq:031}
\bar{q}=\frac{\hbar c q}{k_B T}, \ \ \bar{k}=\frac{\hbar c k}{k_B T},\ \   \bar{q}'=\frac{\hbar c q'}{k_B T}
\ee
and the reduced angles $ \bar {\theta} _ {\kk} = \theta _ {\kk} / \varepsilon $, $ \bar {\theta} _ {\qq '} = \theta _ {\qq'} / \varepsilon $, then take the mathematical limit $ \varepsilon \to 0 $ for fixed reduced variables $ \bar {q }, \bar {k}, \bar {q} ', \bar {\theta} _ {\kk}, \bar {\theta} _ {\qq'} $; we have to find an equivalent of $ \mathcal {A} ^ 2 $ and expand the energy difference in the argument of the Dirac $ \delta $ function up to the order $ \varepsilon ^ 3 $, which explicitly involves the $ \gamma $ curvature parameter in equation (\ref {eq:002}). In references \cite {Annalen2017, EPL}, the angular mean of $ \mathcal {A} ^ 2 \delta $ is thus reduced to a difficult double integral. We have since realized that it is reducible to an extremely simple analytic expression {(see \ref {ann:zero})}: 
%alice032 
\be
\label{eq:032}
\int\dd^2\hat{k}\int\dd^2\hat{q}' [\mathcal{A}(\qq,\qq';\kk,\kk')]^2 \delta(\epsilon_{\kk}+\epsilon_{\kk'}-\epsilon_{\qq}-\epsilon_{\qq'})
\underset{\varepsilon\to 0}{\stackrel{\scriptsize \overline{\rm var}\ \mbox{fixed}}{\sim}}
\frac{(4\pi)^2}{3|\gamma|} \frac{k_B T}{\rho^2} (1+u)^4 \frac{\bar{k}'}{\bar{q}\bar{q}'\bar{k}} [\min(\bar{q},\bar{q}',\bar{k},\bar{k}')]^3
\ee
where the Gr\"uneisen parameter $ u = \partial \ln c / \partial \ln \rho $ is deduced from the equation of state of the Fermi gas at zero temperature. Let us introduce a dimensionless time variable using the typical Landau-Khalatnikov $ \Gamma _ {\rm th} $ thermal damping rate as in references \cite {EPL, Annalen2017}: 
%alice033 
\be
\label{eq:033}
\bar{t}=\Gamma_{\rm th} t \ \ \mbox{with}\ \ \Gamma_{\rm th}=\left(\frac{1+u}{2\pi}\right)^4 \left(\frac{k_B T}{mc^2}\right)^7 \frac{mc^2}{|\gamma|\hbar}
\left(\frac{mc}{\hbar\rho^{1/3}}\right)^6
\ee
We thus obtain the small angle limit ($ k_B T / mc ^ 2 \to 0 $) of linearized kinetic equations: 
%alice034 
\begin{multline}
\label{eq:034}
\frac{\dd}{\dd \bar{t}} \delta n_q = - \frac{4\pi}{3} \int_0^{+\infty} \dd\bar{q}' \int_0^{\bar{q}+\bar{q}'} \dd\bar{k} \frac{\bar{q}'\bar{k}\bar{k}'}{\bar{q}}
[\min(\bar{q},\bar{q}',\bar{k},\bar{k}')]^3 \\
\times \left[
  \delta n_q \frac{\bar{n}_{k}^\ell \bar{n}_{k'}^\ell (1+\bar{n}_{q'}^\ell)}{\bar{n}_{q}^\ell} 
+ \delta n_{q'} \frac{\bar{n}_{k}^\ell \bar{n}_{k'}^\ell (1+\bar{n}_{q}^\ell)}{\bar{n}_{q'}^\ell}
- \delta n_{k} \frac{\bar{n}_{q}^\ell \bar{n}_{q'}^\ell (1+\bar{n}_{k'}^\ell)}{\bar{n}_{k}^\ell}
- \delta n_{k'}\frac{\bar{n}_{q}^\ell \bar{n}_{q'}^\ell (1+\bar{n}_{k}^\ell) }{\bar{n}_{k'}^\ell}
\right]
\end{multline}
where $ \bar {k} '= \bar {q} + \bar {q}' - \bar {k} $ as in equation (\ref {eq:030}) and where the thermal occupation numbers must be those of the linearized phonon spectrum $ \epsilon_{\kk} \simeq \hbar ck $: 
%alice035 
\be
\label{eq:035}
\bar{n}_k^\ell = \frac{1}{\eee^{\beta(\hbar c k-\mu_\phi)}-1} = \frac{1}{\eee^{\bar{k}-\nu}-1}
\ee
as indicated by the exponent $ \ell $. The diagonal part of the integral equation (\ref {eq:034}) can be written $ - \bar {\Gamma} _q \delta n_q $, where $ \bar {\Gamma} _q = \Gamma_q / \Gamma_ {\rm th} $ is the reduced decay rate of the wave number phonons $ q $; thanks to the breakthrough in equation (\ref {eq:032}), we obtain a much more explicit expression of the rate than in references \cite {EPL, Annalen2017}: 
%alice036 
\be
\label{eq:036}
\boxed{
\bar{\Gamma}_q = \frac{4\pi}{3\bar{q}\bar{n}_q^\ell} \int_0^{+\infty}\dd\bar{q}' \bar{q}'(1+\bar{n}_{q'}^\ell)\int_0^{\bar{q}+\bar{q}'}\dd\bar{k}\, \bar{k} \bar{n}_k^\ell \bar{k}'\bar{n}_{k'}^\ell [\min(\bar{q},\bar{q}',\bar{k},\bar{k}')]^3
}
\ee
We can go even further and obtain a single integral after integration on $ \bar {k} $ (see \ref {ann:I}). 

Let's continue calculating the component $ \Delta C (\tau) $ of zero long-time limit of the correlation function of $ \dd \hat {\theta} / \dd t $. In its expression (\ref {eq:023}), we now know the operator $ M $, defined in reduced form by the right-hand side of equation (\ref {eq:034}). We now have to determine the low temperature limit $ k_B T / mc ^ 2 \to 0 $ of the components $ \Delta A_ \qq $ of the source vector, taken at fixed $ \bar {q} $ {and $ \nu $}. A fairly direct calculation \footnote {A simplifying trick is to isolate in $ A_ \qq $ a component proportional to $ \epsilon_ \qq $, namely $ (2u / \hbar N) \epsilon_ \qq $ (where {the eigenenergy} $ \epsilon_ \qq $ is not yet linearized here), which brings out the term $ 2u / (\hbar N) $ in $ \lambda_ \epsilon $; the rest of $ A_ \qq $ and $ \lambda_ \epsilon $, as well as $ \lambda _ {\rm un} $, are already of order $ k_B T \varepsilon ^ 2 $, which is the leading order in equation (\ref {eq:037}). As expected, the given values of $ X (\nu) $ and $ Y (\nu) $ express the fact that the vector $ Q _ {\rm d} (0) \Delta \vec {A} $ written to the order $ k_B T \varepsilon ^ 2 $, with coordinates $ \propto \bar {n} _q ^ \ell (1+ \bar {n} _q ^ \ell) (\bar {q} ^ 3 -X (\nu) -Y (\nu) \bar {q}) $, is orthogonal to the vectors of coordinates $ \bar {q} $ and $ \bar {q} ^ 0 = 1 $, reflecting the absence of fluctuations of the energy and the number of phonons.} starting from equations (\ref {eq:149}, \ref {eq:017a}, \ref {eq:017b}, \ref {eq:024}) gives 
%alice037 
\be
\label{eq:037}
\Delta A_\qq \underset{\varepsilon\to 0}{\stackrel{\bar{q}\ \scriptsize\mbox{fixed}}{\sim}} \frac{k_B T\gamma}{4\hbar N}\left(\frac{k_B T}{mc^2}\right)^2\left(\frac{\partial\ln|\gamma|}{\partial\ln\rho}-2u\right)[\bar{q}^3-X(\nu)-Y(\nu)\bar{q}]
\ee
with the coefficients $ X (\nu) $ and $ Y (\nu) $, originating from $ \lambda_ \epsilon $ and $ \lambda _ {\rm un} $, are functions of $ \nu $ written in compact form as 
%alice038 
\bea
\label{eq:038a}
X(\nu) &=& \frac{\langle\langle\bar{q}^2\rangle\rangle\,\langle\langle\bar{q}^3\rangle\rangle-\langle\langle\bar{q}\rangle\rangle\,\langle\langle\bar{q}^4\rangle\rangle}{\langle\langle\bar{q}^2\rangle\rangle-\langle\langle\bar{q}\rangle\rangle^2} \\
\label{eq:038b}
Y(\nu) &=& \frac{\langle\langle\bar{q}^4\rangle\rangle-\langle\langle\bar{q}\rangle\rangle\,\langle\langle\bar{q}^3\rangle\rangle}{\langle\langle \bar{q}^2 \rangle\rangle - \langle\langle \bar{q} \rangle\rangle^2}
\eea
using the notation 
%alice039 
\be
\label{eq:alice039}
\langle\langle\bar{q}^n\rangle\rangle=\frac{\int_0^{+\infty} \dd\bar{q}\,\bar{n}_q^\ell(1+\bar{n}_q^\ell)\bar{q}^{n+2}}{\int_0^{+\infty} \dd\bar{q}\,\bar{n}_q^\ell(1+\bar{n}_q^\ell)\bar{q}^{2}} = \frac{(n+2)! g_{n+2}(\eee^\nu)}{2 g_2(\eee^\nu)}\ \ \ \forall n\in\mathbb{N}
\ee
Here, $ g_ \alpha (z) $ is the usual Bose function (or polylogarithm). 

One last remark separates us from the result: for the Landau-Khalatnikov processes, we can give a hermitian form to linearized kinetic equations (\ref {eq:034}) by taking as variables $ \psi _ {\bar {q}} $ rather than $ \delta n_q $, with 
%alice040 
\be
\label{eq:040}
\bar{q}\,\delta n_q= [\bar{n}_q^\ell(1+\bar{n}_q^\ell)]^{1/2}\psi_{\bar{q}}
\ee
We then rewrite equation (\ref {eq:034}) in the form of a Schr\"odinger equation in imaginary time, with Dirac notation for fictitious Fourier wave functions on the real half-line ($ \psi _ {\bar {q}} \equiv \langle \bar {q} | \psi \rangle $, $ \bar {q} \in \mathbb {R} ^ + $ being the wave vector of a fictitious particle living in dimension one and $ \langle \bar {q} | \bar {q} '\rangle = \delta (\bar {q} - \bar {q} ') $): 
%alice041 
\be
\label{eq:041}
\frac{\dd}{\dd\bar{t}} |\psi\rangle = - \hat{\mathcal{H}} |\psi\rangle
\ee
The fictitious Hamiltonian $ \hat {\mathcal {H}} $ is a positive Hermitian operator; since $ - \hat {\mathcal {H}} $ and $ M $ have the same spectrum, we deduce that the eigenvalues of $ M $ are real negative. $ \hat {\mathcal {H}} $ is indeed written as the sum of an operator  $ \hat {\Gamma} $ diagonal in the basis of $ | \bar {q} \rangle $, which contains the decay rates (\ref {eq:036}) of the phonons, 
%alice042, 
\be
\label{eq:042}
\hat{\Gamma}|\bar{q}\rangle = \bar{\Gamma}_q |\bar{q}\rangle
\ee
and a non-local-in-$q$  Hermitian integral-kernel operator $ \hat {V} $, which describes the phonon redistribution by collision: 
%alice043 
\be
\label{eq:043}
\hat{\mathcal{H}} = \hat{\Gamma} + \hat{V}
\ee
The matrix elements of $ \hat {V} $ have from equation (\ref {eq:034}) the elegant integral form \footnote {We use the same kind of trick as in note \ref {note:ast}.} 
%alice044 
\bea
\nonumber
\langle\bar{q}|\hat{V}|\bar{q}'\rangle &=& \frac{4\pi}{3} \int_0^{+\infty} \dd\bar{k} \int_0^{+\infty}\dd\bar{k}'\, \delta(\bar{k}+\bar{k}'-\bar{q}-\bar{q}') [\min(\bar{q},\bar{q}',\bar{k},\bar{k}')]^3 \phi(\bar{k})\phi(\bar{k}') \\
&-&\frac{8\pi}{3} \int_0^{+\infty} \dd\bar{k} \int_0^{+\infty} \dd\bar{k}'\, \delta(\bar{k}'+\bar{q}'-\bar{q}-\bar{k}) [\min(\bar{q},\bar{q}',\bar{k},\bar{k}')]^3 \phi(\bar{k}) \phi(\bar{k}')
\label{eq:044}
\eea
with 
%alice045 
\be
\label{eq:045}
\phi(\bar{k}) = \bar{k} [\bar{n}_k^\ell(1+\bar{n}_k^\ell)]^{1/2}
\ee
We give an explicit expression in terms of Bose's functions in \ref {ann:I}. We finally obtain the low-temperature equivalent of the sub-ballistic contribution (\ref {eq:019b}) to the variance of the phase shift of the condensate: 
%alice046 
\be
\label{eq:046}
\boxed{
\mbox{Var}_{\scriptsize\mbox{s-bal}}[\hat{\theta}(t)-\hat{\theta}(0)] \underset{k_B T/mc^2\to 0}{\stackrel{\bar{t}, \nu\, \mbox{\scriptsize fixed}}{\sim}} \frac{1}{N} \frac{16\pi^6|\gamma|^4}{(1+u)^8} \left(\frac{mc^2}{k_BT}\right)^5 \left(\frac{\hbar\rho^{1/3}}{mc}\right)^9\left(\frac{\partial\ln|\gamma|}{\partial\ln\rho}-2u\right)^2 \bar{\mathcal{V}}(\bar{t})
}
\ee
with \footnote {This expression of $ \bar {\mathcal {V} } (\bar {t}) $ comes from the formal integration of $ \int_0 ^ {\bar {t}} \dd \bar {\tau} \, (\bar {t} - \bar {\tau}) \langle \chi | \eee ^ {- \hat {\mathcal {H}} \bar {\tau}} | \chi \rangle $, where $ \Delta \bar {C} (\bar{\tau}) = \langle \chi | \eee ^ {- \hat {\mathcal {H}} \bar {\tau}} | \chi \rangle $ is a dimensionless form of $ \Delta C (\tau) $.} 
%alice047 
\be
\label{eq:047}
\boxed{
\bar{\mathcal{V}}(\bar{t})=\langle\chi|\frac{\eee^{-\hat{\mathcal{H}}\bar{t}}-1+\hat{\mathcal{H}}\bar{t}}{\hat{\mathcal{H}}^2}|\chi\rangle \ \ \mbox{and}\ \ \langle\bar{q}|\chi\rangle=\phi(\bar{q})[\bar{q}^3-X(\nu)-Y(\nu)\bar{q}]
}
\ee
the limit $ \varepsilon \to 0 $ being taken at fixed reduced time $ \bar {t} $ given by equation (\ref {eq:033}) and at fixed phonon fugacity $ \eee ^ \nu $. The rest of this work is devoted to studying the reduced variance $ \bar {\mathcal {V}} (\bar {t}) $ in the collision-dominated regime $ \bar {t} \gg 1 $. 

\subsubsection {Studying $ \bar {\mathcal {V}} (\bar {t}) $ in the rate approximation} 
\label {subsubsec:edvdadt} 

In an exploratory way, we first use the rate approximation proposed in reference \cite {PRAvrdiff}, which consists in keeping in the right-hand side of linearized kinetic equations (\ref {eq:034}) the decay term $ - \bar {\Gamma} _q \delta n_q $ only, which makes them diagonal. This kind of approximation is quite common in solid state physics, see equation (16.9) in reference \cite {AshcroftMermin}; we must of course keep the dependence of the rate on the wave number $ q $, which plays a crucial role. In the fictitious Schr\"odinger formulation (\ref {eq:041}), this amounts to neglecting $ \hat {V} $ in the Hamiltonian (\ref {eq:043}). Replacing $ \hat {\mathcal {H}} $ with $ \hat {\Gamma} $ in equation (\ref {eq:047}), and setting $ \bar {\Gamma} (\bar { q}) = \bar {\Gamma} _q $, we end up with the approximation 
%alice048 
\be
\label{eq:048}
\bar{\mathcal{V}}_{\rm app}(\bar{t}) = \int_0^{+\infty} \dd\bar{q}\, \frac{\langle\bar{q}|\chi\rangle^2}{\bar{\Gamma}(\bar{q})^2}
\left[\eee^{-\bar{\Gamma}(\bar{q})\bar{t}}-1+\bar{\Gamma}(\bar{q})\bar{t}\right]
\ee
Its behavior at long times is dominated by the more or less singular behavior of the integrand at low wave number $ \bar {q } $. Its analysis must therefore distinguish cases of a negative ($ \nu <0 $) or zero ($ \nu = 0 $) phonon chemical potential. 

\paragraph {Case $ \nu <0 $} 
The decay rate $ \bar {\Gamma} (\bar {q}) $ goes to zero quadratically in $ \bar {q} = 0 $, and we have the expansion 
%alice049 
\be
\label{eq:049}
\bar{\Gamma}(\bar{q})\underset{\bar{q}\to 0}{\stackrel{\nu<0}{=}} C(\nu)\bar{q}^2 [1+\alpha(\nu)\bar{q}+O(\bar{q}^2)]
\ee
An integral expression of the coefficients $ C (\nu) $ and $ \alpha (\nu) $, as well as the sub-subdominant term, is given in \ref {ann:I}. We just need to know here that $ C (\nu) $ and $ \alpha (\nu) $ are positive. Since $ \langle \bar {q} | \chi \rangle ^ 2 $ also vanishes quadratically: 
%alice050 
\be
\label{eq:050}
\langle\bar{q}|\chi\rangle^2 \underset{\bar{q}\to 0}{\sim} \bar{q}^2 \bar{n}_0(1+\bar{n}_0) X(\nu)^2
\ee
with $ \bar {n} _0 = (\eee ^ {- \nu} -1) ^ {- 1 } $, we can separate in the integral (\ref {eq:048}) the term $ \bar {\Gamma} (\bar {q}) \bar {t} $, to obtain a diffusive dominant behavior at long times, but we can not separate the constant term $ -1 $ without triggering an infrared divergence: unlike the case (\ref {eq:008}) of the convex acoustic branch, the sub-diffusive term is no longer a mere delay to diffusion, but is itself divergent at long times. A complete mathematical study obtains, from the following behavior at low $ \bar {q} $ under the integral sign in equation (\ref {eq:048}), 
%alice051 
\be
\label{eq:051}
\frac{\langle\bar{q}|\chi\rangle^2}{\bar{\Gamma}(\bar{q})^2} \underset{\bar{q}\to 0}{\stackrel{\nu<0}{=}} \frac{-\bar{A}_{\rm app}(\nu) C(\nu)^{-1/2}-\bar{B}_{\rm app}(\nu)\bar{q}+O(\bar{q}^2)}{\bar{q}^2}
\ee
the particularly rich asymptotic expansion (see \ref {ann:II}) 
%alice052 
\be
\label{eq:052}
\bar{\mathcal{V}}_{\rm app}(\bar{t}) \underset{\bar{t}\to +\infty}{\stackrel{\nu<0}{=}} \bar{D}_{\rm app}(\nu) \bar{t} +\bar{A}_{\rm app}(\nu) (\pi \bar{t})^{1/2} +\bar{B}_{\rm app}(\nu) \ln(\bar{t}^{1/2})+\bar{E}_{\rm app}(\nu) + o(1)
\ee
with 
%alice053 
\bea
\label{eq:053a}
\bar{D}_{\rm app}(\nu)  &=& \int_0^{+\infty} \dd\bar{q}\, \frac{\langle\bar{q}|\chi\rangle^2}{\bar{\Gamma}(\bar{q})} \\
\label{eq:053b}
\bar{A}_{\rm app}(\nu)  &=&  \frac{-X(\nu)^2 \bar{n}_0(1+\bar{n}_0)}{C(\nu)^{3/2}}\\
\label{eq:053c}
\bar{B}_{\rm app}(\nu)  &=& \frac{-2 X(\nu)^2 \bar{n}_0(1+\bar{n}_0)}{C(\nu)^2} \left[\frac{Y(\nu)}{X(\nu)} -\alpha(\nu)-\left(\frac{1}{2}+\bar{n}_0\right)\right]\\
\nonumber
\bar{E}_{\rm app}(\nu) &=& -\int_0^{\bar{q}_c} \dd\bar{q}\, \left[\frac{\langle\bar{q}|\chi\rangle^2}{\bar{\Gamma}(\bar{q})^2}+ \frac{\bar{A}_{\rm app}(\nu)C(\nu)^{-1/2}+\bar{q}\bar{B}_{\rm app}(\nu)}{\bar{q}^2}\right] - \int_{\bar{q}_c}^{+\infty} \dd\bar{q}\,\frac{\langle\bar{q}|\chi\rangle^2}{\bar{\Gamma}(\bar{q})^2}\\
&& - \bar{A}_{\rm app}(\nu) C(\nu)^{-1/2} \left[\frac{1}{\bar{q}_c}-\frac{1}{2}\alpha(\nu)\right] + \bar{B}_{\rm app}(\nu)\left\{\ln[\bar{q}_cC(\nu)^{1/2}]+\frac{1}{2}\gamma_{\rm Euler}\right\}
\label{eq:053d}
\eea
In equation (\ref {eq:053d}), $ \gamma _ {\rm Euler} = - 0.577 \, 215 \ldots $ is the Euler constant and $ \bar {q} _c> 0 $ is an arbitrary cut-off; the value of $ \bar {E} _ {\rm app} (\nu) $ does not depend on it, as can be verified by taking the derivative with respect to $ \bar {q} _c $. The four coefficients of expansion (\ref {eq:052}) are represented as a function of the reduced chemical potential $ \nu $ of the phonons in figure \ref {fig:dabe} (red solid line). They exhibit when $ \nu \to 0 ^ - $ a divergent behavior, which we compensate by multiplying them by well chosen powers of $ \thf | \nu | $; similarly, we multiply the last three coefficients by a well chosen power of $ \eee ^ \nu $ in order to compensate for their divergence in $ \nu = - \infty $. The plotted functions are therefore bounded. 

\begin {figure} [t] 
\centerline {\includegraphics [width = 0.4 \textwidth, clip =] {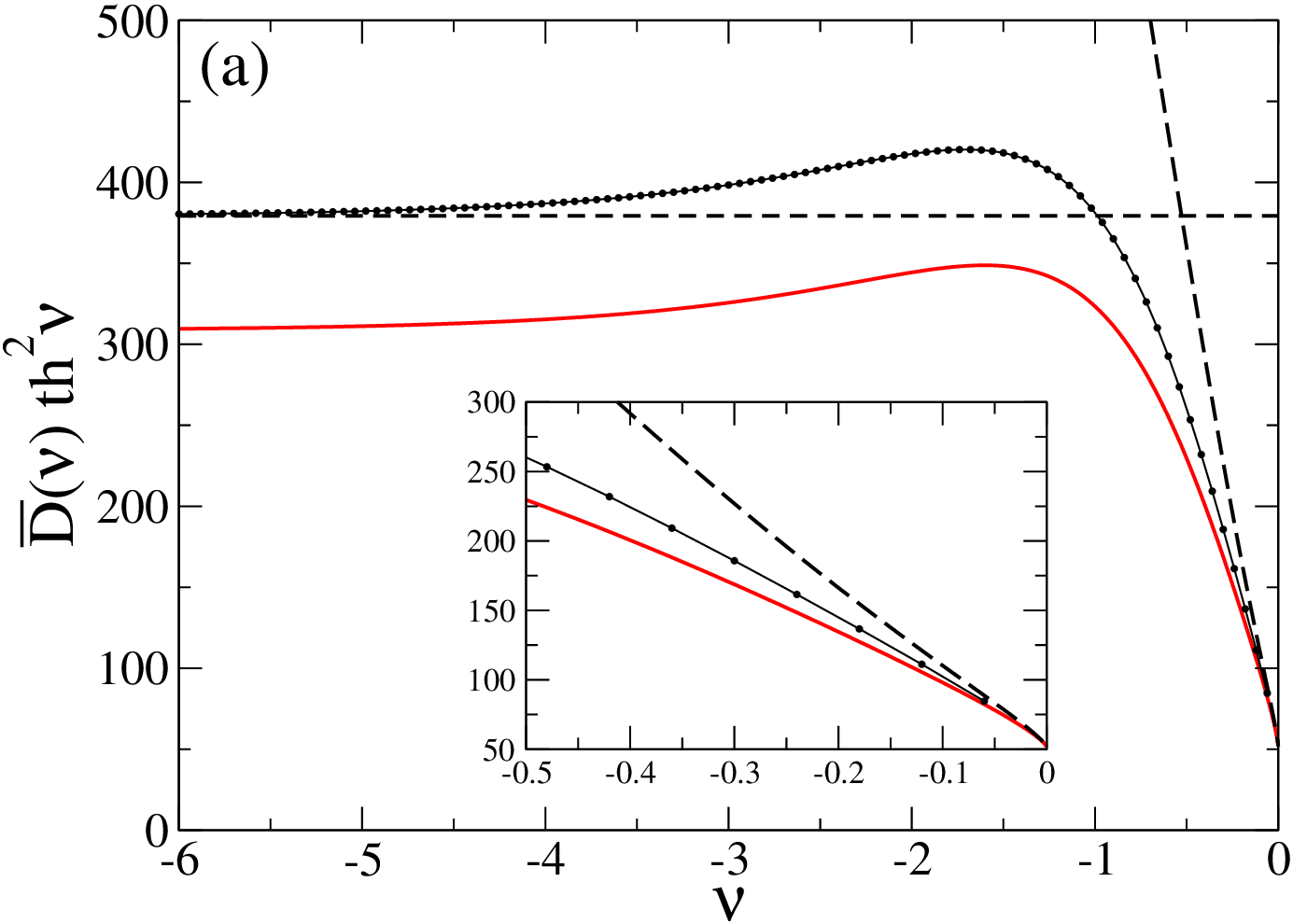} \hspace {1cm} \includegraphics [width = 0.4 \textwidth, clip =] {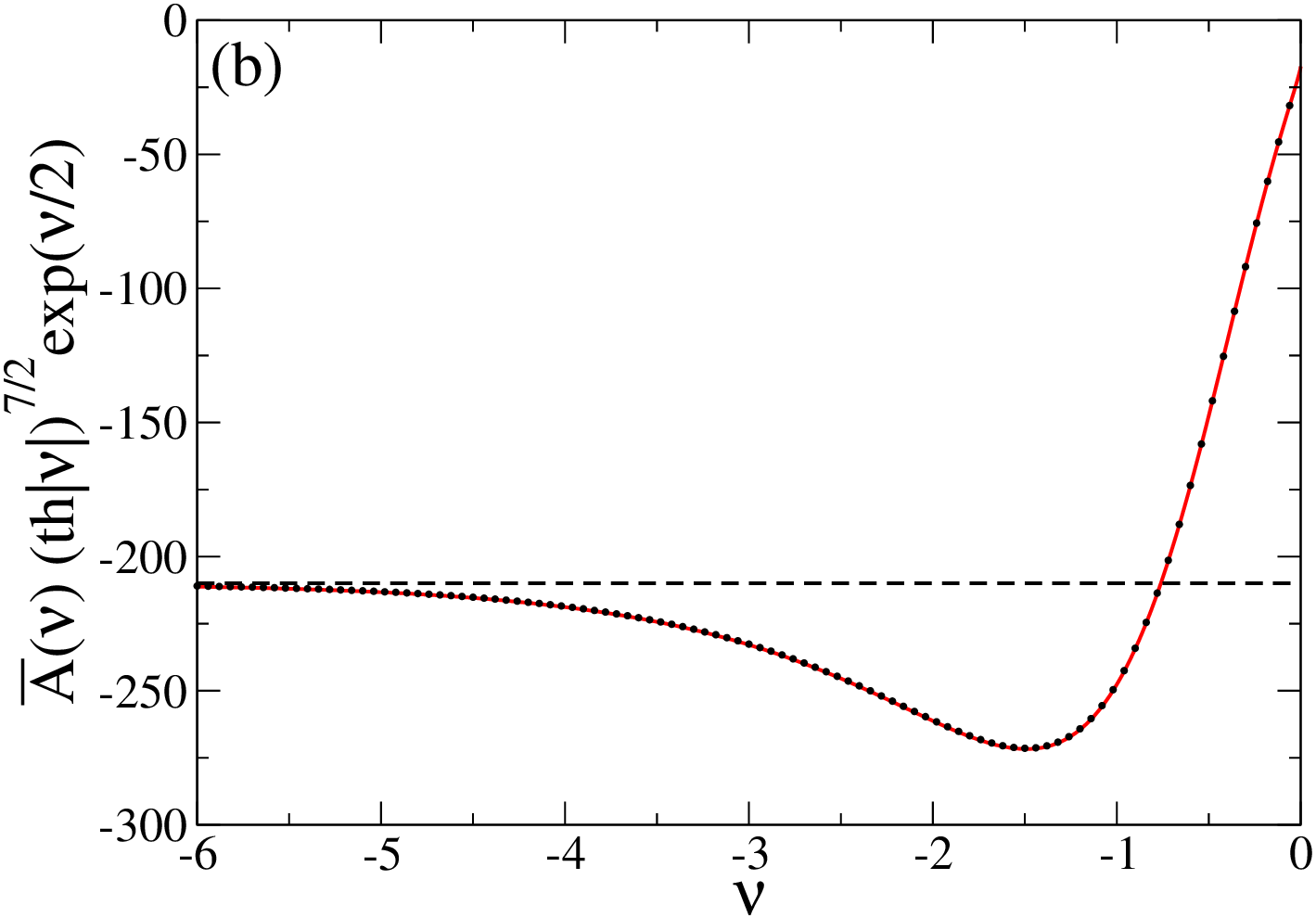} } 
\centerline {\includegraphics [width = 0.4 \textwidth, clip =] {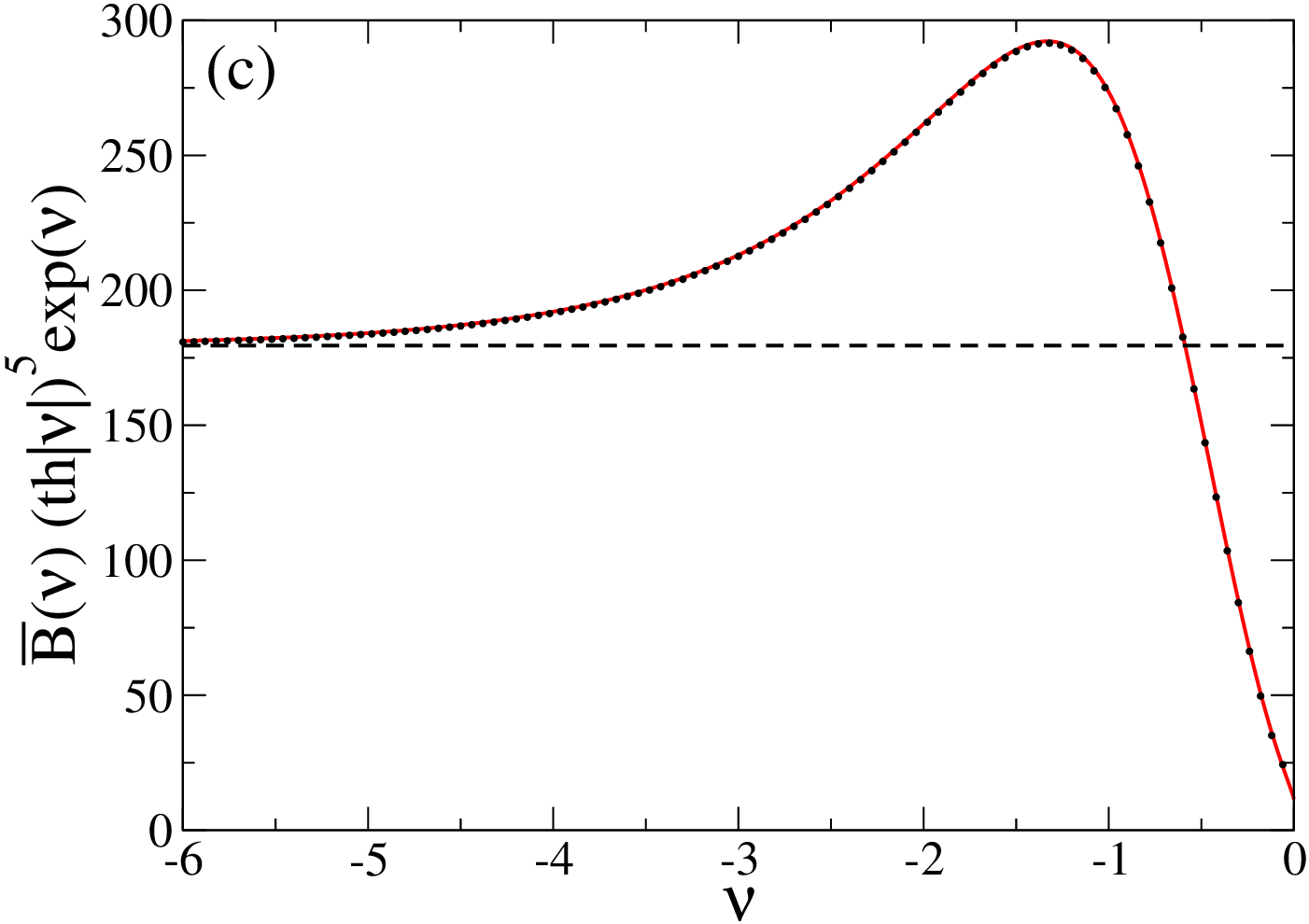} \hspace {1cm} \includegraphics [width = 0.4 \textwidth, clip =] {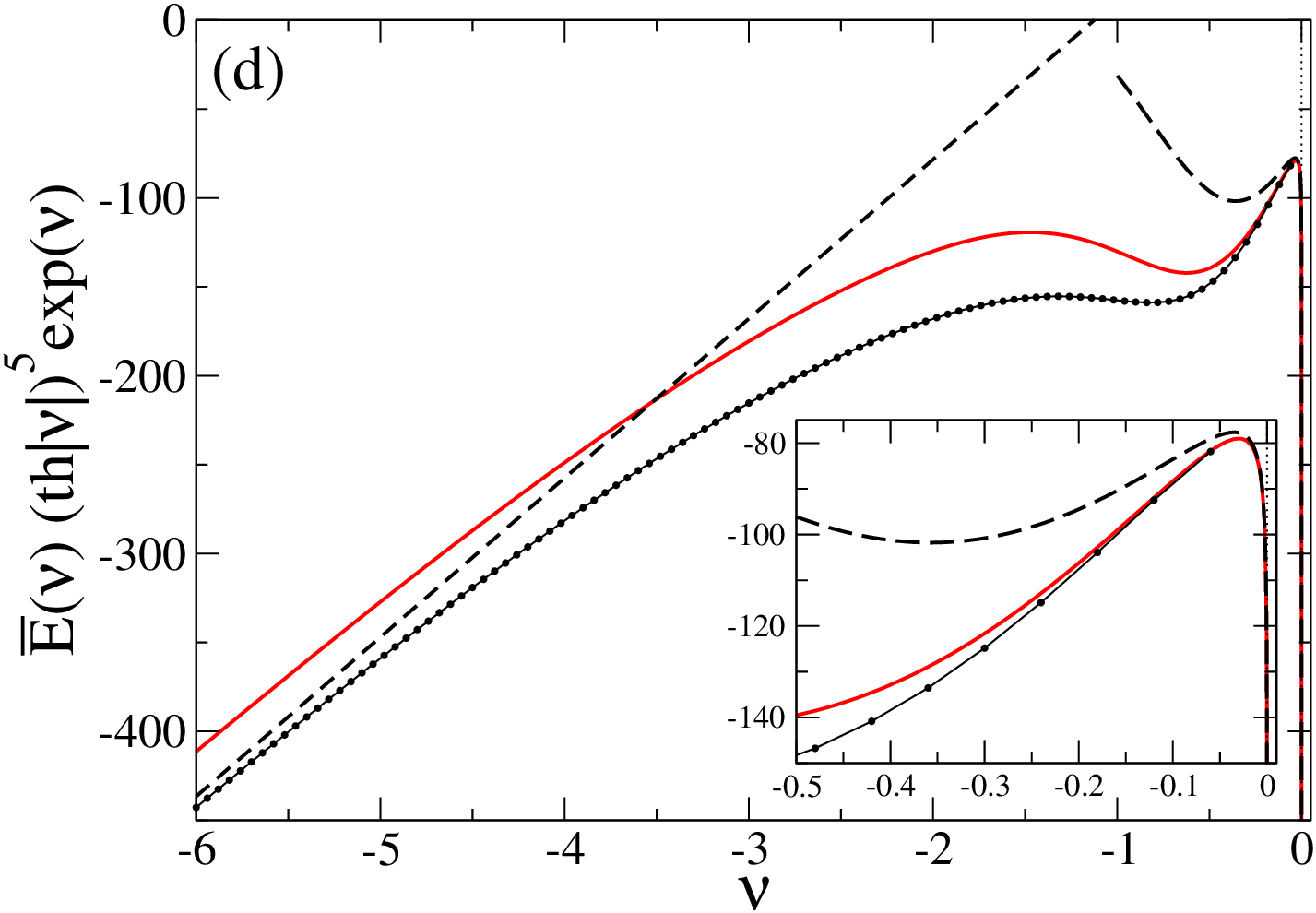}} 
\caption {Coefficients $ \bar {D} $, $ \bar {A} $, $ \bar {B} $ and $ \bar {E} $ from the long-time expansion of the 
sub-ballistic variance (\ref {eq:047}) of the condensate phase shift, 
see equations (\ref {eq:062}) and (\ref {eq:052}), depending on the reduced chemical potential $ \nu = \mu_ \phi / k_B T $ of the phonon gas. Here the acoustic branch has a concave start and the collisional processes between phonons taken into account are those of Landau-Khalatnikov at small angles. Black discs (sometimes with a thin line to guide the eye): exact results obtained numerically. Red solid line: results (\ref {eq:053a}, \ref {eq:053b}, \ref {eq:053c}, \ref {eq:053d}) in the rate approximation. Dashed: limiting behaviors (\ref {eq:078a}, \ref {eq:078b}) and (\ref {eq:084a}, \ref {eq:084b}, \ref {eq:084c}, \ref {eq:084d}) predicted analytically. The coefficients are multiplied by the powers of $ \thf | \nu | $ and $ \eee ^ {\nu} $ giving them finite and non-zero limits in $ \nu = 0 $ and $ \nu = - \infty $. Insets are magnifications. 
\label {fig:dabe} 
} 
\end {figure} 

\paragraph {Case $ \nu = 0 $} 
The divergence of the phase diffusion coefficient in $ \nu = 0 $ suggests a superdiffusive behavior of the sub-ballistic variance at zero phonon chemical potential, {at least in the rate approximation}. Now, the decay rate of the phonons vanishes cubically in $ \bar {q} = 0 $ \cite {Eckstein,EPL, Annalen2017}: 
%alice054 
\be
\label{eq:054}
\bar{\Gamma}(\bar{q})\underset{\bar{q}\to 0}{\stackrel{\nu=0}{=}} C_0 \bar{q}^3[1+\alpha_0\bar{q}+O(\bar{q}^2)]
\ee
with the explicit forms obtained in \ref {ann:I} (the value of $ C_0 $ is in agreement with references \cite {EPL, Annalen2017}, that of $ \alpha_0 $ is new; \ref {ann:I} also gives the coefficient of the sub-sub-dominant term): 
%alice055 
\be
\label{eq:055}
C_0=\frac{16\pi^5}{135} \ \ \mbox{and}\ \ \alpha_0=-\frac{15}{8\pi^2}-\frac{45\zeta(3)}{4\pi^4}
\ee
and we obtain an abnormally rapid sub-ballistic spreading of the condensate phase (see \ref {ann:II}): 
%alice056 
\be
\label{eq:056}
\bar{\mathcal{V}}_{\rm app}(\bar{t})\underset{\bar{t}\to +\infty}{\stackrel{\nu=0}{=}} \bar{d}_0^{\rm app} \bar{t}^{5/3} + \bar{d}_1^{\rm app}\bar{t}^{4/3} +\bar{d}_2^{\rm app}\bar{t}\ln\bar{t} + O(\bar{t})
\ee
We give here explicitly only the coefficient of the dominant term, 
%alice057 
\be
\label{eq:057}
\bar{d}_0^{\rm app} = \frac{3X_0^2\Gamma(1/3)}{10 C_0^{1/3}} \ \ \mbox{with}\ \ X_0=\lim_{\nu\to 0^-}X(\nu) = -\frac{360\pi^4}{7} 
\frac{\pi^2\zeta(3)-7\zeta(5)}{\pi^6-405\zeta(3)^2}
\ee
which depends only on the dominant term in the low $ \bar {q} $ expansion: 
%alice063 
\be
\label{eq:063}
\frac{\langle\bar{q}|\chi\rangle^2}{\bar{\Gamma}(\bar{q})^2}\underset{\bar{q}\to 0}{\stackrel{\nu=0}{=}} \frac{X_0^2}{C_0^2\bar{q}^6}+O(1/\bar{q}^5)
\ee

\paragraph {Case $ \nu $ infinitesimal nonzero: connection between superdiffusive and diffusive regime} 
If $ \nu $ is nonzero but very close to zero, $ | \nu | \ll 1 $, it is expected that $\bar{\mathcal {V}}_{\rm app} (\bar {t}) $ present first a superdiffusive behavior $ \bar {t} ^ {5/3} $, like that of (\ref {eq:056}) for $ \nu = 0 $, before reconnecting at long enough time to the diffusive behavior (\ref {eq:052}) expected for $ \nu <0 $. The existence of these two temporal regimes results from the presence at low $ | \nu | $ of two quite distinct scales of variation with $ \bar {q} $ of the occupation numbers $ \bar {n} _q ^ \ell $, that is $ \bar {q} = | \nu | $ and $ \bar {q} = 1 $, see equation (\ref {eq:035}). Due in particular to the presence of $ \bar {n} _q ^ \ell $ in the denominator of equation (\ref {eq:036}), the decay rate $ \bar {\Gamma} (\bar {q}) $ also presents these two scales. At long times $ \bar {t} \gg 1 $ but not too long, integral (\ref {eq:048}) is dominated by values of $ \bar {q} $ such that $ | \nu | \ll \bar {q} \ll $ 1 and the sub-ballistic spreading of the condensate phase is super-diffuse; at very long times, the values $ \bar {q} \ll | \nu | $ dominate and spreading is diffusive. Mathematically, to describe the crossover between both sides of the $ \bar {q} = | \nu | $ scale, we perform in integral (\ref {eq:048}) the change of variable 
%alice058 
\be
\label{eq:058}
\bar{q}=|\nu| Q
\ee
and let $\nu$ go to $0$ for fixed $Q$. The calculation also gives the low-$\bar{q}$ uniform approximations
%alice059 
\bea
\label{eq:059a}
\bar{n}_q^\ell &  \underset{\nu\to 0^-}{\stackrel{Q \scriptsize \ \mbox{fixed}}{=}} &  \frac{1}{|\nu|(Q+1)} -\frac{1}{2} + O(\nu) \\
\label{eq:059b}
\bar{\Gamma}(\bar{q}) & \underset{\nu\to 0^-}{\stackrel{Q \scriptsize \ \mbox{fixed}}{=}} & |\nu|^2 C(\nu) Q^2 (Q+1) [1 + |\nu| \Phi(Q) +O(|\nu|\ln|\nu|)^2]
\eea
where $ \Phi (Q) $, independent of $ \nu $, has an explicit expression given in \ref {ann:II} and $ C (\nu) \underset {\nu \to 0 ^ -} {\sim} | \nu | C_0 $ as one might expect. Under the exponential in equation (\ref {eq:048}), we must rescale the time as follows: 
%alice060 
\be
\label{eq:060}
\Theta=C(\nu)|\nu|^2 \bar{t}
\ee
and we obtain the sought phase-shift expression connecting the two spreading regimes $ \propto \Theta ^ {5/3} $ for $ \Theta \ll 1 $ and $ \propto \Theta $ for $ \Theta \gg 1 $: 
%alice061 
\be
\label{eq:061}
\bar{\mathcal{V}}_{\rm app}(\bar{t}) \underset{\nu\to 0^-}{\stackrel{\Theta \scriptsize \ \mbox{fixed}}{\sim}} 
\frac{X(\nu)^2}{C(\nu)^2|\nu|^3} I(\Theta)\ \ \mbox{with}\ \  I(\Theta)=\int_0^{+\infty} \dd Q\, \frac{\eee^{-Q^2(1+Q)\Theta}-1+Q^2(1+Q)\Theta}{Q^2(1+Q)^4}
\ee
The integral in equation (\ref {eq:061}) admitting {\sl a priori} no simple expression, we have plotted it in figure \ref {fig:2}. 

\begin {figure} [t] 
\centerline {\includegraphics [width = 0.4 \textwidth, clip =] {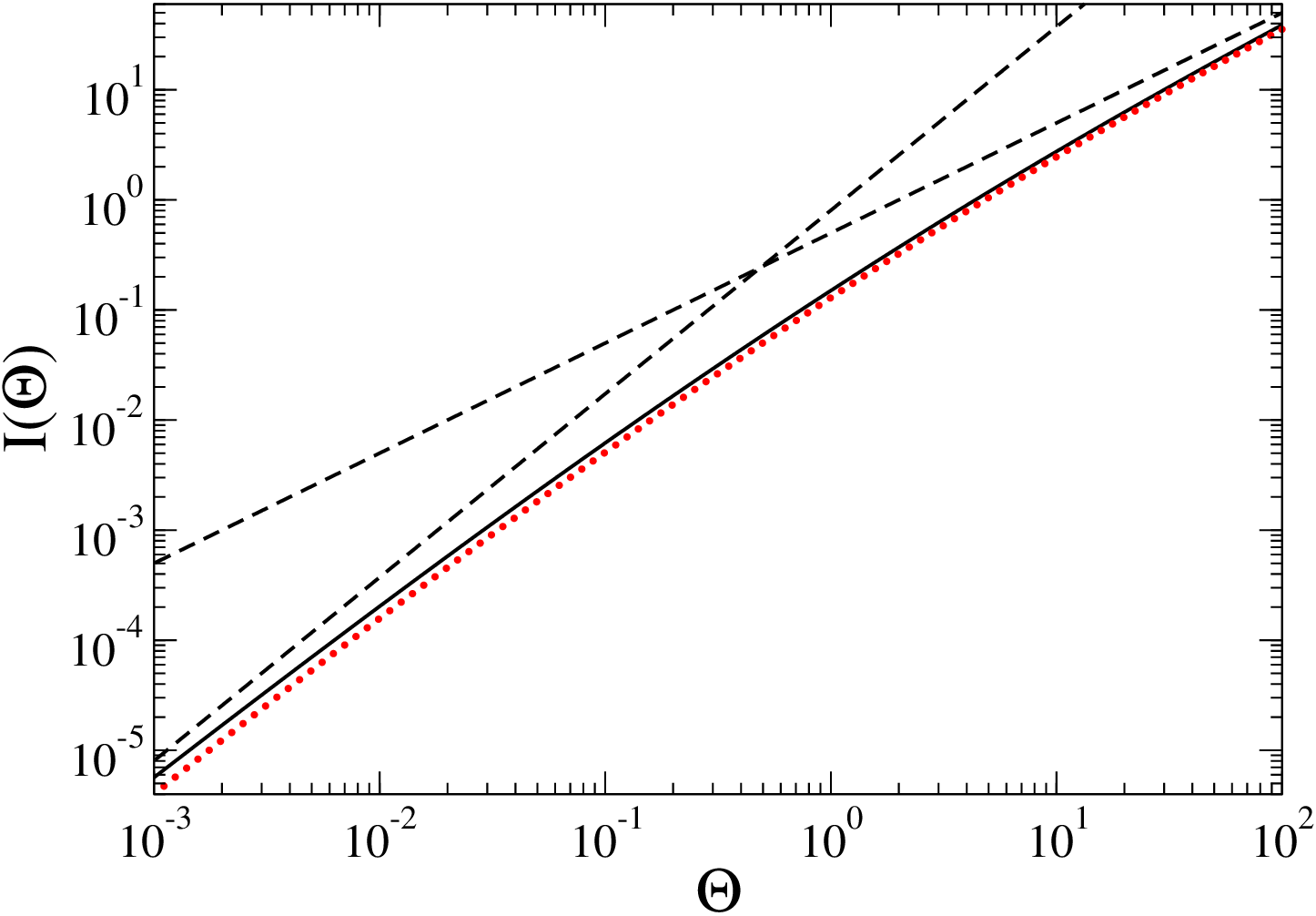}} 
\caption {Interpolation law (solid line) connecting the superdiffusive regime ($ \Theta \ll 1 $) and the diffusive regime ($ \Theta \gg 1 $) of the sub-ballistic spreading of the condensate phase at very low reduced phonon chemical potential ($ \nu = \mu_ \phi / k_B T \to 0 ^ - $ to $ \Theta $ fixed) due to small-angle Landau-Khalatnikov collisions. The renormalized time $ \Theta $ is that of equation (\ref {eq:060}), and the plotted function is the integral $ I (\Theta) $ in the right-hand side of equation (\ref {eq:061}). {The interpolation law, initially obtained in the rate approximation (\ref {eq:048}), is in fact a low $ \nu $ exact equivalent of the reduced sub-ballistic variance $ \bar {\mathcal {V}} (\bar {t}) $ as shown in section \ref {subsubsec:edvslfe}, see equation (\ref {eq:148}).} Dashed: limiting behaviors $ I (\Theta) \sim \frac {3} {10} \Gamma (1/3) \Theta ^ {5/3} $ for $ \Theta \to 0 $ and $ I (\Theta) \sim \frac { 1} {2} \Theta $ for $ \Theta \to + \infty $. Red disks: numeric results for $ \nu = -1 / 10 $ from the exact expression (\ref {eq:069}) {of $ \bar {\mathcal {V}} (\bar {t}) $ } and subjected to the same rescaling as in equation (\ref {eq:061}) (we give $ C (\nu = -1 / 10) = 2.596 \, 303 \dots $ and $ X (\nu = - 1/10) = - 75.909 \, 694 \ldots $); {the fact that the red disks are close to the solid line shows the success of the rate approximation, and more generally of the interpolation law, for very small non-zero $ \nu $.} 
\label {fig:2}} 
\end {figure} 

\paragraph {Moral of the study} 
The exotic behaviors (\ref {eq:052}) and (\ref {eq:056}) of the condensate phase shift predicted in the thermodynamic limit by the rate approximation for low-temperature $ T \to 0 $ Landau-Khalatnikov collisions are of course only valid  at time scales $ 0 (1 / T ^ 9) $, as stated in section \ref {subsec:edtc}. Their origin is clear and results from the combination of two effects: 
\begin {itemize} 
\item the decay rate of low wavenumber phonons tends to zero faster than in Beliaev-Landau damping, as $q^2$ or even $q^3$ for $\nu=0$, instead of $q$. 
\item the existence of a second undamped mode of the linearized kinetic equations, associated with a new quantity conserved by Landau-Khalatnikov collisions, the number of phonons $ N_ \phi $, has a spectacular effect for $ \nu = 0 $: the function $ \langle \bar{q}|\chi\rangle^2 $ no longer tends to zero when $ \bar {q} \to 0 $, while this function (or its equivalent in equation (49) of reference \cite {PRAvrdiff}) was vanishing quadratically in the Beliaev-Landau case. This is due to the fact that the new left eigenvector of $ M $, namely $ \vecl {\rm un} $, has constant coefficients (these are the weights of the occupation numbers $ n_ \qq $ in $ N_ \phi $) whereas the preexisting eigenvector, namely $ \vec {\epsilon} $, has coefficients linear in $ q $ at low $ q $ (these are the weights of $ n_ \qq $ in the energy). 
\end {itemize} 
It remains to be seen what part of truth the rate approximation contains. 

\subsubsection {Studying $ \bar {\mathcal {V}} (\bar {t}) $ on the exact form} 
\label {subsubsec:edvslfe} 

Let's start from the expression (\ref {eq:047}) (exact at sufficiently low temperature) of the sub-ballistic variance of the condensate phase shift. We first carried out a numerical study by discretizing and truncating the half-line $ \bar {q} \in [0, + \infty [$, and diagonalising the corresponding matrix of $ \hat {\mathcal {H} } $. \footnote {In the most accurate calculations, we took a truncation $ \bar {q} _ {\rm max} = 40 + | \nu | $, and we extrapolated linearly to a step $ \dd \bar { q} = 0$ from $ \dd \bar {q} =  0.004$ and $ \dd \bar {q} =  0.006$ or $ 0.008$. The case $ \nu \neq 0 $ but $ | \nu | \ll 1 $ is hard to study because one has to have $ \dd \bar {q} <| \nu | / 10 $ for good convergence.}

\begin {figure} [t] 
\centerline {\includegraphics [width = 0.4 \textwidth, clip =] {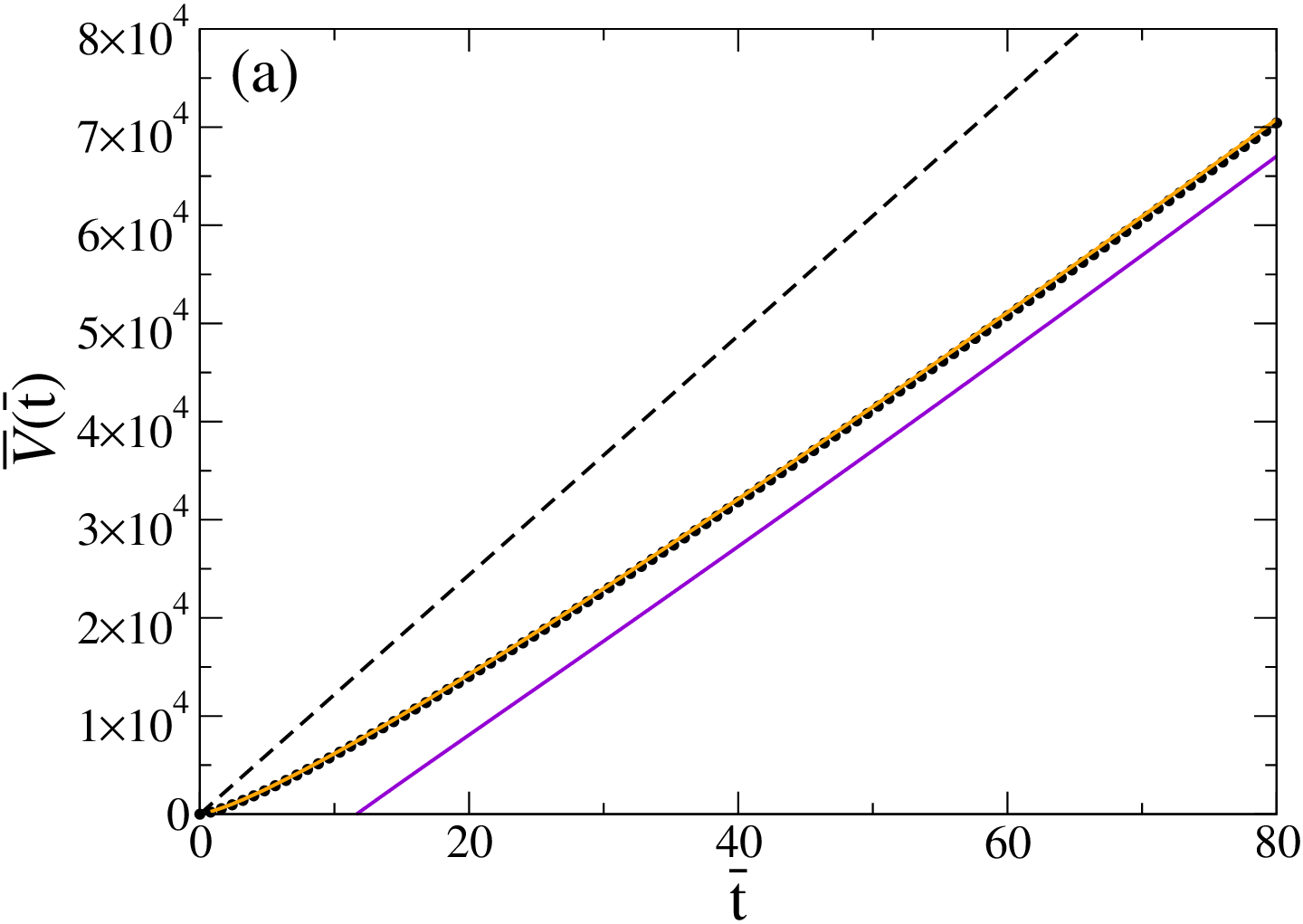} \\\includegraphics [width = 0.4 \textwidth, clip =] {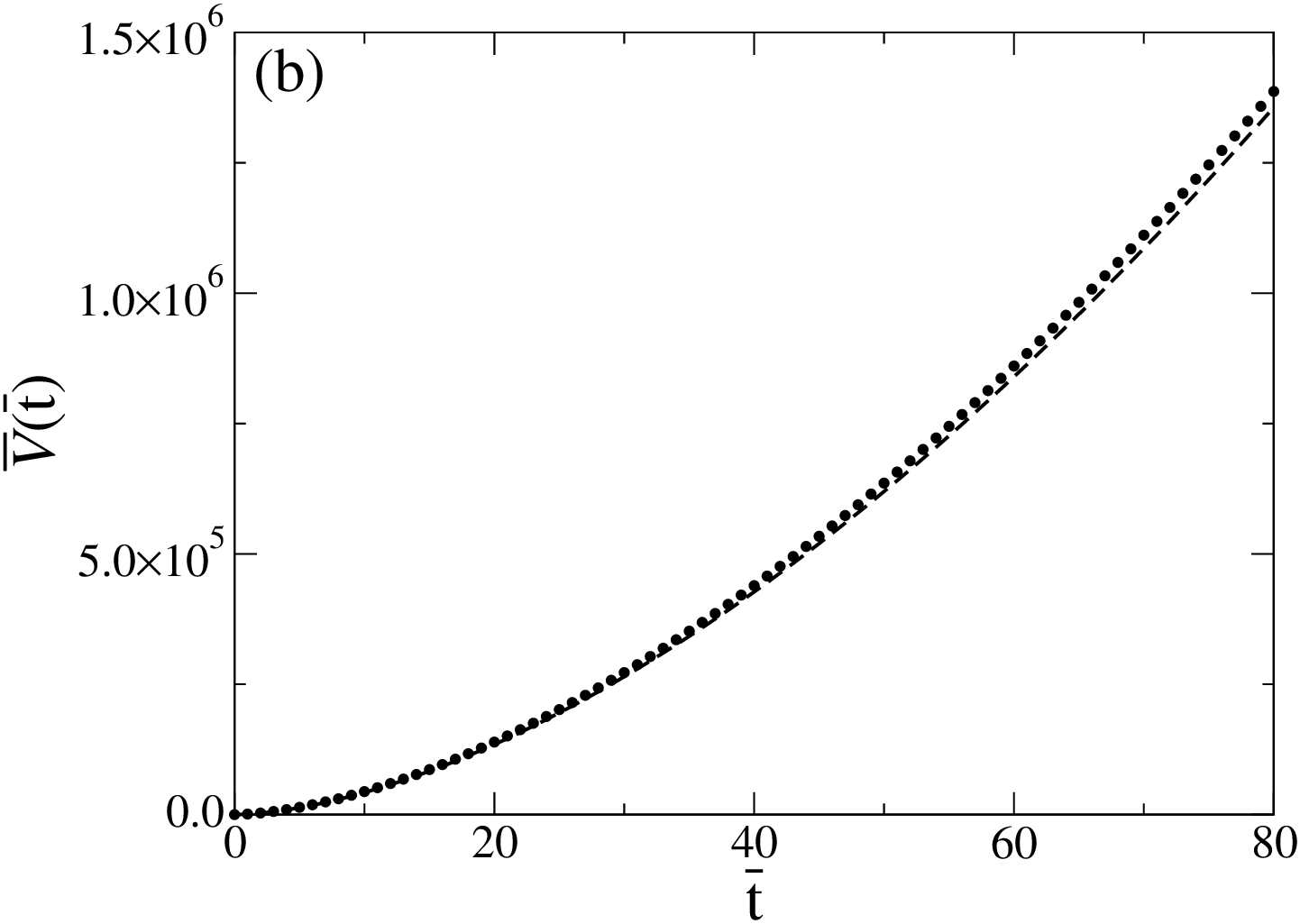}} 
\caption {Reduced sub-ballistic variance $ \bar {\mathcal {V}} (\bar {t}) $ of the condensate phase shift as a function of time {due to small-angle Landau-Khalatnikov collisions}, obtained by numerical diagonalization of $ \hat {\mathcal {H}} $ in equation (\ref {eq:047}), for a reduced phonon chemical potential (a) $ \nu = -1 / 2 $, (b) $ \nu = $ 0. Black disks: numerical results. Dashed: (a) exact diffusive part $ \bar {D} \bar {t} $; (b): asymptotic equivalent $ \bar {d} _0 ^ {\rm app} \bar {t} ^ {5/3} $ predicted by the rate approximation, see equations (\ref {eq:056}, \ref {eq:057}). Purple straight-looking full line in (a): asymptotic expansion (\ref {eq:062}), exact within $ o (1) $. Orange full line in (a): a hybrid approximation of $ \bar {\mathcal {V}} (\bar {t}) $, given by the sum of the numeric diffusive part $ \bar {D} \bar {t} $ and the sub-diffusive part in the rate approximation. 
\label {fig:varred} 
} 
\end {figure} 

\paragraph {Case $ \nu <0 $} 
At long times, we find for $ \bar {\mathcal {V}} (\bar {t}) $ a law of the same form as equation (\ref {eq:052}): 
%alice062 
\be
\label{eq:062}
\boxed{
\bar{\mathcal{V}}(\bar{t})  \underset{\bar{t}\to +\infty}{\stackrel{\nu<0}{=}} \bar{D}(\nu)\bar{t} +\bar{A}(\nu) (\pi\bar{t})^{1/2} +\bar{B}(\nu)\ln(\bar{t}^{1/2}) + \bar{E}(\nu) + o(1)
}
\ee
with the numerically obtained coefficients represented by black disks in figure \ref {fig:dabe}. For the sake of illustration, we plot as a function of time in figure \ref {fig:varred}, the numerically obtained sub-ballistic variance $ \bar {\mathcal {V}} (\bar {t}) $, its diffusive part $ \bar {D} (\nu) \bar {t} $, which approximates it rather badly, and its complete asymptotic expression (\ref {eq:062}). The latter, even if it is better, makes an error of order $ 1 / \bar {t} ^ {1/2} $ (see note \ref {note:rac} of \ref {ann:II}), which tends rather slowly to zero. We therefore also represent a very efficient hybrid approximation combining the exact diffusive part and the sub-diffusive part in the rate approximation. But let's go back to figure \ref {fig:dabe}. Remarkably, the rate approximation is in perfect agreement with the numerical results for the $ \bar {A} $ and $ \bar {B} $ coefficients of intermediate orders in time. On the other hand, for the coefficients $ \bar {D} $ and $ \bar {E} $ of extreme orders in time, it quantitatively fails, except in the limit $ \nu \to 0 ^ - $ where it seems to give exact equivalents (see the magnification in the insets). In particular, the phase diffusion coefficient does diverge when $ \nu \to 0 ^ - $. 

\paragraph {Case $ \nu = 0 $} 
The numerical computation fully confirms the superdiffusive spreading law $ \bar {\mathcal {V}} (\bar {t}) \propto \bar {t} ^ {5 / 3} $ predicted by the rate approximation, including the value (\ref {eq:057}) of the coefficient of $ \bar {t} ^ {5/3} $, as can be judged qualitatively in figure \ref {fig:varred}b, and quantitatively on its variant (not reproduced here) plotting $ \bar {\mathcal {V}} (\bar {t}) / \bar {t} ^ {5/3} $ as a function of $ \bar {t} ^ {- 1/3} $ and extrapolating quadratically to $ \bar {t} ^ {- 1/3} = 0 $. 

\paragraph {Case $ \nu $ very close to $ 0 $} 
Finally, for $ \nu = -1 / 10 $, the crossover of $ \bar {\mathcal {V}} (\bar {t}) $ from a superdiffusive spreading to diffusive spreading is fairly well described by the interpolating law (\ref {eq:061}) derived from the rate approximation, as shown by inclusion of the numerical results in the rescaled figure \ref {fig:2}. 

\paragraph {The reasons for success} 
It remains to understand analytically why the rate approximation is so good, and even accurate for some quantities ($ \bar {A}, \bar {B} $) or in some cases ($ \nu \to 0 ^ - $). For that, let's start from the form (\ref {eq:043}) of the fictitious Hamiltonian $ \hat {\mathcal {H}} $ and let us interpret the two terms $ \hat {\Gamma} $ and $ \hat { V} $: 
\begin {itemize} 
\item operator $ \hat {\Gamma} $ is diagonal in the wave number space of the fictitious particle, i.e.\ in the continuous basis $ | \bar {q} \rangle $; the associated eigenvalue $ \bar {\Gamma} (\bar {q}) $ is a positive and increasing function of $ \bar {q} $, vanishing quadratically in $ \bar {q} = 0 $ (at least for $ \nu <$ 0, 
see equation (\ref {eq:049})) and tending to $ + \infty $ when $ \bar {q} \to + \infty $, as shown by the equivalent in \ref {ann:II} (even for $ \nu = 0 $): 
%alice064 
\be
\label{eq:064}
\bar{\Gamma}(\bar{q}) \underset{\bar{q}\to+\infty}{\sim} \frac{16\pi}{3} g_5(\eee^\nu)\,\bar{q}^2
\ee
where $ g_5 $ is a Bose function. $ \hat {\Gamma} $ thus has the properties required to be considered formally as a kinetic energy operator for the fictitious particle, even if the particle is always moving forward ($ \bar {q } \geq 0 $).
\item operator $ \hat {V} $ is not diagonal in the basis $ | \bar {q} \rangle $ but its matrix elements $ \langle \bar {q} | \hat {V} | \bar {q} '\rangle $ quickly go to zero when $ \bar {q}' \to + \infty $ for fixed $ \bar {q} $ and have no divergence at low $ \bar {q} '$ (they also tend to zero there), as we can verify on the explicit expression given in \ref {ann:I}. We can therefore formally consider $ \hat {V} $ as a short-range external potential for the fictitious particle, even if it has the oddity of not being local in position. \footnote {\label {note:distinguishedo} More precisely, $ \langle \bar {q} | \hat {V} | \bar {q} '\rangle $ is the sum of two contributions. The first one tends uniformly to zero when $ \bar {q} $ and $ \bar {q} '$ tend to $ + \infty $; qualitatively, this corresponds to a separable potential model $ \langle \bar {q} | u \rangle \langle u | \bar {q} '\rangle $ with $ \langle u | u \rangle <+ \infty $. The second one has a finite limit $ \tilde {U} (\bar {Q}) $ when $ \bar {q} $ and $ \bar {q} '\to + \infty $ at fixed $ Q = \bar {q } - \bar {q} '$; qualitatively, this corresponds to a potential $ U (x) $ local in the position space of the fictitious particle, $ \tilde {U} (Q) $ being its Fourier transform. Here $ \tilde {U} (Q) $ is rapidly decreasing. These statements are justified in \ref {ann:I}, see (\ref {eq:116a}, \ref {eq:116b}).} 
\end {itemize} 

Moreover, we know that the spectrum of $\hat {\mathcal {H}} $ is non-negative: $ \hat {\mathcal {H}} $ comes {\sl in fine} from a linearization of kinetic equations, and small initial deviations from thermal occupation numbers can not diverge exponentially in time. The fictitious potential $ \hat {V} $ therefore does not lead to the formation of strictly bound states for the fictitious particle; it gives rise, however, to two discrete (normalizable) eigenstates of $ \hat {\mathcal {H}} $ of zero eigenvalue, so exactly located at the threshold, \footnote {This is a simple rewrite of equation (\ref {eq:013}) according to the change of function (\ref {eq:040}).} 
%alice065 
\be
\label{eq:065}
\hat{\mathcal{H}}|\phi\rangle=0 \ \ \mbox{and}\ \ \hat{\mathcal{H}} (\hat{\bar{q}} |\phi\rangle)=0
\ee
consequences of conservation of energy and phonon number by Landau-Khalatnikov processes (the function $ \langle \bar {q} | \phi \rangle $ is that of equation (\ref {eq:045}) and $ \hat {\bar {q}} $ is the fictitious wave number operator). 
Hence the following important conclusion: the other eigenstates of $ \hat {\mathcal {H}} $ are the stationary scattering states of the fictitious particle on the potential $ \hat {V} $. Each scattering state is marked by the wave number $ \bar {q} $ of the incoming wave; the corresponding ket $ | \psi _ {\bar {q}} \rangle $, the sum of the incoming wave and the scattered wave, is written in the Fourier space 
%alice066 
\be
\label{eq:066}
\langle \bar{q}'|\psi_{\bar{q}}\rangle = \delta(\bar{q}-\bar{q}') + \langle\bar{q}'|\psi_{\bar{q}}^{\rm diff}\rangle
\ee
The first term is a Dirac $ \delta $ function, the second is a regular function of $ \bar {q} '$ except at $ \bar {q}' = \bar {q} $ where it diverges like $ (\bar {q} '- \bar {q}) ^ {- 1} $, according to the usual wave scattering theory \cite {livrediff}, which also shows that the eigenenergy of $ | \psi _ {\bar {q}} \rangle $ reduces to the (here fictitious) kinetic energy of the incoming wave: 
%alice067 
\be
\label{eq:alice067}
\hat{\mathcal{H}} |\psi_{\bar{q}}\rangle = \bar{\Gamma}(\bar{q}) |\psi_{\bar{q}}\rangle 
\ee
The spectrum of $ \hat {\mathcal {H}} $ is thus reduced to the twice degenerate discrete zero eigenvalue and to the non-degenerate continuous spectrum $ \bar {\Gamma} (\bar {q}) $: 
%alice068 
\be
\label{eq:068}
\mbox{Spec}\,\hat{\mathcal{H}} = \{0\,;0\} \cup \{ \bar{\Gamma}(\bar{q})\,\vert\, \bar{q}\in\mathbb{R}^{+*}\}
\ee
what a careful numerical study confirms. 

Still according to scattering theory, normalized steady states as in equation (\ref {eq:066}) are orthogonal, $ \langle \psi _ {\bar {q}} | \psi _ {\bar {q} '} \rangle = \delta (\bar {q} - \bar {q}') $. By injecting a closure relation on the $ | \psi _ {\bar {q}} \rangle $ into expression (\ref {eq:047}) of the sub-ballistic variance of the condensate phase shift, we obtain the exact writing 
%alice069 
\be
\label{eq:069}
\boxed{
\bar{\mathcal{V}}(\bar{t}) = \int_0^{+\infty} \dd\bar{q}\, \frac{|\langle\chi|\psi_{\bar{q}}\rangle|^2}{\bar{\Gamma}(\bar{q})^2}
\left[\eee^{-\bar{\Gamma}(\bar{q})\bar{t}}-1+\bar{\Gamma}(\bar{q})\bar{t}\right]
}
\ee
The rate approximation (\ref {eq:048}) thus amounts to neglecting the scattered wave in $ | \psi _ {\bar {q}} \rangle $ (the second term in equation (\ref {eq:066})). This beautiful reinterpretation allows us to go beyond and estimate the first correction to the approximation $ \langle \chi | \psi _ {\bar {q}} \rangle \simeq \langle \chi | \bar {q } \rangle $}, treating the effect of $ \hat {V} $ in the Born approximation: \footnote {We say ``estimate" because the Born approximation does not give in the exact way the first correction to the rate approximation. Indeed, the matrix elements $ \langle \bar {k} | \hat {V} | \bar {k} '\rangle $, for $ \bar {k} $ and $ \bar {k}' = O (1) $, are infinitesimal only if at least one of $ \bar {k} $ and $ \bar {k} '$ is. Now the next order of Born's expansion of $ \langle \bar {k} | \psi _ {\bar {q}} \rangle $, namely $ \langle \bar {k} | \hat {G}_0 \hat {V} \hat {G} _0 \hat {V} | \bar {q} \rangle $, where $ \bar {k} = O (1) $ and $ \bar {q} = o (1) $, gives rise to matrix elements $ \langle \bar {k} | \hat {V} | \bar {k} '\rangle $ of order unity since the function $ \bar {k}' \mapsto \langle \bar {k} '| \hat {V} | \bar {q} \rangle $ has a width of order unity as shown by equation (\ref {eq:072}); this next order thus provides a relative correction of order unity to the Born term $ \langle \bar {k} | \hat {G} _0 \hat {V} | \bar {q} \rangle $.} 
%alice070 
\be
\label{eq:070}
|\psi_{\bar{q}}\rangle \simeq [1+\hat{G}_0 (\bar{\Gamma}(\bar{q})+\ii 0^+)\hat{V}] |\bar{q}\rangle
\ee
where $ \hat {G} _0 (z) = (z- \hat {\Gamma}) ^ {- 1} $, for $ z \in \mathbb {C} \setminus \mathbb {R} ^ + $, is the resolvent operator of the unperturbed Hamiltonian $ \hat {\Gamma} $. \footnote {Equation (\ref {eq:070}) becomes exact if we replace $ \hat {V} $ with the matrix $ \hat {T} (z) = \hat {V} + \hat {V} \hat {G} (z) \hat {V} $ where $ \hat {G} (z) = (z- \hat {\mathcal {H}}) ^ {- 1} $ is the resolvent operator of the complete Hamiltonian and $ z = \bar {\Gamma} (\bar {q}) + \ii 0 ^ + $.} 
It follows that 
%alice071 
\be
\label{eq:071}
\langle\chi|\psi_{\bar{q}}\rangle-\langle\chi|\bar{q}\rangle \simeq \int_0^{+\infty} \dd\bar{k}\, \frac{\langle\chi|\bar{k}\rangle \langle\bar{k}|\hat{V}|\bar{q}\rangle}{\bar{\Gamma}(\bar{q})-\bar{\Gamma}(\bar{k})+\ii 0^+}
\ee
The correction is generally of order unity, hence the quantitative failure of the rate approximation on the diffusion coefficient $ \bar {D} (\nu) $ for arbitrary $ \nu $ (see figure \ref {fig:dabe}a). In the special case of low wavenumbers $ \bar {q} \to 0 $, on the other hand, the matrix element of $ \hat {V} $ tends to zero uniformly in $ \bar {k} $: 
%alice072 
\be
\label{eq:072}
\langle\bar{k}|\hat{V}|\bar{q}\rangle \underset{\bar{q}\to 0}{\stackrel{\bar{k}=O(1)}{\sim}} [\min(\bar{q},\bar{k})]^3 \mathcal{W}(\bar{k},\nu)
\ee
The function $ \mathcal {W} (\bar {k}, \nu) $ is given in \ref {ann:I}. It suffices here to know that it is uniformly bounded on $ \mathbb {R} ^ + \times \mathbb {R} ^ - $: 
%alice073 
\be
\label{eq:073}
\mathcal{W}(\bar{k},\nu) = O(1)
\ee
In particular, it has a finite and non-zero limit in $ \bar {k } = 0 $ (even for $ \nu = 0 $). We deduce the following: 
\begin {itemize} 
\item for $ \nu <0 $ fixed, \footnote {\label {note:dom} Since the integral over $ \bar {k} $ in (\ref {eq:071}) is dominated by $ \bar {k} \ll \bar {k} _c \ll 1 $, where $ \bar {k} _c $ is arbitrary but fixed, we approximate $ \mathcal {W} (\bar {k}, \nu) $ by $ \mathcal {W} (0, \nu) $, $ \langle \chi | \bar {k} \rangle $ by $ - \bar {k} [\bar { n} _0 (1+ \bar {n} _0)] ^ {1/2} X (\nu) $, $ \bar {\Gamma} (\bar {k}) $ by $ C (\nu) \bar {k} ^ 2 $ and, of course, $ \bar {\Gamma} (\bar {q}) $ by $ C (\nu) \bar {q} ^ 2 $. After the change of variable $ \bar {k} = \bar {q} K $, we find the integral $ \bar {q} ^ 3 \int_0 ^ {\bar {k} _c / \bar {q} } \dd K \, K [\min (1, K)] ^ 3 / (K ^ 2-1- \ii 0 ^ +) = O (\bar {q} ^ 3 \ln \bar {q}) $.} 
%alice074 
\be
\label{eq:074}
\langle\chi|\psi_{\bar{q}}\rangle-\langle\chi|\bar{q}\rangle \underset{\bar{q}\to 0}{=} O(\bar{q}^3\ln\bar{q})
\ee
Since $ \langle \chi | \bar {q} \rangle $ vanishes linearly in this limit, the rate approximation introduces a relative error of order $ \bar {q} ^ 2 \ln \bar {q} $ on $ | \langle \chi | \psi _ {\bar {q}} \rangle | ^ 2 $ in equation (\ref {eq:069}). Also $ | \langle \chi | \psi _ {\bar {q}} \rangle | ^ 2 / \bar {\Gamma} (\bar {q}) ^ 2 $ exactly has the Taylor expansion (\ref {eq:051}). The expressions (\ref {eq:053b}, \ref {eq:053c}) of the resulting coefficients $ \bar {A} (\nu) $ and $ \bar {B} (\nu) $ are exact. We write them here: 
%alice075 
\begin{equationarray}{lcl}
\label{eq:075a}
\bar{A}(\nu)  &=&  \frac{-X(\nu)^2 \bar{n}_0(1+\bar{n}_0)}{C(\nu)^{3/2}}\\
\label{eq:075b}
\bar{B}(\nu)  &=& \frac{-2 X(\nu)^2 \bar{n}_0(1+\bar{n}_0)}{C(\nu)^2} \left[\frac{Y(\nu)}{X(\nu)} -\alpha(\nu)-\left(\frac{1}{2}+\bar{n}_0\right)\right]
\end{equationarray}
On the other hand, expressions {(\ref {eq:053a})} and (\ref {eq:053d}) do not depend only on the behavior of $ \langle \bar {q} | \chi \rangle $ in the neighborhood of $ \bar {q} = 0 $; they do not give the exact value of {$ \bar {D} (\nu) $} and $ \bar {E} (\nu) $. 
\item for $ \nu = 0 $, \footnote {We can proceed as in note \ref {note:dom} except that $ \langle \chi | \bar {k} \rangle \underset {\bar {k} \to 0} {\to} -X_0 $ and $ \bar {\Gamma} (\bar {k}) $ vanishes cubically. We find the integral $ \bar {q} \int_0 ^ {\bar {k} _c / \bar {q}} \dd K [\min (1, K)] ^ 3 / (K ^ 3-1 - \ii 0 ^ +) = O (\bar {q}) $.} 
%alice085 
\be
\label{eq:085}
\langle\chi|\psi_{\bar{q}}\rangle-\langle\chi|\bar{q}\rangle \underset{\bar{q}\to 0}{=} O(\bar{q})
\ee
Since $ \langle \chi | \bar {q} \rangle $ tends to a finite and non-zero limit in $ \bar {q} = 0 $, the rate approximation introduces a relative error $ \bar {q} $ on $ | \langle \chi | \psi _ {\bar {q}} \rangle | ^ 2 $. The expansion (\ref {eq:063}) is only affected to subdominant order. The same goes for equation (\ref {eq:056}). We thus keep the exact result predicted by the rate approximation (see (\ref {eq:057}) for the expression of $ X_0 $ and (\ref {eq:055}) for that of $ C_0 $): 
%alice076 
\be
\label{eq:076}
\boxed{
\bar{\mathcal{V}}(\bar{t})\underset{\bar{t}\to +\infty}{\stackrel{\nu=0}{=}} \bar{d}_0 \bar{t}^{5/3}+O(\bar{t}^{4/3})
\ \ \mbox{with}\ \ \bar{d}_0=\frac{3X_0^2\Gamma(1/3)}{10 C_0^{1/3}}=913. 184\, 011\ldots
}
\ee
\item for $ \nu $ infinitesimal non-zero, we saw in the rate approximation that {the crossover of} $ \bar {\mathcal {V}} (\bar {t}) $ {from superdiffusive spreading to diffusive spreading} is dominated by the wave numbers $ \bar {q} \simeq | \nu | $. We therefore set $ \bar {q} = | \nu | Q $ as in equation (\ref {eq:058}) and $ \bar {k} = | \nu | K $ in the integral (\ref {eq:071}), then set $ \nu $ to zero for fixed $ Q $ and $ K $. Keeping the leading order in expression (\ref{eq:047}) of $ \langle \bar {k} | \chi \rangle $ and in equations (\ref {eq:059a}, \ref {eq:059b}), and approximating $ \mathcal {W} (\bar {k}, \nu) $ by $ \mathcal {W} (0,0) $, we find that 
%alice077 
\be
\label{eq:077}
\langle\chi|\psi_{\bar{q}}\rangle-\langle\chi|\bar{q}\rangle \underset{\nu\to 0^-}{\stackrel{Q \scriptsize \ \mbox{fixed}}{\simeq}}
\frac{X_0}{C_0} \mathcal{W}(0,0) |\nu| \int_0^{+\infty} \dd K \frac{\frac{K}{K+1} [\min(K,Q)]^3}{K^2(1+K)-Q^2(1+Q)-\ii 0^+} = O(|\nu|)
\ee
Since $ \langle \chi | \bar {q} \rangle \sim -X_0 Q / (1 + Q) $ in this limit, the relative and absolute errors introduced on $ | \langle \psi _ {\bar {q}} \chi \rangle | ^ 2 $ by the rate approximation are of order one in $ | \nu | $. The rate approximation therefore gives the exact values of the diffusion coefficient $ \bar {D} $ and the constant term $ \bar {E} $ in the limit $ \nu \to 0 ^ - $ to leading order in $ | \nu | $: 
%alice078 
\bea
\label{eq:078a}
\bar{D}(\nu)  &=&  \int_0^{+\infty} \dd\bar{q}\, \frac{|\langle\chi|\psi_{\bar{q}}\rangle|^2}{\bar{\Gamma}(\bar{q})} 
\underset{\nu\to 0^-}{=} \frac{X(\nu)^2}{2 C(\nu) |\nu|} +O\left(\frac{\ln|\nu|}{|\nu|}\right) \\
\label{eq:078b}
\bar{E}(\nu) &\underset{\nu\to 0^-}{=}& \frac{2 X(\nu)^2}{C(\nu)^2 |\nu|^3} \left[\ln(C(\nu)|\nu|^2)-\frac{29}{12}+\gamma_{\rm Euler}\right]+ 
O\left(\frac{\ln |\nu|}{|\nu|^4}\right)
\eea
as we can see on the magnifications around $ \nu = 0 $ in the insets of figures \ref {fig:dabe}a and \ref {fig:dabe}d. We obtained these results by performing the change of variable $ \bar {q} = | \nu | Q $ in equation (\ref {eq:053a}) and in the first integral of equation (\ref {eq:053d}) and replacing the integrand with an equivalent for $ \nu \to 0 $ for fixed $ Q $, see equation (\ref {eq:059b}), thus making a relative error $ O (| \nu |) $ {of the same order as in the rate approximation}. In equation (\ref {eq:053d}), we also replace $ \bar {A} _ {\rm app} (\nu), \bar {B} _ {\rm app} (\nu) $ and $ \alpha (\nu) $ by equivalents and we neglect the second integral $ \int _ {\bar {q} _c} ^ {+ \infty} $ of bounded contribution. \footnote {We take $ C (\nu) ^ {- 1/2} \bar {A} _ {\rm app} (\nu) \sim -X (\nu) ^ 2 / [| \nu | ^ 2C (\nu) ^ 2] $, $ \bar {B} _ {\rm app} (\nu) \sim 4X (\nu) ^ 2 / [C (\nu) ^ 2 | \nu | ^ 3 ] $ and $ \langle \bar {q} | \chi \rangle ^ 2 / \bar {\Gamma} (\bar {q}) ^ 2 \sim X (\nu) ^ 2 / [| \nu | ^ 4 C (\nu) ^ 2 Q ^ 2 (1 + Q) ^ 4] $. We also take $ \alpha (\nu) \sim 1 / | \nu | $, as suggested by (\ref {eq:059b}) and as confirmed by an explicit calculation (see \ref {ann:I}). In all these expressions, the relative error is $ O (| \nu |) $. It remains to say that $ \int_0 ^ {\bar {q} _c / | \nu |} \dd Q \, \left [\frac {1} {Q ^ 2 (1 + Q) ^ 4} - \frac {1} {Q ^ 2} + \frac {4} {Q} \right] = \frac {13} {3} + 4 \ln \frac {\bar {q} _c} {| \nu |} + O (| \nu |) $.} 
{In the same way, we deduce from equation (\ref {eq:077}) that the interpolation law (\ref {eq:061}) between the superdiffusive and diffusive sub-ballistic spreadings, derived from the rate approximation, is in fact a low $ \nu $ exact equivalent of $ \bar {\mathcal {V}} (\bar {t}) $ for fixed $ \Theta = C (\nu) | \nu | ^ 2 \bar {t} $, which we write here:} 
%alice148 
\be
\boxed{
\label{eq:148}
\bar{\mathcal{V}}(\bar{t}) \underset{\nu\to 0^-}{\stackrel{\Theta \scriptsize \ \mbox{fixed}}{\sim}} 
\frac{X(\nu)^2}{C(\nu)^2|\nu|^3} \int_0^{+\infty} \dd Q\, \frac{\eee^{-Q^2(1+Q)\Theta}-1+Q^2(1+Q)\Theta}{Q^2(1+Q)^4}
}
\ee
\end {itemize} 

\paragraph {Case $ \nu \to - \infty $} 
Finally, another exact result can be obtained in the $ \nu \to - \infty $ limit where the phonon gas is nondegenerate. The thermal occupation numbers of the modes then tend exponentially to zero: 
%alice079 
\be
\label{eq:079}
\bar{n}_k^\ell \underset{\nu\to -\infty}{\sim} \eee^\nu \eee^{-\bar{k}}
\ee
As we see on the expressions (\ref {eq:036}) of its diagonal part and (\ref {eq:044}) of its non diagonal part, the fictitious Hamiltonian (\ref {eq:043}) tends to zero in the same way: 
%alice080 
\be
\label{eq:080}
\hat{\mathcal{H}} \underset{\nu\to-\infty}{\sim} \eee^\nu \hat{\mathcal{H}}_\infty
\ee
where the $ \hat {\mathcal {H}} _ {\infty} $ operator is independent of $ \nu $. This also applies to the fictitious ket $ | \chi \rangle $ in equation (\ref {eq:047}): 
\footnote {We give $ X_ \infty = \lim _ {\nu \to- \infty} X (\nu) = - 120 $ and $ Y_ \infty = \lim _ {\nu \to- \infty} Y (\nu) = 60 $.} 
%alice081 
\be
\label{eq:081}
|\chi\rangle \underset{\nu\to-\infty}{\sim} \eee^{\nu/2} |\chi_\infty\rangle \ \ \mbox{with}\ \ \langle\bar{q}|\chi_{\infty}\rangle = \bar{q}\,\eee^{-\bar{q}} (\bar{q}^3-X_\infty -Y_\infty \bar{q})
\ee
If you cleverly redefine the time and the sub-ballistic variance of the condensate phase shift as follows, $ \eee ^ \nu \bar {t} = \bar {t} _ \infty $ and $ \bar {\mathcal {V}} _ \infty (\bar {t} _ \infty) = \eee ^ \nu \bar {\mathcal {V}} (\bar {t}) $, the reduced variance $ \bar {\mathcal {V}} _ \infty (\bar {t} _ \infty) $ depends only on $ \bar {t} _ \infty $ and no more on $ \nu $ in the nondegenerate limit: 
%alice082 
\be
\label{eq:082}
\bar{\mathcal{V}}_\infty(\bar{t}_\infty)= \langle\chi_\infty|\frac{\eee^{-\hat{\mathcal{H}}_\infty\bar{t}_\infty}-1+\hat{\mathcal{H}}_\infty\bar{t}_\infty}{\hat{\mathcal{H}}_\infty^2}|\chi_\infty\rangle
\ee
Now it must have at long times $ \bar {t} _ \infty $ an asymptotic expansion of the same form as (\ref {eq:062}): 
%alice083 
\be
\label{eq:083}
\bar{\mathcal{V}}_\infty(\bar{t}_\infty)\underset{\bar{t}_\infty\to +\infty}{=} \bar{D}_\infty\bar{t}_\infty +\bar{A}_\infty(\pi\bar{t}_\infty)^{1/2} +\bar{B}_\infty\ln(\bar{t}_\infty^{1/2}) + \bar{E}_\infty + o(1)
\ee
with analytically known $ \bar {A} _ \infty, \bar {B} _ \infty $ and numerically computable $ \bar {D} _ \infty, \bar {E} _ \infty $ constant coefficients. \footnote {We give $ \bar {D} _ \infty \simeq $ 379, $ \bar {E} _ \infty \simeq $ 101, $ \bar {A} _ \infty = - (5!) ^ 2 / (16 \pi / 3) ^ {3/2} \simeq -210 $ and $ \bar {B} _ \infty = 2 (5!) ^ 2 (1+ \alpha_ \infty) / (16 \pi / 3) ^ 2 \simeq 180 $, knowing that $ \alpha_ \infty = \lim _ {\nu \to- \infty} \alpha (\nu) = 3/4 $ after (\ref {eq:105b}).} If we go back to the time variable $ \bar {t} $ and to the variance $ \bar {\mathcal {V}} (\bar {t}) $ in equation (\ref {eq:083}), we must find the limiting behavior of equation (\ref {eq:062}) when $ \nu \to - \infty $, up to a relative error $ O (\eee ^ \nu) $ as in equations (\ref {eq:079}, \ref {eq:080}, \ref {eq:081}), hence 
%alice084 
\bea
\label{eq:084a}
\bar{D}(\nu) &=&  \bar{D}_\infty [1+O(\eee^\nu)] \\
\label{eq:084b}
\bar{A}(\nu) &=&  \eee^{-\nu/2} \bar{A}_\infty [1+O(\eee^\nu)] \\
\label{eq:084c}
\bar{B}(\nu) &=&  \eee^{-\nu} \bar{B}_\infty [1+O(\eee^\nu)] \\
\label{eq:084d}
\bar{E}(\nu) &=&  \eee^{-\nu} (\bar{E}_\infty + \bar{B}_\infty \nu/2) [1+O(\eee^\nu)]
\eea
These predictions are plotted {as dashed lines} in figure \ref {fig:dabe}. 

\section {Conclusion} 
\label {sec:conclusion} 

In a three-dimensional gas of $ N $ spin-1/2 fermions at equilibrium, that is spatially homogeneous, unpolarized and isolated from its environment, a condensate of pairs accumulates during time $ t $ a phase shift $ \hat {\theta} (t) - \hat {\theta} (0) $. The variance of this phase-shift at times long compared to the collision time between thermal phonons of the system was studied in this paper. The gas is close to the thermodynamic limit, but remains of finite size. Its temperature is non-zero but extremely low, sufficiently to justify the use of quantum hydrodynamics. We had to distinguish two cases, depending on the strength of the contact interactions between opposite spin fermions. 

In the first case, the phonon dispersion relation is convex at low wavenumber $ q $. The dominant collisional processes are the three phonon Beliaev-Landau ones  $ \phi \leftrightarrow \phi + \phi $, with a time scale $ \propto \hbar (mc ^ 2) ^ {5/2} \epsilon _ {\rm F} ^ {3/2} (k_B T) ^ {- 5} $, where $ c $ is the speed of sound at zero temperature and $ \epsilon _ {\rm F} $ the Fermi energy of the gas. The variance of the condensate phase shift is asymptotically the sum of a ballistic term $ \propto t ^ 2 $ and a subleading contribution $ 2 D (t-t_0) $ corresponding to a phase diffusion with a time delay $ t_0 $. The ballistic term results from the variations from one realization of the experiment to the other of the conserved quantities, the total energy $ E $ and the total number $ N $ of fermions in the gas; it therefore vanishes if the system is prepared in the microcanonical ensemble. It is insensitive to the initial fluctuations of other observables because the phonon gas has an ergodic quantum dynamics. On the other hand, the coefficients $ D $ and $ t_0 $, of which we give analytic expressions, do not depend on the statistical ensemble in the thermodynamic limit. This situation is very similar to that of a condensate in a weakly-interacting Bose gas \cite {PRAYvanEmiliaAlice}. 

In the second case, the phonon dispersion relation is concave at low wavenumber. The situation is then much richer, the collisional processes having several relevant time scales. At the time scale $ \propto \hbar (mc ^ 2) ^ 6 (k_B T) ^ {- 7} $, the dominant processes are the four-phonon Landau-Khalatnikov ones at small-angle (collisions $\qq + \qq '\to \kk + \kk' $ with angles $ O (k_BT / mc ^ 2) $ between the wave vectors). We limit ourselves here to times $ t $ negligible compared to $ \hbar (mc ^ 2) ^ 8 (k_BT) ^ {- 9} $ to exclude other processes, and to isotropic thermal phonon distributions in $ \qq $ space for simplicity (the small-angle Landau-Khalatnikov collisions, the way we treat them, do not thermalise the direction of $ \qq $ \cite {JETPKhalatnikov}). The total number of phonons $ N_ \phi $ is thus effectively a constant of motion and the phonons exhibit, contrary to the previous case, quasi-stationary distributions of nonzero chemical potential $ \mu_ \phi $, $ \mu_ \phi <0 $. The variance of the phase shift is then asymptotically the sum of a ballistic term $ \propto t ^ 2 $, a diffusive term $ 2Dt $, exotic sub-sub-leading terms $ 2A (\pi t) ^ {1 / 2} + 2B \ln (t ^ {1/2}) $ and a constant term. The coefficient of the ballistic term results, as in the first case, from the fluctuations of the conserved quantities, here $ N $, $ E $ and $ N_ \phi $; it vanishes in a generalized microcanonical ensemble setting $ N $, $ E $ and $ N_ \phi $. The coefficients $ A $ and $ B $ depend only on the infrared physics of the phonon gas, i.e., wave numbers $ q \ll q _ {\rm th} $ where $ q _ {\rm th} = k_B T / \hbar c $ is the thermal wave number. Their exact value can thus be obtained analytically by a simple rate approximation reducing to their diagonal part, i.e.\ to their decay terms, the linearized kinetic equations on the fluctuations of the occupation numbers of the phonon modes. The coefficient $ D $ and the constant term are dominated by this infrared physics only when $ \mu_ \phi / k_B T \to 0 $; within this limit, we obtain analytically their leading, divergent behavior. When the chemical potential of the phonons is zero, $ \mu_ \phi = 0 $, the variance of the condensate phase shift is asymptotically the sum of the eternal ballistic term, of exotic superdiffusive terms (with non-integer exponents) $ d_0 t ^ {5/3} + d_1 t ^ {4/3} $ and sub-subleading terms that are $ O (t \ln t) $. The exact value of the coefficient $ d_0 $ is obtained analytically by the rate approximation. One can likewise obtain a very small non-zero $ \mu_ \phi / k_B T $ interpolating law describing the crossover, for the sub-ballistic variance, from the superdiffusive spreading $ \propto t ^ {5/3} $ to the diffusive spreading $ \propto t $, the regime change occurring  at a time $ t \propto \hbar \epsilon _ {\rm F} ^ 6 | \mu_ \phi | ^ {- 3} (k_B T) ^ {- 4} $.

In all cases, we have used centrally the quantum generalization of the second Josephson relation, which links the time derivative of the condensate-of-pairs phase operator $ \hat {\theta} $ to the occupation-number operators of the phonon modes \cite {CRAS2016}. In all cases, we find that the coefficient of the ballistic term is $ O (1 / N) $ for normal fluctuations of the conserved quantities, and that all the coefficients of the sub-ballistic terms ($ D $ and $ D t_0 $ in the convex case, $ D $, $ A $, $ B $, {the constant term}, $ d_0 $ and $ d_1 $ in the concave case) are $ \propto 1 / N $. In the concave case, new (and strong) results on the Landau-Khalatnikov damping rate of phonons have been obtained {(see in particular equation (\ref {eq:036}))}, for $ \mu_ \phi <0 $ of course but for $ \mu_ \phi = 0 $ too; being only intermediate results here, they have been relegated to an appendix. 

A natural extension of our work on the condensate-of-pairs phase spreading would be to perform the analysis at longer time scales {$ \hbar (mc ^ 2) ^ 8 (k_B T) ^ {- 9} $} where the number of phonons is no longer conserved, to take into account the coupling to fermionic BCS quasi-particles, i.e.\ the thermally broken pairs of fermions \cite {phorot0, phorot1}, even to treat the harmonically trapped case as it is done for bosons in reference \cite {CRASharm}. Let us also point out, as an emerging and promising research subject, the study of temporal coherence at nonzero temperature of an isolated low-dimensional (Bose or Fermi) gas \cite {SvistunovJETP}; physics is quite different from our three-dimensional case in the thermodynamic limit since there is no condensate anymore, but some of the theoretical tools of this work may be reused.

\section * {Acknowledgments} 
We warmly thank Alice Sinatra and Hadrien Kurkjian for their contributions at a preliminary stage of our study, as well as Boris Svistunov for his encouragement {and Jacques Villain for his questions}. 

\appendix 
\section {Analytical calculation of the Landau-Khalatnikov angular average in equation (\ref {eq:032})} 
\label {ann:zero} 

{ 
To justify equation (\ref {eq:032}) and therefore the very simple expression of the Landau-Khalatnikov damping rate in equation (\ref {eq:036}), 
it must be shown that the function of the reduced incoming and outgoing wave numbers (\ref {eq:031}) defined 
on the hyperplane $ \bar {k} + \bar {k} '= \bar {q} + \bar {q}' $ by 
%alice134 
\be
\label{eq:134}
F(\bar{q},\bar{q}';\bar{k},\bar{k}') \equiv \frac{\bar{q}'\bar{k}}{2\bar{q}\bar{k}'|w|} \int_0^{\pi} \dd\phi\, \int_0^{\pi/2}\dd\alpha\, \sin(2\alpha) 
\Theta\left(\frac{-w}{u}\right) \left|\mathcal{A}_{\rm red}\right|^2
\ee
is simply 
%alice135 
\be
\label{eq:135}
F(\bar{q},\bar{q}';\bar{k},\bar{k}') = \frac{4\pi}{3} \frac{[\min(\bar{q},\bar{q}',\bar{k},\bar{k}')]^3}{\bar{q}\bar{q}'\bar{k}\bar{k}'}
\ee
We have included here the results of reference \cite {EPL} with almost the same notations. 
In particular, $ \Theta $ is the Heaviside function and 
%alice136 
\bea
\label{eq:136a}
u &=& \frac{1}{\bar{k}'} \left[\bar{q}(\bar{k}\sin^2\alpha-\bar{q}'\cos^2\alpha)+\bar{q}'\bar{k}(1-\sin2\alpha\cos\phi)\right] \  ; \
w=\frac{1}{4} \left(\bar{q}^3+\bar{q}'^3-\bar{k}^3-\bar{k}'^3\right) \\
\label{eq:136b}
\mathcal{A}_{\rm red} &=& \frac{1}{\bar{q}'\left(\frac{\cos^2\alpha}{(\bar{q}+\bar{q}')^2}-\frac{3u}{4w}\right)}-\frac{1}{\bar{k}\left(\frac{\sin^2\alpha}{(\bar{q}-\bar{k})^2}-\frac{3u}{4w}\right)} -\frac{1}{\frac{\bar{q}'\cos^2\alpha-\bar{k}\sin^2\alpha+u}{(\bar{q}'-\bar{k})^2} -\frac{3u}{4w}\bar{k}'}
\eea
The function $F$ inherits symmetry properties of the four-phonon coupling amplitude $ \mathcal {A} (\qq, \qq '; \kk, \kk ') $ of the effective Hamiltonian (\ref {eq:026}): $ F (\bar {q}, \bar {q}'; \bar {k}, \bar {k} ') = F (\bar {q} ',\bar {q}; \bar {k}, \bar {k}') = F (\bar {q}, \bar {q} ';\bar {k}', \bar {k}) = F (\bar {k}, \bar {k} '; \bar {q}, \bar {q}') $. It is also a positively homogeneous function of degree $ -1 $, $ F (\lambda \bar {q}, \lambda \bar {q} '; \lambda \bar {k}, \lambda \bar {k}') = \lambda ^ {- 1} F (\bar {q}, \bar {q} '; \bar {k}, \bar {k}') $ for all $ \lambda> 0 $, as can be seen easily on the form (\ref {eq:134}). So one just needs to prove (\ref {eq:135}) on the fundamental domain $ 0 <\bar {k} <\bar {q} <1/2 $ and $ \bar {k} '+ \bar {k } = \bar {q} '+ \bar {q} = $ 1, to which we now restrict ourselves. 

On this fundamental domain, $ w = - \frac {3} {4} (\bar {q} - \bar {k}) (1- \bar {k} - \bar {q}) $ is negative and, introducing $ X = \tan \alpha $, one has: 
%alice137 
\be
\label{eq:137}
u>0 \Longleftrightarrow s X -\frac{s'}{X} > \cos \phi \Longleftrightarrow X>X_0(\phi)=\frac{\cos\phi +\sqrt{4s s'+\cos^2\phi}}{2s} 
\ \mbox{with}\ s=\frac{1}{2(1-\bar{q})}>0 \ \mbox{and}\ s'=\frac{\bar{q}-\bar{k}}{2\bar{k}}>0
\ee
As suggested in a note of reference \cite {Annalen2017}, we can perform a partial fraction decomposition of the reduced amplitude $ \mathcal {A} _ {\rm red} $ for the variable $ \cos \phi $: 
%alice138 
\be
\label{eq:138}
\mathcal{A}_{\rm red} = \frac{1}{\sin 2\alpha} \sum_{i=1}^{3} \frac{b_i}{a_i-\cos\phi}
\ee
with 
%alice139 
\begin{equationarray}{lclclclclcl}
\label{eq:139a}
a_1 &\!\!\!\!\!=\!\!\!\!\!& \frac{1}{2(1-\bar{q})}&& a_1' &\!\!\!\!\!=\!\!\!\!\!& -\frac{(\bar{q}-\bar{k})(2-\bar{k}-\bar{q})}{2(1-\bar{q})} && b_1 &\!\!\!\!\!=\!\!\!\!\!& \frac{(1-\bar{k})(\bar{q}-\bar{k})(1-\bar{k}-\bar{q})}{\bar{k}(1-\bar{q})^2} \\
a_2 &\!\!\!\!\!=\!\!\!\!\!&\frac{1-2\bar{k}}{2\bar{k}(\bar{q}-\bar{k})} && a_2' &\!\!\!\!\!=\!\!\!\!\!& -\frac{\bar{q}-\bar{k}}{2\bar{k}} && b_2 &\!\!\!\!\!=\!\!\!\!\!& -\frac{(1-\bar{k})(\bar{q}-\bar{k})(1-\bar{k}-\bar{q})}{\bar{k}^2(1-\bar{q})} \\
a_3 &\!\!\!\!\!=\!\!\!\!\!& \frac{1-2\bar{k}}{2[1+\bar{k}^2-\bar{k}(3-\bar{q})]} && a_3'&\!\!\!\!\!=\!\!\!\!\!& \frac{(\bar{q}-\bar{k})(2-\bar{k}-\bar{q})}{2[1+\bar{k}^2-\bar{k}(3-\bar{q})]} && b_3 &\!\!\!\!\!=\!\!\!\!\!& -\frac{(1-\bar{k})(\bar{q}-\bar{k})(1-\bar{k}-\bar{q})^2}{\bar{k}(1-\bar{q})[1+\bar{k}^2-\bar{k}(3-\bar{q})]}
\end{equationarray}
However, we must take the opposite strategy of this note by first integrating over the angle $ \alpha $ rather than over the angle $ \phi $: 
%alice140 
\be
\label{eq:140}
F(\bar{q},1-\bar{q};\bar{k},1-\bar{k}) = \frac{\bar{k}(1-\bar{q})}{3\bar{q}(1-\bar{k})(\bar{q}-\bar{k})(1-\bar{k}-\bar{q})} \sum_{i,j=1}^{3} b_i b_j I(a_i,a_i';a_j,a_j')
\ee
where we introduced the function 
%alice141 
\be
\label{eq:141}
I(a,a';\check{a},\check{a}') \equiv \int_0^{\pi}\dd\phi\, \int_{X_0(\phi)}^{+\infty} \dd X\, \frac{X}{(a X^2-X\cos\phi+a')(\check{a}X^2-X\cos\phi +\check{a}')}
\ee
Second helpful observation, the coefficients $ a_i $ and $ a_i '$ obey the remarkable and unexpected relations 
%alice142 
\be
\label{eq:142}
(a_i a_j'-a_j a_i')^2+(a_i-a_j) (a_i'-a_j')=0 \ \ \ \forall i,j\in \{1,2,3\}
\ee
So we calculate the function $ I (a, a '; \check {a}, \check {a}') $ only on the manifold 
%alice143 
\be
\label{eq:143}
\mathbb{V}=\{ (a,a',\check{a},\check{a}') \in \mathbb{R}^4 \,\vert\, (a\check{a}'-a'\check{a})^2+(a-\check{a})(a'-\check{a}')=0\}
\ee
The partial fraction decomposition of the integrand of (\ref {eq:141}) over real numbers vis-\`a-vis the variable $ X $ then exhibits a global factor with a 
particularly simple dependence on $ \phi $, since it is $ \propto \sin ^ {- 2} \phi $. \footnote {In the general case, we get a more complicated factor 
$ [(a \check {a} '- a' \check {a}) ^ 2+ (a- \check {a}) (a '- \check {a} ') \cos ^ 2 \phi] ^ {- 2} $, and the rest of the computation fails.} 
We are thus reduced, after integration over $ X $, to 
%alice144 
\be
\label{eq:144}
I(a,a';\check{a},\check{a}')=\frac{J(a,a')-J(\check{a},\check{a}')-\frac{1}{2}K(a,a')+\frac{1}{2}K(\check{a},\check{a}')}{a\check{a}'-a'\check{a}}
+2 \frac{a'J(a,a')-\check{a}'J(\check{a},\check{a}')}{a'-\check{a}'}
\ee
The functions of two variables that we introduced, 
%alice145 
\bea
\label{eq:145a}
J(a,a') &\equiv& \int_\eta^{\pi-\eta} \frac{\dd\phi}{\sin^2\phi} \frac{\cos\phi}{\sqrt{4aa'-\cos^2\phi}}
\left[\atan \frac{2a\Lambda}{\sqrt{4aa'-\cos^2\phi}}-\atan\frac{2aX_0(\phi)-\cos\phi}{\sqrt{4aa'-\cos^2\phi}}\right] \\
\label{eq:145b}
K(a,a') &\equiv& \int_\eta^{\pi-\eta} \frac{\dd\phi}{\sin^2\phi} \ln\frac{a X_0^2(\phi)-X_0(\phi)\cos\phi+a'}{a}
\eea
contain positive regularization parameters $ \Lambda $ and $ \eta $ which are set respectively to $ + \infty $ and $ 0 ^ + $ at the end of the calculations. It remains to integrate equations (\ref {eq:145a}, \ref {eq:145b}) by parts, taking the derivative of the non-rational functions $ \atan $ and $ \ln $ to get rid of them. \footnote {In equation (\ref {eq:145a}), the arc tangent and square root functions are defined on the complex plane. A primitive of $ \cos \phi / (\sin ^ 2 \phi \sqrt {4aa '- \cos ^ 2 \phi}) $ is $ \sqrt {4aa' - \cos ^ 2 \phi} / [(1 -4aa ') \sin \phi] $. The square root $ \sqrt {4aa '- \cos ^ 2 \phi} $ that it contains multiplies with that derived from the derivative of $ \atan $: the integration by parts thus makes disappear in a single blow the two inconvenient square root and arc tangent functions.} The fully integrated terms diverge as 
$ D_J (a, a ') / \eta + O (\eta) $ and $ D_K (a, a') / \eta + O (\eta) $ when $ \eta \to 0 $; the expression of the coefficients $ D_J (a, a ') $ and $ D_K (a, a') $ does not matter since their contributions exactly compensate in the finite-limit function $ I (a, a '; \check {a } \check {a} ') $. The remaining integrals converge when $ \eta \to 0 $: they are Hadamard finite parts $ \mathrm {Pf} \, J $ and $ \mathrm {Pf} \, K $ of the integrals $ J $ and $ K $. They are integrals of rational functions $ f (\phi) $ of $ \cos \phi $ and $ X_0 (\phi) $, which are in principle reducible to elliptic integrals \cite {Gradshteyn}; actually, the symmetry trick replacing their integrand $ f (\phi) $ by $ [f (\phi) + f (\pi- \phi)] / 2 $ (which we can always do for a integral on $ [0, \pi] $) removes the square root $ \sqrt {4ss' + \cos ^ 2 \phi} $ coming from $ X_0 (\phi) $ and reduces us to integrals of rational fractions of $ \cos \phi $, 
%alice146 
\bea
\label{eq:146a}
\mathrm{Pf}\,J(a,a') &=& -\frac{a-s}{1-4a a'} \int_0^\pi \dd\phi\, \frac{a'(a's+as')-\frac{1}{2}(a'+s')\cos^2\phi}{(as'+a's)^2+(a-s)(a'+s')\cos^2\phi} \\
\label{eq:146b}
\mathrm{Pf}\,K(a,a') &=& -(a-s)(a'+s') \int_0^\pi \dd\phi\, \frac{\cos^2\phi}{(as'+a's)^2+(a-s)(a'+s')\cos^2\phi}
\eea
which are easily deduced from $ \int_0 ^ \pi \dd \phi \, [1-C (a, a ') \cos ^ 2 \phi] ^ {- 1} = \pi [1-C (a, a ')] ^ {- 1/2} $, where $ C (a, a') $ has the same value in $ \mathrm {Pf} J (a, a ') $ and $ \mathrm { Pf} K (a, a ') $. \footnote {Again, the situation is simpler than expected: $ (i) $ since $ a_1-s = 0 $ and $ a_2 '+ s' = 0 $, we have $ \mathrm {Pf} J (a_1, a_1 ') = \mathrm {Pf} K (a_1, a_1') = \mathrm {Pf} K (a_2, a_2 ') = 0 $ and $ C (a_1, a_1') = C (a_2, a_2 ') = $ 0; $ (ii) $ $ 1-C (a_3, a_3 ') = [(1- \bar {k} - \bar {q}) / (1+ \bar {k} - \bar {q})] ^ 2 $ is a perfect square.} We thus obtain, after replacing $ J $ and $ K $ in (\ref {eq:144}) by their finite part $ \mathrm {Pf} \, J $ and $ \mathrm { Pf} \, K $, an essentially explicit expression of $ I (a, a '; \check {a}, \check {a}') $: 
%alice147 
\be
\label{eq:147}
I(a,a';\check{a},\check{a}')=\frac{\mathrm{Pf}\,J(a,a')-\mathrm{Pf}\,J(\check{a},\check{a}')-\frac{1}{2}\mathrm{Pf}\,K(a,a')+\frac{1}{2}\mathrm{Pf}\,K(\check{a},\check{a}')}{a\check{a}'-a'\check{a}}
+2 \frac{a'\mathrm{Pf}\,J(a,a')-\check{a}'\mathrm{Pf}\,J(\check{a},\check{a}')}{a'-\check{a}'}
\ee
The computation of $ I (a, a '; \check {a}, \check {a} ') $ we just exposed fails in the particular case $ (a, a') = (\check {a}, \check {a} ') $ and does not allow us {\sl a priori} to get the diagonal terms $ i = j $ in equation (\ref {eq:140}). A simple trick, however, bypasses the obstacle: 
just take $ (a, a ') = (a_i, a_i') $ and make $ (\check {a}, \check {a} ') $ tend to $ (a_i, a_i ') $ in (\ref {eq:147}) while remaining on the manifold $ \mathbb {V} $ of equation (\ref {eq:143}). In practice, we take $ (\check {a}, \check {a} ') = (a_i + \epsilon, a_i' + \lambda_i \epsilon) $ with $ \epsilon \to 0 $ and $ \lambda_i = [\sqrt {1-4 a_i a_i '} - (1-2a_ia_i')] / (2a_i ^ 2) $, that is $ \lambda_1 = \lambda_2 = - (\bar {k} - \bar {q}) ^ 2 $ and $ \lambda_3 = - (\bar {k} - \bar {q}) ^ 2 / (1-2 \bar {k}) ^ 2 $. 
We finally get $ F (\bar {q}, 1- \bar {q}; \bar {k}, 1- \bar {k}) = 4 \pi \bar {k} ^ 2 / [3 \bar {q} (1- \bar {k}) (1- \bar {q})] $ as in equation (\ref {eq:135}) restricted to the fundamental domain $ 0 <\bar {k} <\bar {q} <1/2 $, which was to be demonstrated. 
} 

\section {Expressions and analytical results on the Landau-Khalatnikov decay rate of phonons and on the integral kernel of the corresponding linearized kinetic equations} 
\label {ann:I} 

To make the notation lighter, we omit here the bars above $ \bar {q}, \bar {q} ', \bar {k} $: wave numbers are implicitly expressed in units of $ k_B T / \hbar c $ ($ T $ gas temperature, $ c $ speed of sound).

\subsection {Case of the decay rate $ \bar {\Gamma} _q $} 

Equation (\ref {eq:036}) gives $ \bar {\Gamma} _q $ as a double integral (instead of a quadruple integral in references \cite {EPL, Annalen2017}). The inner integral on $ k $ can however be performed analytically, which reduces $ \bar {\Gamma} _q $ to the following single integral: 
%alice086 
\be
\boxed{
\label{eq:086}
\bar{\Gamma}_q=\frac{8\pi}{3q\bar{n}_q^\ell} \int_0^{+\infty} \dd q'\, \frac{q'(1+\bar{n}_{q'}^\ell)}{\eee^{q+q'-2\nu}-1}
F(\min(q,q'),\max(q,q'))
}
\ee
with 
%alice087 
\bea
\nonumber
F(q,q') &=& \frac{1}{12} q^3 q'^2 (q'+3q) -\frac{q^5}{20} (q+q') -3 q^3 q'[g_2(\eee^{\nu-q})+g_2(\eee^{\nu-q'})] \\
\nonumber
&+& 6 q^2(2q'-q)[g_3(\eee^{\nu-q'})-g_3(\eee^{\nu-q})] + 12 q (3q-2q') [g_4(\eee^{\nu-q})+g_4(\eee^{\nu-q'})] \\
\nonumber
&+& 24 (q+q') [g_5(\eee^\nu)-g_5(\eee^{\nu-q-q'})] +24 (q'-4q) [g_5(\eee^{\nu-q'})-g_5(\eee^{\nu-q})] \\
\label{eq:087}
&-& 120 [g_6(\eee^{\nu})+g_6(\eee^{\nu-q-q'}) - g_6(\eee^{\nu-q})-g_6(\eee^{\nu-q'})]
\eea
where $ g_ \alpha (z) $ is the usual Bose function (or polylogarithm) with index $ \alpha $, $ g_ \alpha (z) = \sum_ {n = 1} ^ {+ \infty} z ^ n / n ^ \alpha $ (for $ | z | <1 $). 
We can show that the integrand of (\ref {eq:086}) is a function of $ q '$ of class $ C ^ 2 $, its third derivative being discontinuous in $ q' = q $. 

Let's briefly explain how equation (\ref {eq:087}) was obtained. The first step is to write explicitly the $ \min $ in equation (\ref {eq:036}). Since the integral on $ k $ is invariant by the change of variable $ k \to q + q'-k $ (its integrand is invariant under the exchange of $ k $ and $ k '= q + q'-k $), we can restrict it to the interval $ [0, (q + q ') / 2] $ at the price of a multiplication by a factor $ 2 $. Then $ k \leq k '$, $ k' $ can never be the $ \min $ and only four cases remain: 
\begin {itemize} 
\item [(1)] $ q '\geq q, k > q $: the contribution to $ \bar {\Gamma} _q $ is 
%alice088 
\be
\label{eq:088}
\bar{\Gamma}_1(q) = \frac{8\pi q^2}{3\bar{n}_q^\ell} \int_q^{+\infty} \dd q'\, q'(1+\bar{n}_{q'}^\ell) f_1(q,q')
\ \ \ \mbox{with}\ \ \ f_1(q,q')=\int_q^{(q+q')/2} \dd k\, k k' \bar{n}_k^\ell \bar{n}_{k'}^\ell = \frac{1}{2} \int_q^{q'} \dd k\, k k' \bar{n}_k^\ell \bar{n}_{k'}^\ell
\ee
where we have this time used the symmetry $ k \leftrightarrow k '$ to double the integration interval. 
\item [(2)] $ q '\geq q, k <q $: the contribution to $ \bar {\Gamma} _q $ equals 
%alice090 
\be
\label{eq:090}
\bar{\Gamma}_2(q) = \frac{8\pi q^2}{3\bar{n}_q^\ell} \int_q^{+\infty} \dd q'\, q'(1+\bar{n}_{q'}^\ell) f_2(q,q')
\ \ \ \mbox{with}\ \ \ 
f_2(q,q')=\frac{1}{q^3} \int_0^{q} \dd k\, k^4 k' \bar{n}_k^\ell \bar{n}_{k'}^\ell
\ee
\item [(3)] $ q' \leq q, k < q '$: the contribution to $ \bar {\Gamma} _q $ equals 
%alice092 
\be
\label{eq:092}
\bar{\Gamma}_3(q)=\frac{8\pi q^2}{3\bar{n}_q^\ell} \int_0^q \dd q'\, q'(1+\bar{n}_{q'}^\ell) f_3(q,q')
\ \ \ \mbox{with}\ \ \ f_3(q,q')=\frac{1}{q^3} \int_0^{q'} \dd k\, k^4 k' \bar{n}_k^\ell \bar{n}_{k'}^\ell
\ee
\item [(4)] $ q' \leq q, k> q '$: the contribution to $ \bar {\Gamma} _q$ is  
%alice094 
\be
\label{eq:094}
\bar{\Gamma}_4(q)=\frac{8\pi q^2}{3\bar{n}_q^\ell} \int_0^q \dd q'\, q'(1+\bar{n}_{q'}^\ell) f_4(q,q')
\ \ \ \mbox{with}\ \ \ f_4(q,q')=\frac{q'^3}{q^3} \int_{q'}^{(q+q')/2} \dd k\, k k' \bar{n}_k^\ell \bar{n}_{k'}^\ell
\ee
\end {itemize} 

The second step is to evaluate the functions $f_n(q,q')$. We find that 
%alice096 
\be
\label{eq:096}
f_3(q,q') = \left(\frac{q'}{q}\right)^3 f_2(q',q) \ \ \ \mbox{and}\ \ \ f_4(q,q')=\left(\frac{q'}{q}\right)^3 f_1(q',q)
\ee
It remains to calculate $ f_1 $ and $ f_2 $. Let's explain with $ f_1 $ how to proceed. One performs a partial fraction decomposition of the integrand of $ f_1 (q, q ') $ with respect to the variable $ X = \eee ^ k $ (the $ k $ monomial in the numerator is treated as a constant) after translation of the integration variable by $\nu$ to get 
%alice097 
\be
\label{eq:097}
f_1(q,q')=\frac{1/2}{\eee^{q+q'-2\nu}-1} \int_{q-\nu}^{q'-\nu} \dd k\, \left[\frac{(k+\nu)(q+q'-\nu-k)}{\eee^k-1}-\frac{(k+\nu)(q+q'-\nu-k)}{\eee^{k-(q+q'-2\nu)}-1}\right]
\ee
In the integral of the second term, we make the change of variable $ k \to q + q'-2 \nu-k $ to bring it back to 
%alice098 
\be
\label{eq:098}
f_1(q,q')=\frac{1}{\eee^{q+q'-2\nu}-1} \left[\frac{1}{2}\int_{q-\nu}^{q'-\nu} \dd k\, (q+q'-\nu-k)(\nu+k) +\int_{q-\nu}^{q'-\nu} \dd k\, \frac{(k+\nu)(q+q'-\nu-k)}{\eee^k-1}\right]
\ee
that we know how to calculate since we know a primitive of $ k ^ n / (\eee ^ k-1) $ on $ [0,1], \forall n \in \mathbb {N} $: \footnote {One may take the derivative of equation (\ref {eq:099}) with respect to $ k $ and  use $ \frac {\dd} {\dd t} [g_ \alpha (\eee ^ {- t})] = - g _ {\alpha-1} (\eee ^ {- t}) $ and $ g_0 (\eee ^ {- t}) = 1 / (\eee ^ t-1) $.} 
%alice099 
\be
\label{eq:099}
\int\dd k \frac{k^n/n!}{\eee^k-1} = -\sum_{s=0}^{n} \frac{k^s}{s!} g_{n+1-s}(\eee^{-k})
\ee

\subsection {Case of the integral kernel of the hermitian form of the linearized kinetic equations} 

Let us start from the expression (\ref {eq:044}) of the integral kernel as the sum of two double integrals $ I $ and $ II $. In the first integral, we integrate trivially over $ k '$ thanks to the Dirac $ \delta $ function ($ k' = q + q'-k $ and $ k \leq q + q '$); by writing \footnote {Bose's law obeys $ 1 + \bar {n} _k ^ \ell = \eee ^ {k- \nu} \bar {n} ^ \ell_k $ so $ \phi (k) \phi (k ') = \eee ^ {(k + k') / 2- \nu} kk '\bar {n} _k ^ \ell \bar {n} _ {k'} ^ \ell $ in full generality.} 
%alice100 
\be
\label{eq:100}
\phi(k)\phi(k')\stackrel{{\rm in}\, I}{=} k(q+q'-k)\, \eee^{(q+q')/2-\nu} \bar{n}_k^\ell \bar{n}_{k'}^\ell
\ee
we get back to the same integral over $ k $ as in $ \bar {\Gamma} _q $. In the second double integral, we choose $ q \geq q '$ (without loss of generality given the Hermitian symmetry) and we integrate trivially over $ k' $ thanks to the Dirac $ \delta $ function ($ k '= k + q-q '\geq k $) to obtain
%alice101 
\be
\label{eq:101}
\langle q|\hat{V}_{II}|q'\rangle \underset{q'\leq q}{=} -\frac{8\pi}{3} \int_0^{+\infty} \dd k\, k(k+q-q') \eee^{k+\frac{q-q'}{2}-\nu} \bar{n}_k^\ell \bar{n}_{k+q-q'}^{\ell} [\min(q',k)]^3
\ee
It remains to distinguish $ k \leq q' $ and $ k \geq q '$, to perform a partial fraction decomposition of the integrand vis-\`a-vis $ X = \eee ^ k $ then use (\ref {eq:099}). We finally find \footnote {The inversion between $ \max $ and $ \min $ in equation (\ref {eq:102c}) is not a typo but is an arbitrary historical choice.} 
%alice102 
\bea
\label{eq:102a}
\langle q|\hat{V}|q'\rangle &=& \langle q|\hat{V}_I |q'\rangle + \langle q|\hat{V}_{II} |q'\rangle \\
\label{eq:102b}
\langle q|\hat{V}_{I}|q'\rangle &=& \frac{4\pi}{3\sh\left(\frac{q+q'}{2}-\nu\right)} \, F(\min(q,q'),\max(q,q')) \\
\label{eq:102c}
\langle q|\hat{V}_{II}|q'\rangle &=&  -\frac{4\pi}{3\sh \left|\frac{q-q'}{2}\right|} \, G(\max(q,q'),\min(q,q'))
\eea
where the function $ F $ is that of equation (\ref {eq:087}) and the function $ G $ is given by 
%alice103 
\bea
\nonumber
G(q,q') &\!\!\!\!\!=\!\!\!\!\!& 3q q'^3[g_2(\eee^{\nu-q})-g_2(\eee^{\nu-q'})] + 6 q'^2 (2q+q') [g_3(\eee^{\nu-q})-g_3(\eee^{\nu-q'})] \\
\nonumber
&\!\!\!\!\!+\!\!\!\!\!& 12 q'(2q+3q') [g_4(\eee^{\nu-q})-g_4(\eee^{\nu-q'})] +24 (q-q') [g_5(\eee^\nu)-g_5(\eee^{\nu+q'-q})] \\
\label{eq:103a}
&\!\!\!\!\!+\!\!\!\!\!& 24 (q+4q') [g_5(\eee^{\nu-q})-g_5(\eee^{\nu-q'})] +120 [g_6(\eee^\nu)+g_6(\eee^{\nu-q})-g_6(\eee^{\nu-q'})-g_6(\eee^{\nu-q+q'})]
\eea
Considered as a function of $ q '$, $ \langle q | \hat {V} | q '\rangle $ is continuous but has a kink in $ q' = q $ because of contribution $ II $. 

\subsection {Applications} 

Let us give without proof some Taylor or asymptotic expansions of the decay rate $ \bar {\Gamma} _q $ and the integral kernel {$ \langle q | \hat {V} | q '\rangle $}. 

\subsubsection {$ \bar {\Gamma} _q $ at low $ q $} 

We have to distinguish the cases $ \nu <0, \nu = 0 $ and $ \nu $ infinitesimal. 
\begin {itemize} 
\item Case $ \nu <0 $: $ \bar {\Gamma} _q $ has the Taylor expansion 
%alice104 
\be
\boxed{
\label{eq:104}
\bar{\Gamma}(q)\underset{q\to 0}{\stackrel{\nu<0}{=}} C(\nu)q^2 [1+\alpha(\nu)q+\beta(\nu)q^2+O(q^3)]
}
\ee
with 
%alice105 
\bea
\label{eq:105a}
C(\nu) &=& \frac{4\pi}{3\bar{n}_0} J_0(\nu) \\
\label{eq:105b}
\alpha(\nu) &=& \bar{n}_0+1+\frac{J_1(\nu)}{J_0(\nu)} \\
\label{eq:105c}
\beta(\nu) &=& (\bar{n}_0+1) \left[\frac{1}{2}+\frac{J_1(\nu)}{J_0(\nu)}\right] + \frac{J_2(\nu)}{2 J_0(\nu)} -\frac{1}{5} \bar{n}_0 \frac{g_2(\eee^\nu)}{J_0(\nu)}
\eea
Here, 
%alice106 
\be
\label{eq:106}
J_n(\nu) = \int_0^{+\infty} \dd k\, \frac{f^{(n)}(k)}{1-\eee^{-k}} \left\{k[g_2(\eee^\nu)-g_2(\eee^{\nu-k})]+2 [g_3(\eee^\nu)-g_3(\eee^{\nu-k})]\right\}
\ee
$ f ^ {(n)} (k) $ being the $ n ^ {\scriptsize \mbox {th}} $ derivative of the $ f (k) = k \bar {n} _k ^ \ell $ function with respect to $ k $. 
\item Case $ \nu = 0 $: $ \bar {\Gamma} _q $ has the Taylor expansion 
%alice107 
\be
\boxed{
\label{eq:107}
\bar{\Gamma}(q)\underset{q\to 0}{\stackrel{\nu=0}{=}} C_0 q^3 [1+\alpha_0 q+\beta_0 q^2+O(q^3)]
}
\ee
with 
%alice108 
\bea
\label{eq:108a}
C_0 &=& \frac{4\pi}{3} J_0(0)=\frac{16\pi^5}{135} \\
\label{eq:108b}
\alpha_0 &=& \frac{J_1(0)-\zeta(2)}{J_0(0)} +\frac{1}{2} = -\frac{15}{8\pi^2} -\frac{45\zeta(3)}{4\pi^4}\\
\label{eq:108c}
\beta_0 &=& \frac{\frac{7}{10}+J_1(0)+J_2(0)}{2 J_0(0)} +\frac{1}{6} = \frac{3}{\pi^4} +\frac{5}{16\pi^2}
\eea
\item Case $ \nu $ infinitesimal: for $ \nu $ tending to zero in $ \bar {\Gamma} _q $ at fixed $ Q = q / | \nu | $, we get the expansion (\ref {eq:059b}) with 
%alice109 
\be
\label{eq:109}
\Phi(Q)=\frac{15}{2\pi^2} \left\{\left[-\frac{1}{4}-\frac{3\zeta(3)}{2\pi^2}\right] Q +\left(1+\frac{1}{Q^3}\right)\ln(1+Q)-\frac{1}{Q^2}+\frac{1}{2Q}-\frac{1}{3}\right\}
\ee
Our motivation for calculating $ \Phi (Q) $ was the following finding: the low $ q $ expansion 
%alice110 
\be
\label{eq:110}
\frac{\eee^{\nu/2} [\bar{n}_q^\ell(1+\bar{n}_q^\ell)]^{1/2} \bar{\Gamma}_q}{\frac{4\pi}{3}q^2 J_0(\nu)} \underset{q\to 0}{\stackrel{\nu\scriptsize\ \mbox{fixed}}{=}} 1 +\check{\alpha}(\nu) q + O(q^2)
\ee
exhibits a discontinuous coefficient $ \check {\alpha} (\nu) $ in $ \nu = 0 $. The calculation is simple. On the one hand, $ \check {\alpha} (0) = \alpha_0 $. On the other hand, for $ \nu> 0 $, $ \check {\alpha} (\nu) = \alpha (\nu) - (\bar {n} _0 + 1/2) = 1/2 + J_1 (\nu) / J_0 (\nu) $, and we take the limit $ \nu \to 0 ^ - $ noting that $ J_0 (\nu) $ is continuous 
($ J_0 (\nu) \underset {\nu \to 0 ^ -} {\to} J_0 (0) = 4 \pi ^ 4/45 $) while $ J_1 (\nu) $ is not ($ J_1 (\nu) \underset {\nu \to 0 ^ -} {\to} J_1 (0) +2 \zeta (2) $), \footnote {In other words, to get this limit, it is incorrect to make $ \nu $ tend to zero under the integral sign in $ J_1 (\nu) $, while it is legitimate in $ J_0 (\nu) $. Indeed, for $ k \approx | \nu | $, $ f (k) \simeq k / (k + | \nu |) $ is uniformly bounded but $ f '(k) \simeq | \nu | / (k + | \nu |) ^ 2 $ is not. We must treat separately in $ J_1 (\nu) $, for $ \varepsilon \ll 1 $ fixed, the integration subinterval $ [\varepsilon, + \infty [$ (on which we can make $ \nu \to 0 ^ - $ at fixed $ k $) and the sub-interval $ [0, \varepsilon] $ (on which we must make $ \nu \to 0 ^ - $ at fixed  $ K = k / | \nu | $); the bit $ \int_0 ^ \varepsilon \dd k $ contributes in the limit $ \nu \to 0 ^ - $ by the value $ 2 \zeta (2) = \pi ^ 2/3 $.} so that 
%alice111 
\be
\label{eq:111}
\check{\alpha}(\nu) \underset{\nu\to 0^-}{\to} \check{\alpha}(0^-)=\frac{15}{4\pi^2}-\frac{45\zeta(3)}{4\pi^4} \neq \check{\alpha}(0) = -\frac{15}{8\pi^2}- \frac{45\zeta(3)}{4\pi^4}
\ee
In expansion (\ref {eq:110}) we therefore can not exchange the limits $ q \to 0 $ and $ \nu \to 0 ^ - $. To get the coefficient $ \check {\alpha} (0 ^ -) $, we have to make $ q $ tend to zero faster than $ \nu $ so $ Q \to 0 $ in equation (\ref {eq:109}). To get $ \check {\alpha} (0) $, we have to make $ \nu $ tend to zero faster than $ q $ so $ Q \to + \infty $ in equation (\ref {eq:109}). And one has indeed, 
%alice112 
\be
\label{eq:112}
\Phi(Q)\underset{Q\to 0}{\sim} \check{\alpha}(0^-)Q \ \ \ \mbox{and}\ \ \ \Phi(Q)\underset{Q\to +\infty}{\sim} \check{\alpha}(0) Q
\ee
The function $ \Phi (Q) $ allows one to connect the two limiting cases $ 0 <q \ll | \nu | \ll 1 $ and $ 0 <| \nu | \ll q \ll 1 $. 
\end {itemize} 

\subsubsection {$\bar{\Gamma}_q$ at large $q$} 
We discovered that the asymptotic series in the variable $ 1 / q $ has a finite number of nonzero terms which results in an exponentially small error, for $ \nu $ being zero or $ <0 $: 
%alice113 
\be
\boxed{
\label{eq:113}
\bar{\Gamma}_q\underset{q\to +\infty}{=} \frac{8\pi}{3q} \Big[\sum_{n=0}^3 q^n c_n(\nu) +O\left(q^4\eee^{-q}\right)\Big]
}
\ee
with 
%alice114 
\bea
\label{eq:114a}
c_3 &=& 2\, g_5(\eee^\nu) \\
\label{eq:114b}
c_2 &=& 30\, g_6(\eee^\nu) \\
\label{eq:114c}
c_1 &=& -36\, g_7(\eee^\nu) + 2\int_0^{+\infty} \dd q\, q^4 g_0(\eee^{\nu-q}) [g_2(\eee^{\nu-q}) +q g_1(\eee^{\nu-q})] \\
\label{eq:114d}
c_0 &=& -252\, g_8(\eee^\nu)
\eea

\subsubsection {$ \langle k | \hat {V} | q \rangle $ at low $ q $} 
At $ \nu $ fixed to zero or to a $ <0 $ value, we take $ k = O (1) $ not necessarily fixed when $ q \to 0 $. \footnote {In practice, we treat the cases $ q \to 0 $ at fixed $ k $ and $ q \to 0 $ at fixed $ K = k / q $, then we check that they are well represented by the single expression (\ref {eq:072}, \ref {eq:115}).} We must therefore distinguish the cases $ k <q $ and $ k> q $ in equation (\ref {eq:102a}) applied to $ \langle k | \hat {V} | q \rangle $. We then prove the equivalent (\ref {eq:072}) with 
%alice115 
\bea
\nonumber
\mathcal{W}(k,\nu) &=& \frac{4\pi}{3\sh(\frac{k}{2}-\nu)}  \left\{\frac{k^3}{12} + k[g_2(\eee^\nu)+g_2(\eee^{\nu-k})] + 2 [g_3(\eee^{\nu-k})-g_3(\eee^\nu)] \right\} \\
\label{eq:115}
&-& \frac{4\pi}{3\sh\frac{k}{2}} \left\{ k[g_2(\eee^\nu)-g_2(\eee^{\nu-k})]+2[g_3(\eee^\nu)-g_3(\eee^{\nu-k})]\right\}
\eea
This expression is extended by continuity to $ \mathbb {R} ^ + \times \mathbb {R} ^ - $, the numerator of the first (second) contribution tending quadratically (linearly) to zero where its denominator (always linearly) vanishes. In particular, $ \lim_ {(k, \nu) \to (0,0)} \mathcal {W} (k, \nu) \equiv \mathcal {W} (0,0) = -8 \pi ^ 3/9 $. The function $ \mathcal {W} (k, \nu) $ so extended is bounded. Moreover, when $ \nu \to 0 ^ - $ and $ (k, q) \to (0,0) $ at fixed $ k / | \nu | = K $ and $ q / | \nu | = Q $, we have checked that equation (\ref {eq:072}) always holds, $ \mathcal {W} (k, \nu) $ being replaceable by $ \mathcal {W} (0,0) $. 

\subsubsection {$ \langle q | \hat {V} | q '\rangle $ at large wave number} 
Let's start from equation (\ref {eq:102a}). If $ q '\to + \infty $ for fixed $ q $, $ \langle q | \hat {V} | q' \rangle $ exponentially tends to zero, it is $ O (q '^ 3 \eee ^ {- q / 2}) $. If $ q $ and $ q '$ both tend to infinity, the contributions $ I $ and $ II $ behave differently: 
%alice116 
\bea
\label{eq:116a}
\langle q|\hat{V}_{I}|q'\rangle &\underset{q,q'\to +\infty}{\sim}& \frac{2\pi\,\eee^\nu}{9} q^3 q'^3 \eee^{-(q+q')/2} \left[1+O\left(\frac{\min(q,q')}{\max(q,q')}\right)\right] \\
\label{eq:116b}
\langle q|\hat{V}_{II}|q'\rangle &\underset{q,q'\to +\infty}{\sim}& \frac{-32\pi}{\sh\left|\frac{q-q'}{2}\right|}\left\{|q-q'|[g_5(\eee^\nu)-g_5(\eee^{\nu-|q-q'|})] +5[g_6(\eee^\nu)-g_6(\eee^{\nu-|q-q'|})]\right\}
\eea
This justifies the distinction made in note \ref {note:distinguishedo}. 

\section {Long-time sub-ballistic variance of the condensate phase shift in the concave case} 
\label {ann:II} 

The reduced sub-ballistic variance $ \bar {\mathcal {V}} (\bar {t}) $ of the condensate phase shift in the presence of low-temperature Landau-Khalatnikov processes is given in the exact theory by equation (\ref {eq:069}) and in the rate approximation by equation (\ref {eq:048}). In both cases, it takes the form 
%alice117 
\be
\label{eq:117}
\bar{\mathcal{V}}(\bar{t}) = \int_0^{+\infty}\dd\bar{q}\, \frac{f(\bar{q})}{\bar{\Gamma}(\bar{q})^2} \left[\eee^{-\bar{\Gamma}(\bar{q})\bar{t}}-1+\bar{\Gamma}(\bar{q})\bar{t}\right]
\ee
where $ f (\bar {q}) $ is a regular function, with a Taylor expansion in $ \bar {q} = 0 $. The goal is to calculate the asymptotic expansion of $ \bar {\mathcal {V}} (\bar {t}) $ for $ \bar {t} \to + \infty $. This is done by distinguishing the cases $ \nu <0 $ and $ \nu = 0 $ ($ \nu = \mu_ \phi / k_B T $, where $ \mu_ \phi $ is the chemical potential of the phonon gas) which lead to different behaviors for $ f (\bar {q}) $ and $ \bar {\Gamma} (\bar {q}) $ in the neighborhood of $ \bar {q} = 0 $. In both cases, $ \bar {\Gamma} (\bar {q}) $ is a strictly increasing function of $ \bar {q} $ tending to $ + \infty $ when $ \bar {q} \to + \infty $, \footnote {Actually, as soon as $ \bar {\Gamma} (\bar {q}) $ is a strictly increasing function on a neighborhood of $ \bar {q} = 0 $ and remains elsewhere at a non-zero distance from zero, our results apply.} 
which will make it possible to perform a simplifying change of variable in equation (\ref {eq:117}). 

\subsection {Case $ \nu <0 $} 
$ \bar {\Gamma} (\bar {q}) $ vanishes quadratically in $ \bar {q} = 0 $ (see equation (\ref {eq:049})) and the same for $ f (\bar {q}) $ (see equation (\ref {eq:050}) for the rate approximation and equation (\ref {eq:074}) for the exact theory). We can separate in equation (\ref {eq:117}) the linear-in-time diffusive part (it comes from the term $ \bar {\Gamma} (\bar {q}) \bar {t} $ between brackets). In the rest, we take as the new integration variable 
%alice118 
\be
\label{eq:118}
\Omega=[\bar{\Gamma}(\bar{q})]^{1/2}
\ee
the exponent $ 1/2$ ensuring that $ \Omega $ is to leading order a linear function of $ \bar {q} $ at low $ \bar { q} $. Then $ f (\bar {q}) / \bar {\Gamma} (\bar {q}) ^ 2 $ diverges as $ 1 / \Omega ^ 2 $ at the origin and $ \bar {\mathcal {V} } (t) $ can be written as
%alice119 
\be
\label{eq:119}
\bar{\mathcal{V}}(\bar{t})=\bar{D}\bar{t}+\int_0^{+\infty} \dd\Omega\, \frac{F(\Omega)}{\Omega^2}(\eee^{-\Omega^2\bar{t}}-1)
\ee
with 
%alice120 
\be
\label{eq:120}
\bar{D}=\int_0^{+\infty}\dd\bar{q}\, \frac{f(\bar{q})}{\bar{\Gamma}(\bar{q})} \ \ \ \mbox{and}\ \ \ F(\Omega)=\frac{f(\bar{q})}{\frac{\dd\Omega}{\dd\bar{q}}\bar{\Gamma}(\bar{q})}=\frac{2f(\bar{q})}{[\bar{\Gamma}(\bar{q})]^{1/2}\bar{\Gamma}'(\bar{q})}
\ee
$ F (\Omega) $ has in $ \bar {q} = \Omega = 0 $ a Taylor expansion in powers of $ \bar {q} $ and also in powers of $ \Omega $: 
%alice121 
\be
\label{eq:121}
F(\Omega) \underset{\Omega\to 0}{=} F(0)+\Omega F'(0) +O(\Omega^2)
\ee
with $ F (0) \neq 0 $. Let's introduce an arbitrary cut-off $ \bar {q} _c> 0 $ on the wavenumber $ \bar {q} $ so a cut-off $ \Omega_c = [\bar {\Gamma} (\bar {q} _c)] ^ {1/2} $ on $ \Omega $. In the integral (\ref {eq:119}), the upper contribution has a finite limit when $ \bar {t} \to + \infty $, obtained by neglecting the exponential term in time: this term has an upper bound $ \eee ^ {- \Omega_c ^ 2 \bar {t}} $ and can be neglected without triggering any divergence in the integral ($ \int _ {\Omega_c} ^ {+ \infty} \dd \Omega \, F (\Omega) / \Omega ^ 2 <\infty $). In the lower contribution, we first have to separate out the affine approximation (\ref {eq:121}) from $ F (\Omega) $ before we can neglect the exponential term: \footnote {\label {note:rac} In this case, we make an error $ O (\int_0 ^ {+ \infty} \dd \Omega \, \eee ^ {- \Omega ^ 2 \bar {t}}) = O (\bar {t} ^ {- 1/2}) $, instead of an exponentially small error in time.} 
%alice122 
\begin{multline}
\label{eq:122}
\int_0^{\Omega_c} \dd\Omega\, \frac{F(\Omega)}{\Omega^2}(\eee^{-\Omega^2\bar{t}}-1) \underset{\bar{t}\to+\infty}{=} 
\int_0^{\Omega_c} \dd\Omega\, \frac{F(\Omega)-F(0)-F'(0)\Omega}{\Omega^2} (-1) \\
+F(0) \int_0^{\Omega_c} \dd\Omega\, \frac{\eee^{-\Omega^2\bar{t}}-1}{\Omega^2}
+F'(0) \int_0^{\Omega_c} \dd\Omega\, \frac{\eee^{-\Omega^2\bar{t}}-1}{\Omega}+o(1)
\end{multline}
We then give the results (the notation $ O (X ^ {- \infty}) $ indicates a rapidly decreasing function) 
%alice123 
\bea
\label{eq:123a}
\int_0^X \dd x\, \frac{\eee^{-x^2}-1}{x^2} &\underset{X\to+\infty}{=}& -\pi^{1/2} + \frac{1}{X} +O(X^{-\infty}) \\
\label{eq:123b}
\int_0^X \dd x\, \frac{\eee^{-x^2}-1}{x} &\underset{X\to+\infty}{=}& -\ln X -\frac{1}{2}\gamma_{\rm Euler} + O(X^{-\infty})
\eea
to which we are reduced by the change of variable $ \Omega = x / \bar {t} ^ {1/2} $. We thus find an asymptotic expansion of the form (\ref {eq:052}, \ref {eq:062}) with coefficients 
%alice124 
\bea
\label{eq:124a}
A(\nu)&\!\!\!\!=\!\!\!\!& -F(0) \\
\label{eq:124b}
B(\nu) &\!\!\!\!=\!\!\!\!& -F'(0) \\
\label{eq:124c}
E(\nu)&\!\!\!\!=\!\!\!\!&-\int_0^{\Omega_c}\!\! \dd\Omega\, \frac{F(\Omega)-F(0)-F'(0)\Omega}{\Omega^2} -\int_{\Omega_c}^{+\infty}\!\! \dd\Omega\, \frac{F(\Omega)}{\Omega^2}
+\frac{F(0)}{\Omega_c} -F'(0)\left(\ln\Omega_c +\frac{1}{2}\gamma_{\rm Euler}\right)
\eea
It remains to make the link with the original integration variable $ \bar {q } $ of equation (\ref {eq:117}). The calculation shows that 
%alice125 
\be
\label{eq:125}
\frac{\bar{q}^2 f(\bar{q})}{\bar{\Gamma}(\bar{q})^2} \underset{\bar{q}\to 0}{=} F(0) C(\nu)^{-1/2} +\bar{q} F'(0) + O(\bar{q}^2)
\ee
where $ C (\nu) = \lim _ {\bar {q} \to 0} \bar {\Gamma} (\bar {q}) / \bar {q} ^ 2 $; this equation (\ref {eq:125}), written in the main text as (\ref {eq:051}), is convenient because it gives access to $ F (0) $ and $ F '(0) $ without ever going through the $ \Omega $ variable. We obtain equations (\ref {eq:053b}, \ref {eq:053c}). Returning to the variable $ \bar {q} $ in equation (\ref {eq:124c}) we find 
%alice126 
\begin{multline}
\label{eq:126}
E(\nu)=-\int_0^{\bar{q}_c} \dd\bar{q}\, \left[\frac{f(\bar{q})}{\bar{\Gamma}(\bar{q})^2}- \frac{F(0)C(\nu)^{-1/2}+\bar{q}F'(0)}{\bar{q}^2}\right] - \int_{\bar{q}_c}^{+\infty} \dd\bar{q}\,\frac{f(\bar{q})}{\bar{\Gamma}(\bar{q})^2}\\
+ F(0) C(\nu)^{-1/2} \left[\frac{1}{\bar{q}_c}-\frac{1}{2}\alpha(\nu)\right]-F'(0) \left\{\ln[\bar{q}_cC(\nu)^{1/2}]+\frac{1}{2}\gamma_{\rm Euler}\right\}
\end{multline}
knowing that 
%alice127 
\be
\label{eq:127}
\lim_{\bar{q}\to 0} \frac{1}{[\bar{\Gamma}(\bar{q})]^{1/2}}-\frac{1}{\bar{q}C(\nu)^{1/2}} = -\frac{\alpha(\nu)}{2C(\nu)^{1/2}}  \ \ \ \mbox{and}\ \ \ 
\lim_{\bar{q}\to 0} \ln\bar{q}-\frac{1}{2}\ln\bar{\Gamma}(\bar{q}) = -\frac{1}{2} \ln C(\nu)
\ee
according to equation (\ref {eq:049}). This proves equation (\ref {eq:053d}). 

\subsection {Case $ \nu = 0 $} 

In this case, $ \bar {\Gamma} (\bar {q}) $ vanishes cubically at $ \bar {q} = 0 $ (see equation (\ref {eq:054})) and $ f (\bar {q}) $ does not vanish. To have a new integration variable $ \Omega $ that is linear in $ \bar {q} $ to leading order, we set 
%alice128 
\be
\label{eq:128}
\Omega=[\bar{\Gamma}(\bar{q})]^{1/3}
\ee
Then $ f (\bar {q}) / \bar {\Gamma} (\bar {q}) ^ 2 $ diverges as $ 1 / \Omega ^ 6 $ at the origin, we can no longer pull out a diffusive term and $ \bar {\mathcal {V}} (\bar {t}) $ reads
%alice129 
\be
\label{eq:129}
\bar{\mathcal{V}}(\bar{t})=\int_0^{+\infty} \dd\Omega\, \frac{F(\Omega)}{\Omega^6} \left(\eee^{-\Omega^3\bar{t}}-1+\Omega^3\bar{t}\right)
\ \ \ \mbox{with}\ \ \ F(\Omega)=\frac{f(\bar{q})}{\dd\Omega/\dd\bar{q}}
\ee
In the rate approximation, $ F (\Omega) $ has at $ \Omega = 0 $ a Taylor expansion 
%alice130 
\be
\label{eq:130}
F(\Omega)=F(0)+\Omega F'(0) +\frac{\Omega^2}{2} F''(0) +O(\Omega^3)
\ee
with $ F (0) $ nonzero. We proceed then as for the case $ \nu <0 $. Using the results 
%alice131 
\bea
\label{eq:131a}
\int_0^X \dd x\, \frac{\eee^{-x^3}-1+x^3}{x^n} &\underset{X\to +\infty}{\to}& \frac{1}{3}\Gamma\left(\frac{1-n}{3}\right) = \left\{\begin{matrix}\frac{3}{10}\Gamma(\frac{1}{3}) \ &\mbox{if}&\ n=6 \\ && \\ \frac{3}{4}\Gamma(\frac{2}{3})\ &\mbox{if}&\ n=5\end{matrix}\right.  %}
\\
\label{eq:131b}
\int_0^X \dd x\, \frac{\eee^{-x^3}-1+x^3}{x^4} &\underset{X\to +\infty}{=}& \ln X + O(1)
\eea
we find 
%alice132 
\be
\label{eq:132}
\bar{\mathcal{V}}(\bar{t}) \underset{\bar{t}\to+\infty}{=} \frac{3}{10}\Gamma(1/3) F(0) \bar{t}^{5/3} + \frac{3}{4} \Gamma(2/3) F'(0) \bar{t}^{4/3} +\frac{1}{2} F''(0) \bar{t} \ln(\bar{t}^{1/3}) + O(\bar{t})
\ee
which proves equations (\ref {eq:056}, \ref {eq:057}). In the exact theory, we think that $ F '(0) $ exists but we do not know its value, and we have not proved that $ F''(0)$ exists.

%%%%%%%%%%%%%%%%%%%%%%%%%%%

\end{document}